\documentclass[useAMS,usenatbib]{mn2e}

\usepackage{graphicx}
\usepackage{dcolumn}
\usepackage{bm}
\usepackage{amsmath}
\usepackage{amssymb}

\title[The use of Kepler solver]{The use of Kepler solver in numerical integrations of quasi-Keplerian orbits}
 \author[Deng  et al.]{Chen Deng$^{1}$, 
Xin Wu$^{1,2}$\thanks{E-mail: xinwu@gxu.edu.cn,
wuxin$\_$1134@sina.com}
and Enwei Liang$^{1,2}$ 
\\  $^{1}$School of Physical Science and Technology, Guangxi University,
Nanning 530004, China
\\ $^{2}$Guangxi Key Laboratory for
Relativistic  Astrophysics, Nanning 530004, China}
\begin{document}
\date{Accepted (~~~~~);   Received (~~~~~)}
\pagerange{\pageref{firstpage}--\pageref{lastpage}} \pubyear{2020}
\maketitle \label{firstpage}
\begin{abstract}

A Kepler solver is an analytical method used to solve a two-body
problem. In this paper, we propose a new correction method by
slightly modifying the Kepler solver. The only change to the
analytical solutions is that the obtainment of the eccentric
anomaly relies on the true anomaly that is associated to a unit
radial vector calculated by an integrator. This scheme rigorously
conserves all integrals and orbital elements except the mean
longitude. However, the Kepler energy, angular momentum vector and
Laplace-Runge-Lenz vector for perturbed Kepler problems are
slowly-varying quantities. However, their integral invariant
relations give the quantities high-precision values that directly
govern five slowly-varying orbital elements. These elements
combined with the eccentric anomaly determine the desired
numerical solutions. The newly proposed method can considerably
reduce various errors for a post-Newtonian two-body problem
compared with an uncorrected integrator, making it suitable for a
dissipative two-body problem. Spurious secular changes of some
elements or quasi-integrals in the outer solar system may be
caused by short integration times of the fourth-order Runge-Kutta
algorithm. However, they can be eliminated in a long integration
time of $10^{8}$ years by the proposed method, similar to
Wisdom-Holman second-order symplectic integrator. The proposed
method has an advantage over the symplectic algorithm in the
accuracy but gives a larger slope to the phase error growth.

\end{abstract}


\section{introduction}

In a relative coordinate system, a pure two-body problem in the
solar system is a system with three degrees of freedom. In this
system, the Kepler energy, relative angular momentum vector and
Laplace-Runge-Lenz (LRL) vector are seven integrals of motion.
However, only five of them are independent due to the seven
quantities satisfying two relations (hereafter, the so-called
seven integrals in the pure two-body problem include
\emph{dependent and independent} integrals). They correspond to
five constant orbital elements. In this case, the problem is
integrable and has an analytical solution.

Regardless of whether $N$-body gravitational problems with $N>2$
perform quasi-Keplerian motions (i.e. the Keplerian motions with
small perturbations; orbits for the quasi-Keplerian motions are
quasi-Keplerian orbits), they are consistently non-integrable and
have no analytical solutions. Numerical integration methods are
convenient tools to solve them. Geometric integrators (Hairer et
al. 1999) can preserve one or more physical/geometric properties
of these systems. The properties contain structures, integrals,
symmetries, reversing symmetries and phase-space volumes.
Symplectic integrators (Ruth 1983; Feng 1985; Wisdom $\&$ Holman
1991; Zhong et al. 2010; Mei et al. 2013a, 2013b) are a class of
geometric integration methods that conserve symplectic structures
and phase-space volumes of Hamiltonian systems and have no secular
drift in energy errors. They are particularly suitable for
studying the qualitative properties on the long-term evolution of
Hamiltonian systems because of these advantages. Symmetrical
methods (Quinlan $\&$ Tremaine 1990), extended phase space methods
(Pihajoki 2015; Liu et al. 2016; Luo et al. 2017; Li $\&$ Wu 2017)
and energy-conserving schemes (Chorin et al. 1978; Feng 1985;
Bacchini et al. 2018a, 2018b; Hu et al. 2019) belong to the
geometric integrators. Although non-geometric integrators, such as
Runge-Kutta methods, do not satisfy such geometric properties,
they have wider applications than the geometric integrators. In
addition, they provide more accurate numerical solutions than the
same order geometric integrators (excluded those from the
non-geometric integrators reformed) in a short integration time.
In this case, their numerical results should be reliable.

However, the non-geometric integrators can be reformed as the
geometric ones by means of some particular treatments. One method
includes one or more integrals in a set of differential equations,
thereby enabling the orbits change from the Lyapunov's instability
to the Lyapunov's stability. Thus, the numerical solutions are
consistently confined  to the integral surface in the phase space
and the fast growth of various numerical errors can typically be
suppressed. This technique is the stabilizing method of Baumgarte
(1972, 1973). Its applications are given in (Ascher et al. 1995;
Chin 1995; Avdyushev 2003). Another stabilization path is the
manifold correction scheme of Nacozy (1971), where stabilizing
terms are directly added to the numerical solutions. This approach
applies Lagrange multipliers to take the integrated orbits back to
the original integral hypersurface along the least-squares
shortest path. In this way, the corrected error of an integral is
the square of the uncorrected one. This condition indicates that
the correction solutions do not rigorously satisfy the
integral\footnote{Although the integral is not rigorously
satisfied, such a correction still makes the integral accurate to
the machine double precision if the uncorrected integrator gives
the machine single precision to  the integral.}. Following this
basic idea, several authors focused on the application and
effectiveness of manifold correction methods (Murison 1989; Chin
1995; Zhang 1996; Wu et al. 2006). The steepest descending method
for the approximate consistency of the Kepler energy of the
two-body problem suppresses the fast growth of integration errors
in the semimajor axis (Wu et al. 2007). The approximate
conservation of the seven integrals results in the five constant
elements in the two-body problem (Ma et al. 2008a). In addition to
these manifold correction methods that approximately satisfy the
integrals, methods that rigorously satisfy the integrals have been
developed. For example, the scaling method of Fukushima (2003a)
and the velocity scaling method of Ma et al. (2008b) can exactly
conserve the energy (associated to the semimajor axis) of the
two-body problem. The dual scaling method for the rigorous
consistency of the Kepler energy and LRL vector in the two-body
problem is effective to control the growth of integration errors
in the semimajor axis, eccentricity and longitude of pericenter
(Fukushima 2003b). The rotation method for rigorously satisfying
the angular momentum vector of the two-body problem reduces the
growth of integration errors in the inclination and longitude of
the ascending node (Fukushima 2003c). The linear transformation
method of Fukushima (2004) is the best among Fukushima's manifold
correction methods because it simultaneously and rigorously
satisfies the Kepler energy, angular momentum vector and LRL
vector in the two-body problem. It can intensively reduce the
integration errors in various orbital elements at small additional
computational cost.

Seven integrals of motion, including the total energy, total
momentum vector and total angular momentum vector, are
consistently present for the quasi-Keplerian motions in a
five-body system of the Sun and four outer planets. Unfortunately,
the constancy of the total energy and total angular momentum does
not exhibit good performance (Hairer et al. 1999). Individual
Kepler energies, angular momentum vectors and LRL vectors must be
corrected. However, these quantities are not constant and should
slowly vary in this case. They are called slowly-varying
quantities. No integrals of motion are available in dissipative
and other nonconservative systems. The above correction methods
become useless in these cases. These slowly-varying individual
quantities obtained from their integral invariant relations
(Szebehely $\&$ Bettis 1971; Huang $\&$ Innanen 1983) have higher
accuracies than those directly determined by the integrated
positions and velocities and can be taken as reference values of
correcting the errors. In this way, the above correction methods
(e.g. Fukushima 2003a, 2003b, 2003c, 2004; Wu et al. 2007; Ma et
al. 2008a, 2008b) remain valid. These correction methods are not
limited to treating conservative quasi-Keplerian problems. Wang et
al. (2016, 2018) confirmed that the velocity scaling method of Ma
et al. (2008b) combined with the integral invariant relation is
effective in enhancing the quality of numerical integrations of
non-Keplerian motions of nonconservative elliptic restricted
three-body problems and dissipative circular restricted three-body
problems.

A perturbed two-body problem or each body of the $N$-body problem
in the relative coordinate system performing a quasi-Keplerian
motion is slightly similar to the two-body problem. At this point,
we shall attempt to apply the analytical solvable method of the
two-body problem (i.e. the Kepler solver) to determine the
quasi-Keplerian motion of the perturbed two-body problem or each
body of the $N$-body problem. On this basis, a new manifold
correction method is proposed for the quasi-Keplerian motion. The
solution of each body still uses the Kepler analytical solvable
form. The five orbital elements of each body relative to the
central body that remain invariant in the unperturbed case are
slowly-varying quantities in the perturbed case. They can be
determined by the seven slowly-varying quantities from their
integral invariant relations. The eccentric anomaly is calculated
by the true anomaly between the LRL vector and a radial vector
rather than the Kepler equation. The newly proposed method is
similar to the linear transformation method of Fukushima (2004)
that rigorously satisfies the seven integrals of the Kepler
energy, angular momentum vector and LRL vector in the two-body
problem. However, the two methods are completely different in the
construction mechanisms. Here, someone does not think that the
conservation of several integrals in the two-body problem is
necessary because of the existence of five independent integrals.
The preservation of five independent integrals is the same as that
of the two other dependent integrals from the theory. However,
this condition may be different from the numerical computation.
Thus, numerically keeping the two other dependent integrals
remains vital.

The rest of this paper is organized as follows. Section 2 presents
a new correction method to rigorously satisfy the seven integrals
of the Kepler energy, angular momentum vector and LRL vector in
the pure two-body problem. The performance of several integrators,
including the linear transformation method of Fukushima (2004) and
a second-order symplectic integrator, is checked, and the related
errors are analysed. Section 3 extends the proposed method to
treat the quasi-Keplerian motions of perturbed two-body problems.
A post-Newtonian two-body problem and a dissipative two-body
problem are taken as test models to verify the performance of the
proposed method. Section 4 applies the proposed method to a
five-body system of the Sun and four outer planets. The
second-order symplectic integrator of Wisdom $\&$ Holman (1991)
and the fourth-order explicit symplectic method of Yoshida (1990)
are considered for comparison. Section 5 provides the main
results. The linear transformation method of Fukushima (2004) is
briefly introduced in Appendix A.

\section{New manifold correction method to a Kepler problem}

Firstly, equations of motion, integrals of motion, orbital
elements and analytical solutions for a two-body problem are
introduced. Then, the Kepler solver is slightly modified as a new
manifold correction method for the consistency of the Kepler
energy, angular momentum vector and LRL vector. Finally, the
numerical performance of the proposed method is verified, and the
related error analysis is given.

\subsection{Kepler problem}

A pure Kepler problem is a two-body problem consisting of  a small
body and  a primary body. The motion of the small body relative to
the primary body is expressed as
\begin{eqnarray}
    K=\frac{v^{2}}{2}-\frac{\mu}{r},
 \end{eqnarray}
where $r=|\textbf{r}|$ represents a radial separation,  $v$ is the
magnitude of relative velocity vector $\textbf{v}$, and
$\mu=G(M+m)$. $G$ is a constant of gravity, and $M$ and $m$ are
the masses of the two bodies. The evolution of $(\textbf{r},
\textbf{v})$ with time $t$ satisfies the following relation
\begin{eqnarray}
  \mathbf{\ddot{r}}=-\frac{\mu}{r^3}\textbf{r},
 \end{eqnarray}
which is equivalent to two first-order differential equations
\begin{eqnarray}
 \mathbf{\dot{r}} &=& \textbf{v}, \nonumber \\
 \mathbf{\dot{v}} &=& -\frac{\mu}{r^3}\textbf{r}.
 \end{eqnarray}

In accordance with Equation (2) or (3), $K$ in Equation (1) is an
integral of motion, called as a Kepler energy. An invariant
angular momentum vector  is also found, which can be expressed as
\begin{eqnarray}
    \textbf{L}=\textbf{r}\times\textbf{v}.
 \end{eqnarray}
In fact, it contains three components, indicating the existence of
three integrals of motion. Let $L$ be the magnitude of the angular
momentum vector, $L=|\textbf{L}|$. Three three components of the
LRL vector
\begin{eqnarray}
   \textbf{P}=\textbf{v}\times\textbf{L}-\frac{\mu\textbf{r}}{r}
 \end{eqnarray}
do not vary with time. Take $P$ as the magnitude of the LRL
vector, $P=|\textbf{P}|$. Seven integrals, labelled as $K$,
$P_{x}$, $P_{y}$, $P_{z}$, $L_{x}$, $L_{y}$ and $L_{z}$, are
presented in this Kepler problem. Because the seven quantities
satisfy two relations
\begin{eqnarray}
 \textbf{L}\cdot\textbf{P} &=& 0, \\
 P^2-2KL^2 &=& \mu^2,
 \end{eqnarray}
only five of them are independent.

The five independent integrals correspond to five invariant
elements of an elliptical orbit, namely, semimajor axis $a$,
eccentricity $e$, inclination $I$, longitude of ascending node
$\Omega$ and argument of pericentre $\omega$. The five orbital
elements can be expressed in terms of the seven integrals as
\begin{eqnarray}
      a &=& -\frac{\mu}{2K}, \\
      e &=& \frac{P}{\mu},
\end{eqnarray}
\begin{eqnarray}
      \cos I &=& \frac{L_{z}}{L}, ~~ \sin I=\sqrt{1-\cos^{2} I}, \\
      \sin\Omega &=& \frac{L_{x}}{L \sin I}, ~~ \cos\Omega = -\frac{L_{y}}{L \sin I}, \\
\sin\omega &=& \frac{P_z}{e\mu\sin I}, \nonumber \\
    \cos\omega  &=& \frac{1}{e\mu}(P_{x}\cos\Omega+P_{y}\sin\Omega).
 \end{eqnarray}
The inclination is consistently in the range $0^{\circ}\leq I\leq
 180^{\circ}$, and the other angles are  in the ranges $0^{\circ}\leq
 \Omega<  360^{\circ}$ and $0^{\circ}\leq
 \omega<  360^{\circ}$. The
 location of $\Omega$ in the orbital plane system is given by
 the signs of $L_{x}$ and $-L_{y}$, and that of $\omega$ is determined by
 the signs of $P_z$ and $(P_{x}\cos\Omega+P_{y}\sin\Omega)$. However,
 a  sixth orbital element is the mean anomaly $M$ that depends on
 time in the following form
 \begin{eqnarray}
   M=M_{0}+nt,
 \end{eqnarray}
where $n=\sqrt{\mu/a^3}$ is a mean motion, and $M_{0}$ is the
initial mean anomaly. The mean anomaly and eccentric anomaly $E$
satisfy the Kepler equation
\begin{eqnarray}
  E-e\sin E=M.
\end{eqnarray}
In accordance with Equations (13) and (14), $M_{0}$ is calculated
by
\begin{eqnarray}
 M_0= E_0-e\sin E_0,
\end{eqnarray}
where the initial eccentric anomaly $E_0$ is obtained from the
relations $e\cos E_0=1-r_0/a$ and $e\sin E_0=\textbf{r}_0\cdot
\textbf{v}_0/(a^2n)$ ($\textbf{r}_0$ and $\textbf{v}_0$ are the
initial position and velocity). The Kepler equation (14) is
usually solved by the Newton-Raphson iteration method.

Equation (2) has an analytical solution
\begin{eqnarray}
  \textbf{r} &=& \frac{a}{e\mu}(\cos E-e)\textbf{P}+a\sqrt{1-e^2}\sin E
  \textbf{Q}, \\
  \textbf{v} &=& -\frac{a^2 n}{re\mu}\sin E \textbf{P}+\frac{a^2 n}{r}\sqrt{1-e^2}\cos E
  \textbf{Q}.
\end{eqnarray}
Here, $\textbf{P}$ in Equation (5) is determined by
\begin{eqnarray}
\textbf{P}=e\mu\left(
               \begin{array}{c}
                 \cos\Omega \cos\omega-\sin\Omega \sin\omega \cos I\\
                 \sin\Omega \cos\omega+\cos\Omega \sin\omega \cos I \\
                 \sin \omega \sin I\\
               \end{array}
             \right), \nonumber
\end{eqnarray}
and $r$ and $\textbf{Q}$ are expressed as
\begin{equation}
  r=a(1-e\cos E),
\end{equation}
\begin{eqnarray}
 \textbf{Q}=\left(
              \begin{array}{c}
                 -\cos\Omega \sin\omega-\sin\Omega \cos\omega \cos I \\
                -\sin\Omega \sin\omega+\cos\Omega \cos\omega \cos I \\
                \cos \omega \sin I \\
              \end{array}
            \right).
\end{eqnarray}
Equations (16) and (17) are a Kepler solver of the two-body
problem. This Kepler solver is given completely by the seven
integrals, namely, $K$, $P_{x}$, $P_{y}$, $P_{z}$, $L_{x}$,
$L_{y}$ and $L_{z}$.

The above-mentioned presentations are some basic concepts and
theories of the elliptical motion of the two-body problem in the
solar system dynamics. Additional details can be found from the
book of Murray $\&$ Dermott (1999).

\subsection{Applying the Kepler solver to correct numerical solutions}

A numerical solution $(\textbf{r}^{\star},\textbf{v}^{\star})$ can
be given at each step when a nongeometric integration method, such
as an explict Runge-Kutta integrator, solves Equation (3). The
energy, angular momentum vector and LRL vector at this step cannot
be equal to their initial values, that is,
$K(\textbf{r}^{\star},\textbf{v}^{\star})\neq
K(\textbf{r}_{0},\textbf{v}_{0})$,
$\textbf{L}(\textbf{r}^{\star},\textbf{v}^{\star})\neq
\textbf{L}(\textbf{r}_{0},\textbf{v}_{0})$, and
$\textbf{P}(\textbf{r}^{\star},\textbf{v}^{\star})\neq
\textbf{P}(\textbf{r}_{0},\textbf{v}_{0})$, or simply denoted as
$K^{\star}\neq K_0$, $\textbf{L}^{\star}\neq \textbf{L}_0$, and
$\textbf{P}^{\star}\neq \textbf{P}_0$ because of various numerical
errors. Equivalently, the numerical values of five orbital
elements $a^{\star}$, $e^{\star}$, $I^{\star}$, $\Omega^{\star}$,
and $\omega^{\star}$ are unlike their initial values $a_0$, $e_0$,
$I_0$, $\Omega_0$, and $\omega_0$. Elements $a$, $e$, $I$,
$\Omega$ and $\omega$ in Equations (16) and (17) are given by
$a^{\star}\rightarrow a$, $e^{\star}\rightarrow e$,
$I^{\star}\rightarrow I$, $\Omega^{\star}\rightarrow \Omega$, and
$\omega^{\star}\rightarrow \omega$. When they are adjusted via
$a_0\rightarrow a$, $e_0\rightarrow e$, $I_0\rightarrow I$,
$\Omega_0\rightarrow \Omega$, and $\omega_0\rightarrow
\omega$,\footnote{In this paper, the notation $A\rightarrow B$
indicates substituting $A$ for $B$.} such an adjusted numerical
solution $(\textbf{r}^{*},\textbf{v}^{*})$ should be more accurate
than the non-adjusted numerical solution
$(\textbf{r}^{\star},\textbf{v}^{\star})$. When $E$ is still
solved from Equation (14), the adjusted numerical solution is
completely the same as the analytical solution and the numerical
integrator becomes useless.

To consider the use of this integrator, we provide another method
on calculating the eccentric anomaly $E$. Its calculation needs
the true anomaly $f$. The details of this method are provided as
follows. Obtain a unit radial vector $\mathbf{\hat{r}^{\star}}
=\textbf{r}^{\star}/r^{\star}$, which is given by the integrator
at each step. Because the true anomaly is the angle between the
two unit vectors $\mathbf{\hat{r}^{\star}}$ and
$\mathbf{P}/(e\mu)$, its cosine reads
 \begin{equation}
   \cos f^{*}=\frac{\mathbf{\hat{r}^{\star}}\cdot
   \mathbf{P}_0}{e_0\mu}.
  \end{equation}
Its sine is written as
\begin{eqnarray}
  \sin f^{*}=\frac{\textbf{S}\cdot \mathbf{\hat{r}^{\star}}}{|\textbf{S}|},
\end{eqnarray}
where $\textbf{S}$ is a constant vector:
 \begin{eqnarray}
    \textbf{S}=\textbf{L}_0\times \textbf{P}_0.
\end{eqnarray}
$f^{*}$  denotes a correction value of $f$ that is determined by
Equations (20) and (21). The eccentric anomaly is expressed in
terms of the true anomaly as
\begin{eqnarray}
 \cos E^{*} &=& \frac{\cos f^{*}+e_0}{1+e_0\cos f^{*}}, \\
  \sin E^{*} &=& \frac{(1-e_0\cos E^{*})\sin f^{*}}{\sqrt{1-e^2_{0}}}.
\end{eqnarray}
From an $(i-1)$th step to  an $i$th step, the solution
$(\textbf{r}^{\star},\textbf{v}^{\star})$ obtained from the
adopted integrator is corrected by
\begin{eqnarray}
  \textbf{r}^{*} &=& \frac{a_0}{e_0\mu}(\cos E^{*}-e_0)\textbf{P}_0 \nonumber \\
  && +a_0\sqrt{1-e^2_0}\sin E^{*} \textbf{Q}_0, \\
  \textbf{v}^{*} &=& -\frac{a^2_0 n_0}{r^{*}e_0\mu}\sin E^{*} \textbf{P}_0 \nonumber \\
  && +\frac{a^2_0 n_0}{r^{*}}\sqrt{1-e_0^2}\cos E^{*}
  \textbf{Q}_0,
\end{eqnarray}
where $r^{*}$ in Equation (18) is written as
\begin{equation}
  r^{*}=a_0(1-e_0\cos E^{*}).
\end{equation}

Equations (25) and (26) provide a new manifold correction scheme
with the use of the Kepler solver, called as a projection method
M1. The corrected solution in Equations (25) and (26) is
approximately the same as the analytical solution in Equations
(16) and (17) with $a=a_0$, $e=e_0$, $I=I_0$, $\Omega=\Omega_0$
and $\omega=\omega_0$. A slight difference between the Kepler
solver and the new correction method is the eccentric anomaly
calculated in different methods. For the corrected solution, a
certain numerical integrator must be used to give a value of unit
radial vector $\mathbf{\hat{r}^{\star}}$ in Equations (20) and
(21), and an iteration method is not necessary to calculate the
eccentric anomaly. However, such a computation of the radial
vector for the analytical solution is unnecessary and an iteration
method must be frequently used to solve the eccentric anomaly from
the Kepler equation (14). The corrected solutions rigorously
conserve the seven integrals in the entire course of numerical
integrations, similar to the analytical solutions, whereas the
uncorrected ones do not. However, this finding does not indicate
that the corrected solutions and the analytical ones can achieve
the same numerical accuracy. This condition is because the unit
radial vector is given numerically, and the accuracy of the
corrected solutions is decreased compared with that of the
analytical solutions. This is also based on the fact that the
exact preservation of the seven integrals is a necessary but
insufficient condition for the corrected solutions having high
accuracies. Although the unit radial vector at the beginning of a
correction is the same as that at the end of this correction, it
will be adjusted along the original integral hypersurface in the
next step integration.

The newly proposed method is explicitly unlike the existing linear
transformation method of Fukushima (2004) for rigorously
conserving the Kepler energy, angular momentum vector and LRL
vector. We call the existing method M2, which is briefly described
in Appendix A. The two correction methods have three explicit
differences in their constructions. Firstly, the  proposed method
M1 does not need any scale factor, whereas the method M2 uses
three scale factors. Secondly, the corrected solution appears to
be explicitly and directly dependent on the orbital elements in
the method M1 but not in the method M2. Thirdly, the corrected
solution appears to indirectly depend on the numerical solution
$(\textbf{r}^{\star},\textbf{v}^{\star})$ (except the computation
of $E^{*}$ using the numerical unit radial vector
$\mathbf{\hat{r}^{\star}}$) in M1, whereas it directly comes from
a linear combination of the numerical solution
$(\textbf{r}^{\star},\textbf{v}^{\star})$ in M2.

\subsection{Numerical checks and error analysis}

Whether the new method M1 is effective to intensively suppress the
fast growth of the integration errors in all orbital elements
compared with the uncorrected method needs to be verified. Whether
the new method M1 and the existing method M2 have the same
numerical performance in reducing the integration errors in
various orbital elements  needs to be explored. We perform
numerical tests to answer these questions.

Let us consider the Kepler problem with parameter $\mu=1$ and
initial orbital elements $a_0=2$ AU, $e_0=0.3$, $I_0=20^{\circ}$,
$\Omega_0=50^{\circ}$, $\omega_0=30^{\circ}$, and
$M_0=40^{\circ}$. Take a conventional fourth-order Runge-Kutta
algorithm (RK4) with a fixed time step $h$ being 1/100 of  orbital
period $T$ ($h=T/100$). In Figure 1, the new method M1 drastically
reduces the integration errors of all orbital elements compared
with the uncorrected algorithm RK4. The errors of the five
constant orbital elements for the two projection methods M1 and M2
should be zeros from the theory. However, they may be zeros at
some cases or equal to or near the machine double precision
$\epsilon=10^{-16}$ at some other times from the computation. To
ensure that the logarithms can work well for the zero errors in
Figure 1, we add $\epsilon$ to these errors, that is, $|\Delta
a|+\epsilon\rightarrow |\Delta a|$ in panel (a). The error of mean
longitude $M$ increases in proportion to the square of time for
the uncorrected method RK4, whereas it linearly grows with time
for the corrected schemes M1 and M2. The new method M1 is
approximately the same as the method M2 in controlling the
accumulation of integration errors in all orbital elements.

The two correction methods have the same performance in
suppressing the errors in the positions and velocities, as shown
in Figures 2(a) and 2(b). The rates of error growth in the
velocities and positions for RK4 and its correction methods M1 and
M2 are the same as those in the mean longitudes for the
corresponding algorithms. Here, the estimations of the errors in
the mean longitudes, positions and velocities use the analytical
solutions. The analytical solutions from Equations (16) and (17)
are approximately the same as those given by the Kepler solver of
Wisdom $\&$ Hernandez (2015) or Rein $\&$ Tamayo (2015). When time
$t=10^6$ corresponding to $10^8$ steps ($h=0.01$ in this case but
$h=T/100$ in Figure 2), the Kepler solver of Wisdom $\&$ Hernandez
and the analytical solution method given by Equations (16) and
(17) consume 28.5 and 26.8 s CPU times, respectively. This finding
indicates that the difference in computational effort between the
two analytical solvable methods is small.

The pure two-body problem is a good material to test the numerical
performance of an integrator because it contains known analytical
solutions. The proposed projection method is verified in the above
experiments using RK4 as a basic integrator. To satisfy the later
need, we continue to use the Kepler problem to test other
numerical integration algorithms, including  a five- and six-order
Runge-Kutta-Fehlberg algorithm [RKF5(6)] with a fixed step size,
its new projection method M1', an eighth- and ninth-order
Runge-Kutta-Fehlberg algorithm [RKF8(9)] with variable step
sizes\footnote{The position and velocity are  accurate to 9 and 8
orders, respectively. RKF5(6) also has such a similar meaning.}
and a second-order symplectic integrator S2.

The symplectic method requires that Equation (1) be split into two
parts
\begin{equation}
  K=H_0+H_1, ~~ H_0=\frac{v^2}{2}-\frac{0.999}{r}, ~~
  H_1=-\frac{0.001}{r}.
\end{equation}
$H_0$ is still a main Kepler part, and $H_1$ is a small
perturbation part. They are independently and analytically
solvable. The splitting method is slightly similar to that of
Wisdom $\&$ Holman (1991). Let $\mathcal{A}$ be an operator for
analytically solving $H_0$ and $\mathcal{B}$ be as another
operator for analytically solving $H_1$. This method is simply
written in the form
\begin{equation}
 S2=\mathcal{B}(\frac{h}{2})\circ\mathcal{A}(h)\circ\mathcal{B}(\frac{h}{2}).
\end{equation}
The ratio of the mass for  $H_1$ to the  mass for $H_0$,
$\varepsilon=1/999$, approaches that of the largest planet's mass
to the Sun's mass in the solar system. Such a splitting
Hamiltonian technique will give the established algorithm (29) a
better accuracy than the usually splitting method with the kinetic
energy plus the potential energy. This finding is because the
former truncation energy error is $\mathcal{O}(\varepsilon h^2)$
and the latter one is $\mathcal{O}( h^2)$.

The algorithms about the relative position or velocity errors from
the smallest to the largest in Figures 2(a) and 2(b) are RKF8(9),
the new correction method M1' of RKF5(6), S2, M1 (or M2) of RK4,
RKF5(6) and RK4. The energy errors in panel (c) remain bounded for
S2 because S2 shares an advantage of preserving simplecticity. The
energy errors of RKF8(9) are the smallest but grow with time. The
energy errors of RKF5(6) also grow. This error growth is because
the non-symplecticity of the two algorithms results in the
long-term accumulation of roundoff errors. After $t=1000$
corresponding to 5882 steps, RKF5(6) becomes poorer than S2; this
result can also be observed from the position or velocity errors.
The two projection methods consistently make the energy errors
(not plotted) arrive at the machine precision.

A simple analysis is provided to the rate of error growth for each
of the above algorithms. As claimed by Rein $\&$ Spiegel (2015),
the total error of an integrator comes from four contributions
\begin{equation}
E_{tot}=E_{fl}+E_{rand}+E_{bias}+E_{tr}.
\end{equation}
In the above equation, $E_{fl}$ relates to a computer giving any
calculation a relative error of approximately $2\times 10^{-16}$
in the double floating-point precision, that is, $E_{fl}\propto
W\times10^{-16}$ ($W$ being the number of computations).
$E_{rand}$ is a random error from any calculation involving two
random floating-point numbers. The random Kepler energy error will
grow as $\propto t^{1/2}$, and the random phase errors (e.g. the
mean longitude error or the position error) grow as $\propto
t^{3/2}$ for any integrator. $E_{bias}$ represents an error caused
by floating-point operations with respect to some specific
functions, such as square root, sine or cosine functions. The
errors, namely, $E_{bias}\propto t$ for the Kepler energy error or
 $\propto t^2$ for the phase error, might be biased and grow with time.
The three error contributions depend on floating-point numbers and
are inherent to all integrators. They are usually called roundoff
errors. In addition, $E_{tr}$ is an error associated with the
integrator itself and is also called a truncation error. The
truncation error is $E_{tr, phase}=\mathcal{O}(\frac{h}{T})^{j+1}$
in the phase (solution) and reads $E_{tr,
energy}=\mathcal{O}(\frac{h}{T})^{j}$ in the energy (Hamiltonian)
when the integrator is accurate to an order $j$.

Table 1 of Rein $\&$ Spiegel (2015) showed that the Kepler energy
error grows as $\propto t$ and the phase error does as $\propto
t^2$ for RK4. This fact can be also clearly observed from the
errors of the semimajor axis and mean anomaly in Figures 1(a) and
1(f) and from the errors of the positions and velocities in
Figures 2(a) and 2(b). The rates of phase error growth with time
for RK4 are similarly suited for RKF5(6) and RKF8(9). However,
some differences are found. The truncation phase error of RKF5(6)
with an order of $\mathcal{O}(\frac{h}{T})^{6}$ is smaller than
that of RK4 with an order of $\mathcal{O}(\frac{h}{T})^{5}$ and
that of RKF8(9) is the smallest. In particular, the truncation
phase error of RKF8(9) is $\sim \mathcal{O}(\frac{h}{T})^{9}=
\mathcal{O}(\frac{1}{100})^{9}=\mathcal{O}(10^{-18})$ when a
constant step-size is given. The error is difficult to estimate
because RKF8(9) adopts adaptive time-steps. However, Figure 2 (a)
shows that RKF8(9) approximately makes the position errors remain
at the machine precision until $t=10$. This finding indicates that
the truncation error can be negligible. Although RFK8(9) has an
extremely high precision, it still causes the Kepler energy error
to linearly grow  with time and the phase error to grow as
$\propto t^2$, similar to IAS15 of Rein $\&$ Spiegel (2015). In
addition, this table shows that S2 maintains a bounded energy
error. This result is also shown in Figure 2 (c). In the table,
the phase error is zero when the floating-point precision and
implementation specific errors are ignored. However, the results
under the existence of these errors in Figures 2(a) and 2(b) show
that the position or velocity errors grow linearly with time for
S2. The rates of phase error growth for S2 are also the same for
the projection methods M1 of RK4, M2 of RK4 and M1' of RKF5(6).

To show the phase error growth of these algorithms, we use
Equations (25) and (26) to estimate the position and velocity
errors in accordance with the following forms
\begin{eqnarray}
  \Delta\textbf{r}^{*} &=&  a_0 \Delta E^{*}(-\frac{\sin E^{*}}{e_0\mu}
  \textbf{P}_0 +\sqrt{1-e^2_0}\cos E^{*} \textbf{Q}_0), \\
  \Delta\textbf{v}^{*} &=& -\frac{a^2_0 n_0}{r^{*}}\Delta E^{*}(\frac{\cos E^{*}}{e_0\mu}
   \textbf{P}_0 +\sqrt{1-e_0^2}\sin E^{*}
  \textbf{Q}_0)  \nonumber \\
  && +\frac{a^3_0 n_0}{r^{*2}}(\frac{\sin E^{*}}{e_0\mu}\textbf{P}_0 -\sqrt{1-e_0^2}\cos E^{*}
  \textbf{Q}_0)  \nonumber \\
  && \cdot e_0\sin E^{*} \Delta E^{*},
\end{eqnarray}
where $\Delta E^{*}$ is calculated in terms of the Kepler equation
by
\begin{equation}
  \Delta E^{*}=\frac{\Delta n^{*}}{1-e_0\cos E^{*}} t=-\frac{3}
  {\mu}\frac{(-2K)^{1/2}\delta K}{1-e_0\cos E^{*}} t.
\end{equation}
Then, we have
\begin{eqnarray}
  |\Delta\textbf{r}^{*}| &\leq&  \frac{3a_0}
  {\mu}\frac{(-2K)^{1/2}}{1-e_0} (\frac{1}{e_0\mu} +\sqrt{1-e^2_0}~)|\Delta K| t, \\
  |\Delta\textbf{v}^{*}| &\leq& \frac{3a^2_0e_0 n_0}
  {\mu(1-e_0)}(\frac{1}{e_0\mu} +\sqrt{1-e_0^2}+\frac{1}{a_0(1-e_0)e_0\mu} \nonumber \\
&& +\frac{1}{a_0(1-e_0)}\sqrt{1-e_0^2}~)
\frac{(-2K)^{1/2}}{1-e_0}|\Delta K| t.
\end{eqnarray}

For the projection methods, $|\Delta K|=0$ from the theory,
whereas $|\Delta K|\neq0$ (such as $|\Delta K|\sim 10^{-16}$)
because of the floating-point precision and implementation
specific errors. Consequently, the projection methods lead to a
linear increase of the phase errors. Although eccentric anomaly
$E^{*}$ is obtained from Equations (23) and (24) rather than the
Kepler equation, the new projection method will consistently force
$E^{*}$ in Equations (23) and (24) to satisfy the Kepler equation.

Equations (34) and (35) are  useful to explain the linearly
increasing phase errors of a symplectic integrator (e.g. S2) in
accordance with the boundness of $|\Delta K|$, that is, $|\Delta
K|\leq C$. In addition, they can explain why the phase errors grow
with $t^2$ for a non-symplectic integrator, such as RK4. This is
because the energy errors $|\Delta K|$ for this algorithm grow
linearly with $t$, that is, $|\Delta K|\propto t$.

Which error dominates is difficult to answer. Its answer requires
the consideration of different things in different contexts. In
addition to this, several facts can be concluded from the above
theoretical analysis and numerical checks. The newly proposed
method is extremely successful to rigorously conserve the seven
integrals in the pure Kepler problem. It can intensively suppress
the rapid accumulation of the integration errors in all orbital
elements and cause the phase errors to linearly grow with time.
The proposed method is approximately the same as Fukushima's
method in two points. The application of the proposed method to
quasi-Keplerian motions of perturbed two-body or $N$-body problems
will be discussed in the next sections.

\section{Perturbed two-body problems}

In this section, we mainly focus on the application of the
proposed method in calculating the quasi-Keplerian orbits of
perturbed two-body problems. Some details of the implementation of
the proposed method are described. The numerical performance of
the new method is evaluated using two models of quasi-Keplerian
motions, namely, a post-Newtonian two-body problem and a
dissipative two-body problem.

For the two-body problem with a small perturbation, Equation (2)
becomes
\begin{equation}
    \frac{d\textbf{v}}{dt} = -(\frac{\mu}{r^{3}})\textbf{r}+ \textbf{a},
\end{equation}
where $\textbf{a}$ is a perturbed acceleration. This perturbed
two-body problem is also called the quasi-Keplerian problem. In
this case, the Kepler energy, angular momentum vector and LRL
vector in Equations (1), (4) and (5) are no longer invariant and
slowly vary with time. Thus, the application of the new method
becomes difficult.

\subsection{Integral invariant relations}

Kepler energy $K^{\ast}$, angular momentum vector
$\textbf{L}^{\ast}$ and LRL vector  $\textbf{P}^{\ast}$ are
slowly-varying quantities. They satisfy the following relations
\begin{eqnarray}
  \frac{d\Delta K^{\ast}}{dt} &=& \mathbf{v}\cdot \mathbf{a}, \\
    \frac{d\Delta \mathbf{L}^{\ast}}{dt} &=& \mathbf{r}\times
  \mathbf{a}, \\
  \frac{d\Delta \mathbf{P}^{\ast}}{dt} &=& 2(\mathbf{v}\cdot \mathbf{a})\mathbf{r}
  -(\mathbf{r}\cdot \mathbf{a})\textbf{v}-(\mathbf{r}\cdot
  \mathbf{v})\mathbf{a}.
\end{eqnarray}
Note that $\Delta K^{\ast}=K^{\ast}-K_{0}$,
$\Delta\textbf{L}^{\ast}=\textbf{L}^{\ast}-\textbf{L}_{0}$, and
$\Delta\textbf{P}^{\ast}=\textbf{P}^{\ast}-\textbf{P}_{0}$, where
$K_{0}$, $\textbf{L}_{0}$ and $\textbf{P}_{0}$ are the starting
values of these slowly-varying quantities. Taking zeros as the
initial values of $\Delta K^{\ast}$, $\Delta\textbf{L}^{\ast}$ and
$\Delta\textbf{P}^{\ast}$ can immensely reduce the roundoff errors
in numerical integrations. Equations (37)-(39) are integral
invariant relations of the slowly-varying quantities (Szebehely
$\&$ Bettis 1971).

Equations (36)-(39) are integrated. Thus, a numerical solution
$(\textbf{r}^{\star},\textbf{v}^{\star})$ is obtained. At the same
time, a set of values of the slowly-varying quantities $K^{*}$,
$\textbf{L}^{*}$ and $\textbf{P}^{*}$ are also given. In addition
to them,  another set of values $K^{\star}$, $\textbf{L}^{\star}$
and $\textbf{P}^{\star}$ can be given to these slowly-varying
quantities when the numerical solution
$(\textbf{r}^{\star},\textbf{v}^{\star})$ is directly substituted
into Equations (1), (4) and (5). Which of the two sets of values
from the two different paths are more accurate? The $K^{*}$,
$\textbf{L}^{*}$ and $\textbf{P}^{*}$ values from the integral
invariant relations are. As  indicated in (Huang $\&$ Innanen
1983), numerical solutions consistently keep each integral
constant as much as possible. Because the numerical solutions
$K^{*}$, $\textbf{L}^{*}$ and $\textbf{P}^{*}$ directly come from
the numerical integration of Equations (37)-(39), they naturally
have higher accuracies than the  $K^{\star}$, $\textbf{L}^{\star}$
and $\textbf{P}^{\star}$ values calculated by the coordinates and
velocities $(\textbf{r}^{\star},\textbf{v}^{\star})$. This fact
has been confirmed via many successful examples of manifold
correction to conservative, nonconservative or dissipative systems
(Fukushima 2003a, 2003b, 2003c, 2004; Wang et al. 2016; Wang et
al. 2018).

\subsection{Implementation of the proposed method}

Using the above $K^{*}$, $\textbf{L}^{*}$ and $\textbf{P}^{*}$
values from the integral invariant relations, we obtain the five
slowly-varying orbital elements $a^{*}$, $e^{*}$, $I^{*}$,
$\Omega^{*}$ and $\omega^{*}$ with mean motion $n^{*}$. Then,
$a_0$, $e_0$, $I_0$, $\Omega_0$, $\omega_0$ and $n_0$ in Equations
(20), (21) and (23)-(27) are the values at the beginning of each
step integration rather than those at the initial time. They are
replaced with $a^{*}$, $e^{*}$, $I^{*}$, $\Omega^{*}$,
$\omega^{*}$ and $n^{*}$, respectively. $\textbf{L}_0$ and
$\textbf{P}_0$ in Equation (22) should also give place to
$\textbf{L}^{*}$ and $\textbf{P}^{*}$. In this way, the new
correction scheme M1 given by Equations (25) and (26) can still
work in the perturbed case. Angles $\Omega^{*}$, $I^{*}$ and
$\omega^{*}$ are not necessarily known, but only their sines and
cosines. Such an operation is helpful to save computational cost
and to reduce the roundoff errors.

Summarised from the above demonstrations, the implementation of
the new method in the present case is described as follows.

{\it At an $i$th step, integrate Equations (36)-(39)  using a
certain numerical integrator and obtain the values
$\textbf{r}^{\star}$, $\textbf{v}^{\star}$, $K^{*}$,
$\textbf{L}^{*}$ and $\textbf{P}^{*}$.

Calculate the values $a^{*}$, $e^{*}$, $I^{*}$, $\Omega^{*}$,
$\omega^{*}$ and $n^{*}$ in terms of $K^{*}$, $\textbf{L}^{*}$ and
$\textbf{P}^{*}$.

Give the cosine and sine of  eccentric anomaly $E^{*}$ using
Equations (23) and (24).

Obtain the corrected numerical solution $(\textbf{r}^{*},
\textbf{v}^{*})$ from Equations (25) and (26) with
$a^{*}\rightarrow a_0$, $e^{*}\rightarrow e_0$, $I^{*}\rightarrow
I_0$, $\Omega^{*}\rightarrow \Omega_0$, $\omega^{*}\rightarrow
\omega_0$ and $n^{*}\rightarrow n_0$.

Take $\textbf{r}^{*}\rightarrow\textbf{r}^{\star}$ and
$\textbf{v}^{*}\rightarrow\textbf{v}^{\star}$, and let a
next step integration begin.}

\subsection{Post-Newtonian two-body problem}

Let us consider  a relativistic two-body problem with first-order
post-Newtonian corrections (Newhall et al. 1983) as a test example
of the perturbed quasi-Keplerian problems. It corresponds to the
equations of motion
\begin{eqnarray}
 \ddot{\textbf{r}} &=& -\frac{\mu}{r^3}\textbf{r}+\mathbf{a}_{PN},
 \\
\mathbf{a}_{PN} &=&
 \frac{\mu}{c^2}[(\frac{4\mu}{r}
 -v^2)\frac{\textbf{r}}{r^3}+4\frac{\textbf{r}\cdot
\textbf{v}}{r^3}\textbf{v}], \nonumber
\end{eqnarray}
where $c$ is the velocity of light, and $\mathbf{a}_{PN}$ is a
perturbed acceleration from the post-Newtonian contributions. Such
an equation is derived from a first-order post-Newtonian
Lagrangian approach $\ell$ and truncates second-order and
higher-order post-Newtonian terms.

Although Equation (40) should conserve a conservative energy, this
conservative energy has no way to be exactly written in a detailed
expressional form. A conservative energy of the form $\mathcal{E}
(\textbf{r}, \textbf{v})=\textbf{v}\cdot \mathbf{\wp}-\ell$, where
$\mathbf{\wp}=\partial \ell /\partial\textbf{v}$ is a generalised
momentum, can be conserved approximately using Equation (40).
However, energy $\mathcal{E}$ can be conserved strictly using
equation $\mathbf{\dot{\wp}}=\partial \ell /\partial\textbf{r}$
with $\mathbf{v}$ given by $\mathbf{\wp}=\partial \ell
/\partial\textbf{v}$. It is called as a coherent post-Newtonian
Lagrangian equation (Li et al. 2019; Li et al. 2020). Such a
coherent equation is different from Equation (40) because no terms
are truncated when the coherent equation is derived from the
Lagrangian $\ell$. However, some post-Newtonian terms must be
truncated when Equation (40) is obtained from $\ell$. This
condition explains why the coherent equation can exactly conserve
energy $\mathcal{E}$ but Equation (40) can approximately do.

Equation (40) also approximately conserves a conservative
Hamiltonian $\mathcal{H}(\textbf{r}, \mathbf{\wp})=\textbf{v}\cdot
\mathbf{\wp}-\ell$.\footnote{An explicit difference between
$\mathcal{E}$ and $\mathcal{H}$ is that $\mathcal{E}$ is a
function of $\textbf{r}$ and $\textbf{v}$ and $\mathcal{H}$ is
that of $\textbf{r}$ and $\mathbf{\wp}$. No terms are truncated
when $\mathcal{E}$ is derived from $\ell$, but the higher
post-Newtonian terms must be dropped when $\mathcal{H}$ remains of
the same post-Newtonian orders of $\ell$.}  The two formulations
$\ell$ and $\mathcal{H}$ are not exactly equivalent (Wu et al.
2015; Wu $\&$ Huang 2015). $\mathcal{E}$ and $\mathcal{H}$ are not
either. This condition indicates that Equation (40), coherent
equation and system $\mathcal{H}$ are three different dynamical
problems. These differences among them are negligible for the weak
gravitational solar system, and the three dynamical systems are
approximately the same. However, the differences are large for a
strong gravitational field of compact objects, and the three
dynamical systems are approximately related. Considering that
Equation (40) cannot exactly conserve energy $\mathcal{E}$,
conserving the Kepler energy in Equation (1) is impossible. The
LRL vector in Equation (5) is not invariant. In addition, the
Newtonian angular momentum given by velocity $\mathbf{v}$ in
Equation (4) is not constant, whereas the angular momentum
$\textbf{r}\times\mathbf{\wp}$ defined by momentum $\mathbf{\wp}$
is constant. The constant angular momentum makes the orbital plane
invariable. The motion is limited to the plane because the
post-Newtonian effect is given in the invariable orbital plane. In
this case, inclination $I$ and longitude of ascending node
$\Omega$ are not affected by the post-Newtonian effect and remain
invariant. However, the other orbital elements $a$, $e$ and
$\omega$ are affected and slowly vary with time.

Here, RK4 is still used as a basic integrator. We take parameter
$\mu=1$ and a fixed step size $h=T/120$, where $T$ is an orbital
period. The initial orbital elements are $a=2$ AU, $e=0.1$,
$I=20^{\circ}$, $\Omega=50^{\circ}$, $\omega=30^{\circ}$, and
$M=40^{\circ}$. The speed of light $c=10^4$ corresponds to the
first-order post-Newtonian effect, which is approximate to an
order of $10^{-8}$ (compared with the main Kepler part) in the
solar system (Dubeibe et al. 2017; Huang et al.
2018).\footnote{$c=1$ is taken for a strong gravitational field of
compact objects (Huang $\&$ Wu 2014). In addition,  $c$ has
different values in different unit systems. For example, $c =
172.672/(na)$ when the distance between two main objects and time
are measured in terms of $a$ and $1/n$, respectively. Thus, $c =
22946.5$ for the Sun and Jupiter and $c = 10065.3$ for the Sun and
Earth (Lhotka $\&$ Celletti 2014).} The correction methods M1 and
M2 can naturally preserve the two constant elements $I$ and
$\Omega$ because the conserved angular momentum is strictly kept
in the manifold corrections. In Figure 3, the errors of the
varying elements $a$, $e$ and $\omega$ for the two correction
schemes are decreased by 6 and 7 orders of magnitude compared with
those for RK4. In addition, the errors of $M$ are reduced by 3
orders of magnitude. Thus, the relative position errors in Figure
4 are reduced. Considering that RKF8(9) shows extremely good
accuracy in the above two-body problem under the circumstance of
roundoff errors ignored, it is still used to give the
post-Newtonian problem high-precision reference solutions to
obtain the various errors in Figures 3 and 4. The high-precision
solutions can also be given by IAS15 of Rein $\&$ Spiegel (2015).

We discuss why such a more accurate eccentric anomaly $E^{*}$ is
calculated in terms of Equations (23) and (24). Four possible
choices can be used in
 the calculation of $E^{*}$.

\textbf{Case 1}, a naturally prior choice is to solve $E^{*}$ from
the Kepler equation (14). The Kepler equation reads
\begin{equation}
E^{*}_{i}-e^{*}_{i}\sin E^{*}_{i}=M^{*}_{i-1}
+n^{*}_{i-1}\Delta t,
\end{equation}
where $\Delta t$ is a time step and $M^{*}_{i-1}$ denotes the
value of $M$ at the $(i-1)$th step. This path for the computation
of $E^{*}$ is marked as C1.

\textbf{Case 2}, mean motion $n^{*}_{i-1}$ in Equation (41) takes
an average value of $n^{*}_{i-1}$ at the $(i-1)$th step and
$n^{*}_{i}$ at the $i$th step; namely,
$\bar{n}^{*}_{i-1}=(n^{*}_{i-1}+n^{*}_{i})/2$ because the mean
motion is not invariant. In this case, the Kepler equation is
\begin{equation}
E^{*}_{i}-e^{*}_{i}\sin
E^{*}_{i}=M^{*}_{i-1}+\bar{n}^{*}_{i-1}\Delta t.
\end{equation}
The computation of $E^{*}$ is marked as C2.

\textbf{Case 3}, calculate $E^{*}$ with the numerical solution
$(\textbf{r}^{\star}, \textbf{v}^{\star})$, that is,
\begin{equation}
\cos E^{*}=\frac{a^{*}-r^{\star}}{e^{*}a^{*}}, ~~ \sin
E^{*}=\frac{\textbf{r}^{\star}\cdot
\textbf{v}^{\star}}{e^{*}n^{*}a^{*2}}.
\end{equation}
The method for computing $E^{*}$ is marked as C3.

\textbf{Case 4}, calculate $E^{*}$ using Equations (23) and (24).
The computation of $E^{*}$ along this direction is still marked as
M1.

Let $c$ range from 10 to $10^4$ at an interval of 2. This
condition indicates that the perturbation varies from strong to
weak and is approximately at an interval of [$10^{-8}$,
$10^{-2}$]. For a given value of $c$, the related errors in each
of the four cases are shown in Figure 5. In the numerical
performance, C1 and C2 or M1 and M2 have no explicit differences.
When the perturbation  is extremely small for $c=10^4$, C1 is
approximately consistent with M1. C1 becomes poorer than M1 with
the increase in perturbation and is inferior to the uncorrected
method RK4 when $c = 30$. These tests show that the obtainment of
$E^{*}$ from Equations (23) and (24) is the best choice. This
result is typically suitable for the quasi-Keplerian motions in
the solar system because the perturbation of each planet is
approximately in an interval of $(10^{-5}, 10^{-3})$ compared with
the individual Kepler part.

\subsection{Two-body problem with dissipative force}

As discussed in the Introduction, the velocity scaling method of
Ma et al. (2008b) combined with the integral invariant relation
worked well in nonconservative or dissipative systems (Wang et al.
2016, 2018).  What about the new correction method M1 applied to
dissipative systems?

To answer this question, we take the dissipative two-body problem
considered by Tamayo et al. (2019)
\begin{equation}
    \frac{d\textbf{v}}{dt} = -(\frac{\mu}{r^{3}})\textbf{r}+
    \mathbf{a}_d, ~~  \mathbf{a}_d=-\gamma \mathbf{v},
\end{equation}
where $\gamma$ is a damping parameter, and $\mathbf{a}_d$ is a
perturbed acceleration from a damping force. In this case, the
Kepler energy (1) is not a conserved quantity. Taking the damping
factor $\gamma=2\times 10^{-6}$ and the same initial conditions
and step size $h$ in the above post-Newtonian problem, we still
adopt RK4 and its correction scheme M1 to solve the dissipative
system (44). In addition, we use a fourth-order implicit method
with a symmetric combination of three second-order implicit
midpoint methods
\begin{equation}
    IM4=IM2(\lambda h)\circ IM2((1-2\lambda) h) \circ IM2(\lambda
    h),
\end{equation}
where $\lambda=1/(2-\sqrt[3]{2})$. This construction is based on
the idea of Yoshida (1990).  IM2 and IM4 are symplectic when they
are applied to integrate a Hamiltonian. High-precision reference
solutions are still given by RKF8(9).

As shown in Figures 6(a) and 6(b), the relative errors in the
Kepler energy and position are several orders of magnitude smaller
for M1 than those for RK4 before the integration time  $t=10^5$
corresponding to $5630~T$, where $T$ is the unperturbed period.
The correction method M1 is superior to the fourth-order implicit
method IM4 in the accuracy. The energy errors for IM4 have a
secular growth with integration because of the dissipative force
or roundoff errors. The relative position errors grow as $\propto
t^2$ for RK4, but $\propto t$ for M1 and IM4.

The efficiency regarding the dependence of the relative energy
errors on the computer runtime for each algorithm with integration
time $t=10^4$ (i.e. $563~T$) is plotted in panel (c). The
algorithms RK4, M1 and IM4 use different large fixed time steps
$0.0066\times 563T/10000$, $0.013\times 563T/10000$ and
$0.041\times 563T/10000$, respectively, when a short runtime (e.g.
0.1s) is given. Clearly, the algorithms with the computational
speeds from fast to slow are RK4, M1 and IM4. The algorithms from
high accuracies to low ones are  M1, RK4 and IM4. The time step of
IM4 is larger than those of RK4 and M1. This is an important
reason for IM4 obtaining the poorest accuracies. In terms of
efficiency, an integrator that costs less CPU time has better
efficiency compared with another integrator for a given accuracy.
On this basis,  the efficiencies from good to poor are M1, RK4 and
IM4 for the short runtimes considered. The algorithms have to
adopt different small fixed time steps, such as $h=0.000083\times
563T/10000$ for RK4, $h=0.00014\times 563T/10000$ for M1 and
$h=0.00039\times 563T/10000$ for IM4 when a long runtime (e.g. 8
s) is given. This condition leads to increasing the number of
integration steps and the fast accumulation of roundoff errors for
each integrator. In this case, the three methods have no explicit
differences in the efficiencies.

\section{Multi-body problems}

In this section, we focus on the application of the proposed
method to a five-body problem of the Sun and four outer planets.
For comparison, the Wisdom-Holman method and a fourth-order
symplectic integrator of Yoshida (1990) are considered.

Suppose each planet of an $N$-body gravitational problem in the
solar system moves in a quasi-Keplerian orbit. In the barycentre
coordinate system, this problem is described by the following
Hamiltonian
\begin{equation}
H=\sum^{N-1}_{j=0}\frac{\mathbf{p}^{2}_{j}}{2m_j}-\sum^{N-2}_{s=0}\sum^{N-1}_{j>
s}\frac{Gm_sm_j}{r_{sj}}.
\end{equation}
$j=0$ denotes the Sun, and $j=1,2, \ldots$ correspond to various
planets. The $j$th object has mass $m_j$, position coordinate
$\mathbf{r}_j$ and momentum $\mathbf{p}_j$. This system has seven
conservative integrals, involving the total energy $E$, total
momentum vector $\textbf{p}$ and  total angular momentum vector
$\textbf{L}$:
\begin{eqnarray}
 E &=& H,  \\
\textbf{p} &=& \sum^{N-1}_{j=0}\textbf{p}_j, \\
\textbf{L} &=& \sum^{N-1}_{j=0}\textbf{r}_j \times \textbf{p}_j.
\end{eqnarray}
They all are independent in the $N$-body system. They are also in
the two-body problem in the barycentre coordinate system. However,
the three integrals on the momenta are missing and the three
integrals of the LRL vector are included in the relative
coordinate system. Only five integrals are independent in this
case, as previously indicated.

Hairer et al. (1999)  reported that a five-body integration of the
Sun and four outer planets shows poor numerical performance when
the total energy and  total angular momenta are rigorously
preserved through a projection method. However, the corrections of
individual Kepler energies, angular momenta and LRL vectors work
well in a heliocentric coordinate system (Fukushima 2003a, 2003b,
2003c, 2004; Wu et al. 2007; Ma et al. 2008a, 2008b). Naturally,
the application of the new correction scheme to such an $N$-body
problem should be considered.

In the heliocentric coordinate system, each planet relative to the
Sun has position vector $\mathbf{\tilde{r}}_j$ and  velocity
vector $\mathbf{\tilde{v}}_j$ ($j=1,2, \ldots, N-1$). The
evolution equation of the quasi-Keplerian motion of  each body is
still similar to Equation (36), where the perturbed acceleration
is expressed as
\begin{eqnarray}
  \mathbf{\tilde{a}}_{j}=\sum^{N-1}_{s=1, \neq j} Gm_s
(\frac{\mathbf{\tilde{r}}_s-\mathbf{\tilde{r}}_j}{|\mathbf{\tilde{r}}_s
-\mathbf{\tilde{r}}_j|^3}- \frac{\mathbf{\tilde{r}}_s}
{\tilde{r}^{3}_s}).
\end{eqnarray}
Therefore, the new method M1 fitting for the perturbed two-body
problems is applied to the quasi-Keplerian motions of individual
planets in the multi-body problem.

Taking a five-body problem consisting of the Sun and four outer
planets (Jupiter, Saturn, Uranus and Neptune) as an example, we
consider the application of the new method M1 to this problem. All
data are taken from those in the ephemerides DE431. RK4 uses a
fixed step-size $h=36.525$ days, which is approximately 1/120 of
the orbital period $T$ of Jupiter. For comparison, the
second-order symplectic integrator of Wisdom $\&$ Holman (1991) is
included. We call it the WH method. The use of WH and its
extensions (e.g. Hernandez $\&$ Dehnen 2017) requires that the
Hamiltonian with $N=5$ in Equation (46) have an appropriate
splitting form similar to Equation (28). In the Jacobi coordinate
system, the Kepler part $H_{0}$ and the planet-planet interaction
part $H_{1}$ can be obtained. The ratio of the latter part to the
former part is approximately 1/1047. In this way, the WH
integrator similar to S2 in Equation (29) becomes easily
available.

Figure 7 plots the evolution of some orbital elements and
quasi-integrals in the five-body problem. Clearly, RK4 gives a
secular change to the semi-major axis $a$ of Jupiter in an
integration time of $10^{5}$ years  in Figure 7 (a). The
semi-major axis decreases to 5.198 AU in the integration time.
Such similar changes are also given to eccentricity $e$ and
$z$-direction angular momentum $L_z$ of Jupiter  in an integration
time of $10^{7}$ years in Figure 7 (e). However,  no secular
changes are found in other orbital elements and quasi-integrals
(including those not plotted). Fortunately, the secular changes
can be eliminated using the projection method M1 (similar to WH)
in Figures 7(a) and 7(f). In particular, the semi-major axis $a$,
eccentricity $e$ and  angular momentum $L_z$ of Jupiter remain
bounded until integration time $t$ reaches $10^{8}$ years. For
example, the semi-major axis is consistently limited to a bounded
region between 5.201  and 5.205 AU during the integration time by
M1, which is similar using WH. Similarly, M1 forces the three
other planets' semi-major axes to stay at bounded regions:
$a_S\approx 9.51- 9.59$ AU for Saturn, $a_U\approx 19.10- 19.31$
AU for Uranus and $a_N\approx 29.90 - 30.35$ AU for Neptune. In
other words, the energies between the Sun and planets ($E_{SJ}$,
$E_{SS}$, $E_{SU}$ and $E_{SN}$) in the barycentre coordinate
system are  bounded because of the relation (8). As shown in
Figure 7, the secular changes that exist in some elements or
quasi-integrals for RK4 adopting short integration times are
absent in M1 with long integration times. This fact sufficiently
shows the advantage of the new projection method in typically
suppressing the error growth.

Letting RK4 give place to RKF5(6), we use RKF8(9) as a
high-precision reference integrator to further check the
performance of the new projection method. The above secular
changes to the related elements or quasi-integrals yielded by RK4
lose in each of the algorithms RKF5(6), M1, WH and  RKF8(9). The
above bounded regions of the four planets' semi-major axes are
also kept. Compared with the uncorrected method, the two
correction schemes M1 and M2 have approximately the same
performance in effectively reducing the errors from the related
orbital elements until $t\approx 10^{6}$ years in Figure 8.
Unfortunately, M2 begins to become worse when the integration
exceeds this time, but M1 still works well before the integration
time reaches $10^{8}$ years. Such similar results also occur for
the relative position errors of the four planets in Figure 9 and
for the relative errors of the total energy and the total angular
momenta in Figure 10. The following several points can be observed
from Figures 7-10.

Why do not the related errors in Figures 8 and 9 grow after some
times? For instance, errors $\Delta a_J$ remain at 0.001 for
RKF5(6) after $t\approx 10^{6}$ years and for WH after $t\approx
10^{7}$ years in Figure 8 (a). They also tend to this value for M1
as $t$ approaches $10^{8}$ years. These results are because these
algorithms restrict the individual semi-major axes or energies to
bounded regions. In practice, the same four significant digits
5.200 (i.e. a precision of an order of 0.001) can be given to the
semi-major axis of Jupiter using the algorithms involving RKF8(9).
Thus, the bounded regions or the same significant digits of the
semi-major axes determine that the differences between RKF8(9) and
each of RKF5(6), M1 and WH consistently remain stable at some
times. The Jupiter's relative position errors that remain stable
after $10^{6}$ years in Figure 9 (a) are because they are given
the same three significant digits by RKF5(6) and  RKF8(9). The
higher the accuracy of an algorithm is, the longer the times of
the differences arriving at the stable values will be. This
condition can explain why all errors in Figures 8 and 9 grow with
time although the seven quasi-integrals or the five slowly-varying
orbital elements of each planet are bounded in RKF5(6), M1 and WH.
It also shows that WH is superior to RKF5(6) but inferior to M1.

The correction method M1 strictly satisfies the integral invariant
relations of the seven slowly-varying quantities for each planet
in the heliocentric coordinate system. In particular, it makes the
slowly-varying orbital elements bounded or the related errors
stable at some times in Figures 7-9. However, this condition does
not indicate that it conserves the seven quantities because these
quantities are not constant. Thus, it cannot preserve the
integrals, such as total energy (47) and total angular momenta
(49) in the barycentre coordinate system, as numerically shown in
Figure 10. However, the total momenta (48) are conserved exactly
in any algorithm because the solutions in the Jacobi coordinate
system are transformed into those in the barycentre coordinate
system by the conserved total momenta. The WH symplectic method
can conserve the total energy but cannot conserve the total
angular momenta because of roundoff errors. Although  M1 and WH
have difference in the conservation of the total energy, they have
approximately the same effects on restricting the quasi-integrals
or the slowly-varying elements to the bounded regions in Figure 7.

The preference of M1 over WH in the accuracies in Figures 7-10
sounds naturally reasonable because the order of M1 is at least 5
and that of WH is 2. To basically match with the order of M1, the
Yoshida's fourth-order explicit symplectic integrator Y4 is given
by substituting WH into IM2 in Equation (45). The accuracy of Y4
has an advantage over that of M1 but is poorer than that of
RKF8(9) in Figure 10. In particular, Y4 as a symplectic integrator
shows a linear growth of the energy error, similar to M1 or
RKF8(9). This condition is because Y4  has a truncation energy
error $\mathcal{O}[\varepsilon
(\frac{h}{T})^4]\sim\mathcal{O}[\frac{1}{1047}
(\frac{1}{118.6})^4]\sim \mathcal{O}(10^{-12})$. When the time is
$10^{5}$ years corresponding to $10^{6}$ steps, the total energy
errors of Y4 can remain bounded and change between the orders of
$10^{-14}$ and  $10^{-11}$. If a machine error is accumulated per
step, the roundoff errors grow as $\propto \frac{t}{h} \times
10^{-16}=10^{-10}$ (this result is only a rough estimation to the
roundoff errors). As the integration continues, the roundoff
errors cause the total energy errors of Y4 to linearly grow. The
WH method can restrict the energy errors to the interval
$[10^{-10}, 10^{-8}]$ because its truncation energy error
$\mathcal{O}[\varepsilon
(\frac{h}{T})^2]\sim\mathcal{O}[\frac{1}{1047}
(\frac{1}{118.6})^2]\sim \mathcal{O}(10^{-8})$ is not dominated by
the roundoff errors with an order of $\mathcal{O}(10^{-7})$ in the
integration of $10^{8}$ years corresponding to $10^{9}$ steps. In
fact, the number of integration steps for Y4 is approximately 3
times more than that for WH, and the roundoff errors become
dominant after $10^{5}$ years. To suppress the fast growth of the
roundoff errors, we use a large time step $h^{*}=350$ days for the
fourth-order symplectic method Y4*. As expected, the total energy
errors of Y4* remain bounded and are approximately consistent with
those of WH with the small step size $h=36.525$ days. In this
case, Y4* has a truncation energy error $\mathcal{O}[\varepsilon
(\frac{h}{T})^4]\sim\mathcal{O}[\frac{1}{1047}
(\frac{1}{11.86})^4]\sim \mathcal{O}(10^{-8})$, which is not
governed by the roundoff errors with an order of
$\mathcal{O}(10^{-8})$ in such an integration of $10^{8}$ steps.
In fact, such a similar accuracy can also be yielded by the
second-order WH method with symplectic correctors (Wisdom et al.
1996; Wu et al. 2003) (not plotted) for the large time step.
However, the symplectic correctors need considerable computational
labour because many additional iterations, such as operator
$\mathcal{A}$ in Equation (29), are used. The comparison between
Y4 and Y4* sufficiently shows that the roundoff errors seriously
dominate the numerical errors.

In addition to the above-mentioned two points, the slopes
regarding the error growth with time are different for various
algorithms in Figures 8-10. They should be some combination of
roundoff and truncation errors. As a high-precision reference
integrator, RKF8(9) gives its truncation energy error
$\mathcal{O}(\frac{h}{T})^8\sim\mathcal{O} (\frac{1}{118.6})^8\sim
\mathcal{O}(10^{-17})$ when constant time step $h=36.525$ days is
considered. In fact, it uses variable step-sizes, and its
truncation error is difficultly estimated. As shown in Figures 2
and 10, the outputted results arrive at or  approach the machine
precision in short times. In this sense, the truncation errors are
basically negligible for RKF8(9), and most of the errors are
roundoff errors. RKF8(9), similar to IAS15 (Rein $\&$ Spiegel
2015), makes the total energy errors in Figure 10 and the phase
errors grow as $t$ and $t^2$ because of the roundoff errors as it
did in the above two-body problem. In spite of this, RKF8(9) with
the truncation errors neglected can still be regarded as a
reference algorithm to provide high-precision solutions in an
appropriately long time.

The truncation energy error for RKF5(6) reads
$\mathcal{O}(\frac{h}{T})^5$ $\sim$ $\mathcal{O}
(\frac{1}{118.6})^5$ $\sim$ $\mathcal{O}(10^{-11})$. The roundoff
errors for RKF5(6) lead to the total energy errors growing
$\propto t$  in Figure 10 and relative position errors growing
$\propto t^2$ in Figures 9(a) and 9(c). The position error growth
is considered before the errors remain at  stable values. The
phase errors can be roughly observed from Equation (34). Here,
$a_j$ and $e_j$ of individual planets are approximately constants.
Because individual Kepler energy errors $\Delta K_j \propto t$ for
RKF5(6), individual position errors $|\Delta \mathbf{r}_j| \propto
t^2$. The projection method M1 can appropriately decrease the
growth of errors $\Delta K_j$. Therefore,  $|\Delta \mathbf{r}_j|
\propto t^{3/2}$, as shown in Figure 9. This finding is mainly
from the contribution of random errors to the phase errors of each
planet, similar to that to the phase errors of the two-body
problem (Rein $\&$ Spiegel 2015). The WH symplectic algorithm
should make $\Delta K_j$ bounded if the roundoff errors are
ignored. Equation (34) shows that $|\Delta \mathbf{r}_j| \propto
t$.\footnote{Considering that individual energy errors $\Delta
K_j$ remain bounded for M1, we believe that WH and M1 will have
the same slopes on the phase error growth if the roundoff errors
can be set to 0. However, we have not tested or proved it
experimentally in quadrupole precision.} This result is suitable
for the errors of WH in Figure 8. In a word, the slopes for the
position error growth with time are WH $<$ M1 $<$ RKF5(6) before
the errors tend to stable values in Figure 9. The result on WH
slope $<$ M1 slope in the present case is different from the WH
slope $=$ M1 slope in the two-body problem.

Let us measure the computational efficiencies of the algorithms
RKF5(6), M1, WH, Y4 and RKF8(9). In Figure 11, the same CPU time
indicates that these algorithms (except RKF8(9)) should  use
different fixed time steps when they independently integrate the
five-body problem up to $10^4$ years. These methods need small
computational cost when they take large step-sizes. For example,
when RKF5(6), M1 and WH use the same large step-size $h=16.14$
days and Y4 uses another large step-size $h=48.23$ days, these
constant step-size algorithms have approximately the same CPU time
of 0.5 s. RKF8(9) has only one point, corresponding to CPU time of
0.45 s because it uses adaptive time-steps. In this case, the
efficiencies from high to low are RKF8(9), M1, Y4, RK5(6) and WH.
As the constant step-size methods adopt small step-sizes, they
need many machine labours. Given $h=0.08$ days for  RKF5(6) and
WH, $h=0.1$ days for M1 and $h=0.25$ days for Y4, the constant
step-size methods cost 80 s CPU times. Increasing integration
steps are added, and the roundoff errors become more important
than the truncation errors from the schemes because the adopted
time-steps are smaller. This condition explains why the explicit
differences cannot be observed among the efficiencies of the
constant step-size methods when the runtime spans 30
s.\footnote{We give some details of our codes in the present work.
All codes are edited in Fortran 77 and are suitable for solving
first-order ordinary differential equations. Codes of WH, WH with
correctors and Yoshida's method for $N$-body problems in the solar
system were written in 2001 by the corresponding author Wu. At
that time, he was taking his Ph.D Programme in Nanjing University
of China and edited the codes with many lines in the work of Wu et
al. (2003). Codes of RKF5(6), RKF6(7), RKF7(8) and RKF8(9) with
constant  and adaptive step-sizes were given by  Wu's advisor
Prof. Tian-Yi Huang and Huang's colleagues. They are complicated
and have many lines. Codes of the newly proposed method and
Fukushima's method have been edited by the authors. The new codes
are simple and have tens of lines. The codes except those of
RKF5(6) and RKF8(9) are freely available from the corresponding
author on request.}

\section{Summary}

Using the analytical solutions of a pure two-body problem, a new
projection method is proposed to rigorously conserve the seven
independent and dependent integrals, including the Kepler energy,
angular momentum vector and LRL vector. Unlike the analytical
method that solves the eccentric anomaly from the Kepler equation
with an iterative method, the newly proposed method does not need
any iteration but uses the true anomaly between the constant LRL
vector and a varying radial vector to obtain the eccentric
anomaly. The unit radial vector is given by an integrator. In the
construction mechanism, the proposed method is  typically
different from Fukushima's linear transformation method for the
consistency of the seven integrals. On the one hand, the former
projection method does not use any scale factor, whereas the
latter one uses three scale factors. On the other hand, the former
corrected solutions appear to directly depend on the orbital
elements rather than the numerical solutions, whereas the latter
ones are typically a linear combination of the numerical
solutions. Numerical simulations of the two-body problem show that
the proposed method can successfully give the machine epsilon to
the integration errors in all orbital elements except the mean
longitude at the epoch. In addition, the proposed method and
Fukushima's method have approximately the same numerical
performance. The slope of phase error growth with time for the
proposed method is consistent with that for the second-order
symplectic integrator.

For the quasi-Keplerian motion of a perturbed two-body problem or
each body in an $N$-body problem, the seven quantities slowly vary
with time. This condition is an obstacle to the application of the
proposed method. We simultaneously integrate the time evolution of
the seven slowly-varying quantities (called the integral invariant
relations of these quantities) and the usual equations of motion.
The seven quantities from the direction integration of the
invariant relations are more accurate than those obtained from the
integrated positions and velocities. These high-precision
quantities can determine the five slowly-varying orbital elements,
namely, semimajor axis, eccentricity, inclination, longitude of
ascending node and  argument of pericentre. The eccentric anomaly
is calculated similarly using the method of the two-body problem.
When these values from the integral invariant relations are
substituted into the analytical solutions of the two-body problem,
the solutions of the perturbed two-body problem or each planet in
the $N$-body problem can be adjusted. In the expressional forms,
the corrected solutions resemble the analytical solutions of the
two-body problem. In this way, the proposed method can be
implemented without difficulty.

The post-Newtonian two-body problem numerically confirms that the
proposed method can significantly improve the accuracies by
several orders of magnitude compared with the case without
correction. The proposed projection method and  Fukushima's method
are approximately the same in the numerical performance. It can
also exhibit extremely good correction effectiveness for the
dissipative two-body problem. When the five-body problem of the
Sun and outer planets is taken as a test model, the proposed
method also has an explicit effect on suppressing the fast growth
of numerical errors. In fact, the secular changes of some elements
or quasi-integrals that are caused by short integration times of
the fourth-order Runge-Kutta algorithm can be eliminated in a long
integration time of $10^{8}$ years using the new projection method
similar to the Wisdom-Holman integrator. If RKF5(6) is used as a
basic integrator, Fukushima's method does not work well because of
roundoff errors for such a long-term integration of the five-body
problem. The new correction method has an advantage over the
Wisdom-Holman symplectic integrator in the accuracy  in an
appropriately long integration time, but the former slope of phase
error growth is larger than the latter one. This finding indicates
that the advantage of the new projection method will gradually
lose as the integration time increases (e.g. $10^{9}$ years) in
the five-body integration. The new correction scheme included in
another  high-precision non-symplectic integrator can exhibit
better numerical performance in a long integration. This is an
advantage of this type of correction scheme in the applicability.

The construction of the new correction method is based on the
theory of two-body dynamics in celestial mechanics. It needs a
small amount of additional computational cost, compared with the
uncorrected basic  non-symplectic integrator. In particular, the
application of the proposed method is wider and more convenient
than that of symplectic integrators. The proposed method is
suitable for simulating elliptical or quasi-elliptical orbital
motions of various objects, such as major and minor planets,
satellites and comets. In addition to the Newtonian gravity
interactions, various perturbations involving the $J_2$
perturbation and relativistic post-Newtonian terms (Quinn et al.
1991) are admissible. It is also applicable to the quasi-Keplerian
motions in systems of extrasolar planets. Apart from these
conservative systems, non-conservative or dissipative systems are
fit for the use of the proposed method.

\section*{Acknowledgments}

The authors are grateful to the referee Dr. David M. Hernandez for
many valuable comments and suggestions. This research was
supported by the National Natural Science Foundation of China
(Grant Nos. 11533004, 11973020, 11663005, 11533003, and 11851304),
the Special Funding for Guangxi Distinguished Professors
(2017AD22006), and the Natural Science Foundation of Guangxi
(Grant Nos. 2018GXNSFGA281007 and 2019JJD110006).

\section*{Appendix A}

\subsection*{Projection method of Fukushima}

The linear transformation method of Fukushima (2004) has two
procedures as follows.

Firstly, a single-axis rotation transformation adjusts the
integrated velocity $\textbf{v}^{\star}$ and position
$\textbf{r}^{\star}$ as
\begin{eqnarray}
  \textbf{v}' &=& d\textbf{v}^{\star}+\textbf{s}\times \textbf{v}^{\star}+(\frac{\textbf{s}\cdot
  \textbf{v}^{\star}}{1+d})\textbf{s}, \\
  \textbf{r}' &=& d\textbf{r}^{\star}+\textbf{s}\times \textbf{r}^{\star}
  +(\frac{\textbf{s}\cdot \textbf{r}^{\star}}{1+d})\textbf{s},
\end{eqnarray}
where  vector $\textbf{s}$ and  factor $d$ are
\begin{eqnarray}
  \textbf{s}=\frac{(\textbf{r}^{\star}\times \textbf{v}^{\star})\times
  \textbf{L}^{*}}{|\textbf{r}^{\star} \times \textbf{v}^{\star}||\textbf{L}^{*}|},
~~~~   d=\sqrt{1-\textbf{s}^{2}}.
\end{eqnarray}
In fact, the adjusted solution $(\textbf{r}', \textbf{v}')$ is
perpendicular to the angular momentum vector $\textbf{L}^{*}$.

Secondly, a linear transformation to the above adjusted solution
is
\begin{eqnarray}
  \textbf{r}^{\ast} &=& s_{r}\textbf{r}', \\
  \textbf{v}^{\ast} &=& s_{v}(\textbf{v}'- \alpha \textbf{r}').
\end{eqnarray}
The three factors $s_{r}$, $s_{v}$ and $\alpha$ are determined by
the second adjusted solution $(\textbf{r}^{\ast},
\textbf{v}^{\ast})$, which rigorously satisfies the Kepler energy
$K^{\ast}$, LRL vector $\textbf{P}^{\ast}$ and angular momentum
vector $\textbf{L}^{\ast}$ in the pure two-body problem. They have
explicit expressions
\begin{eqnarray}
  s_{r}=\frac{\textbf{L}^{\ast 2}}{\textbf{F}\cdot \textbf{r}'}, ~~
  \alpha=\frac{\textbf{F}\cdot \textbf{v}'}{\textbf{F}\cdot
  \textbf{r}'}, ~~ \textbf{F}=\textbf{P}^{\ast}+\mu(\frac{\textbf{r}'}{r'}),
\end{eqnarray}
\begin{eqnarray}
  s_{v}=\sqrt{\frac{2K^{\ast}+2\mu / (s_{r}\textbf{r}')}{(\textbf{v}')^{2}-2\alpha (\textbf{r}'\cdot
  \textbf{v}')+\alpha^{2}(r')^{2}}}.
\end{eqnarray}

For the pure Kepler problem, $K^{\ast}$, $\textbf{L}^{\ast}$ and
$\textbf{P}^{\ast}$ take their initial values $K_{0}$,
$\textbf{L}_{0}$ and $\textbf{P}_{0}$, respectively. In this case,
the seven integrals and all orbital elements, except the mean
longitude, are consistently conserved by Equations (54) and (55).
These seven quantities $K^{\ast}$, $\textbf{L}^{\ast}$ and
$\textbf{P}^{\ast}$  are obtained from Equations (37)-(39) in the
perturbed two-body or multi-body problems. They are more accurate
than those obtained from the integrated coordinates and
velocities.

\newpage

\begin{figure*}
\center{
  \includegraphics[scale=0.25]{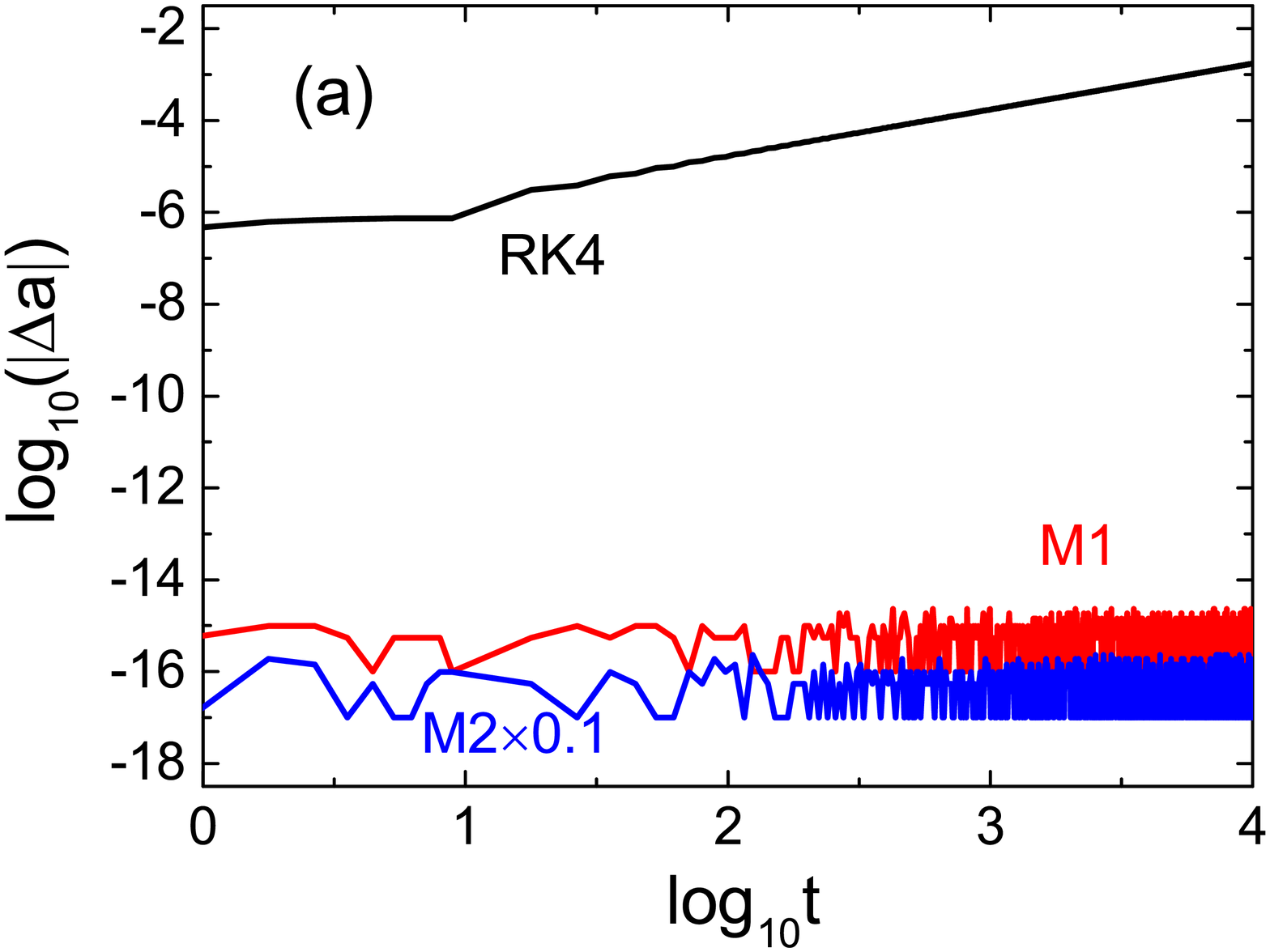}
  \includegraphics[scale=0.25]{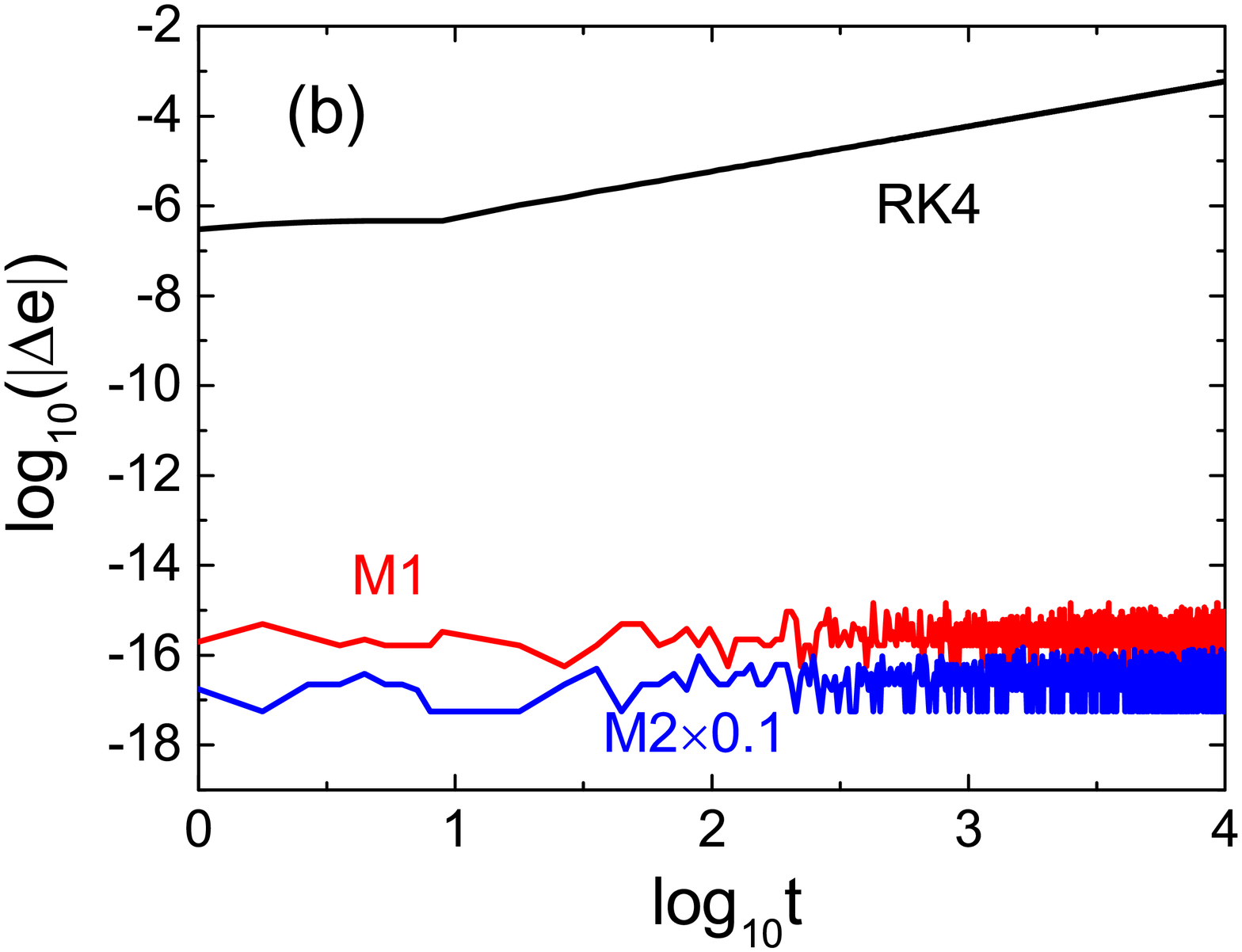}
  \includegraphics[scale=0.25]{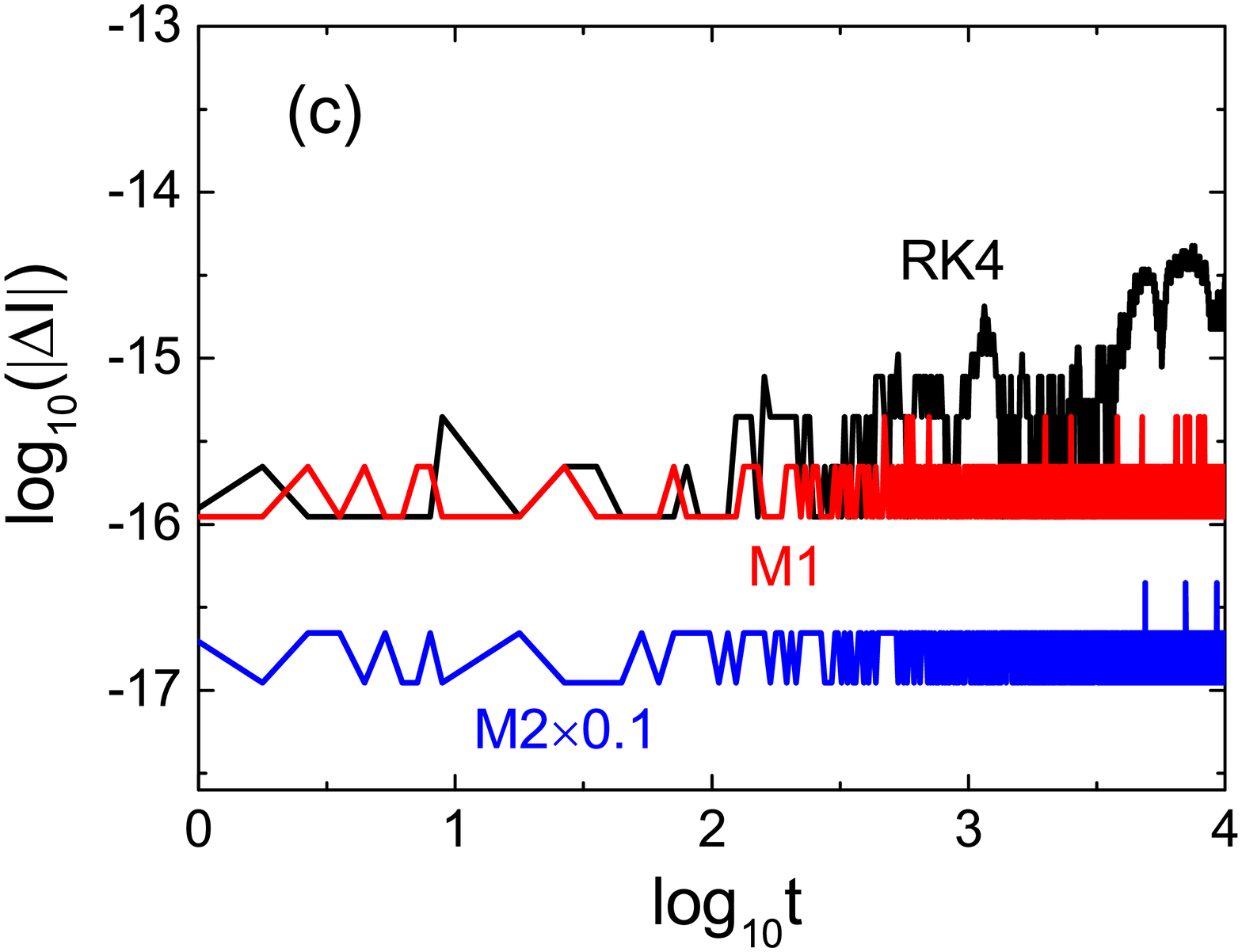}
  \includegraphics[scale=0.25]{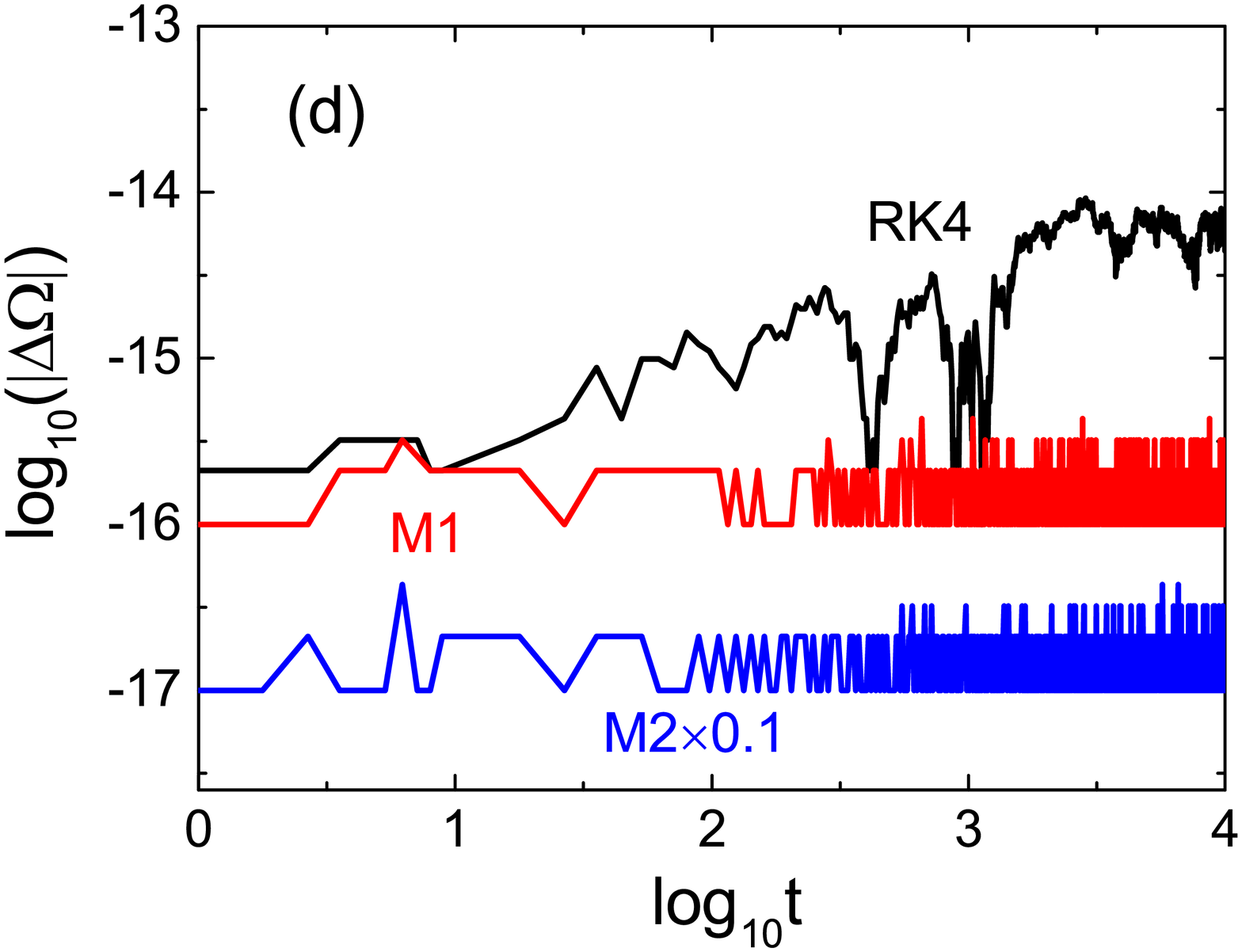}
  \includegraphics[scale=0.25]{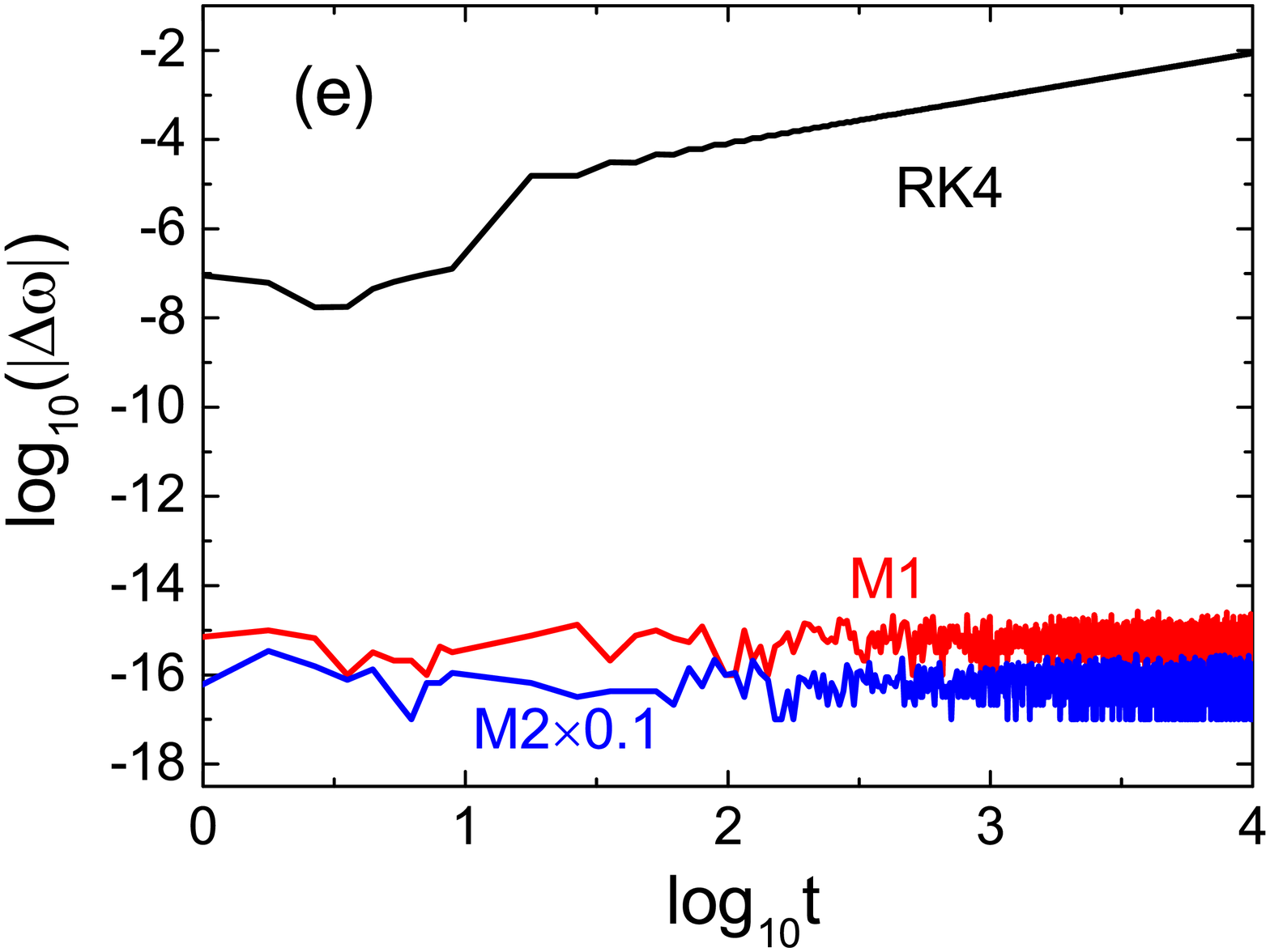}
  \includegraphics[scale=0.25]{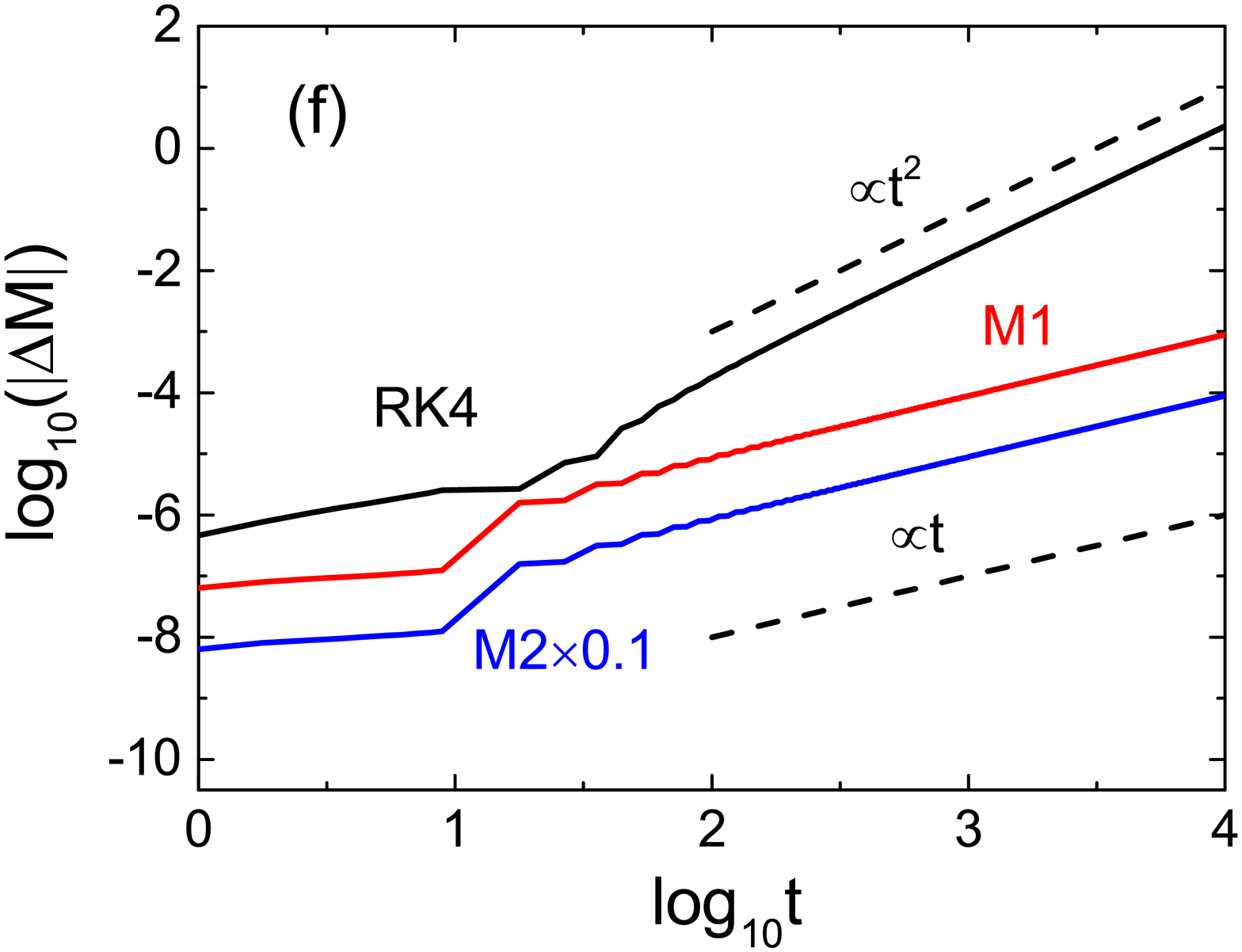}
 \caption{Errors of six orbital elements for the pure Keplerian
orbit. The adopted algorithms are RK4 and its correction methods:
the newly proposed method M1 and  Fukushima's linear
transformation method M2. For M2, all errors are reduced by 10
times. The error of
 mean longitude $M$ increases in proportion to the square of
time for RK4, whereas it linearly increases for M1 and M2. In
fact, the error curve of  M1 basically coincides with that of M2
in each panel.}} \label{fig1}
\end{figure*}

\begin{figure*}
\center{
\includegraphics[scale=0.185]{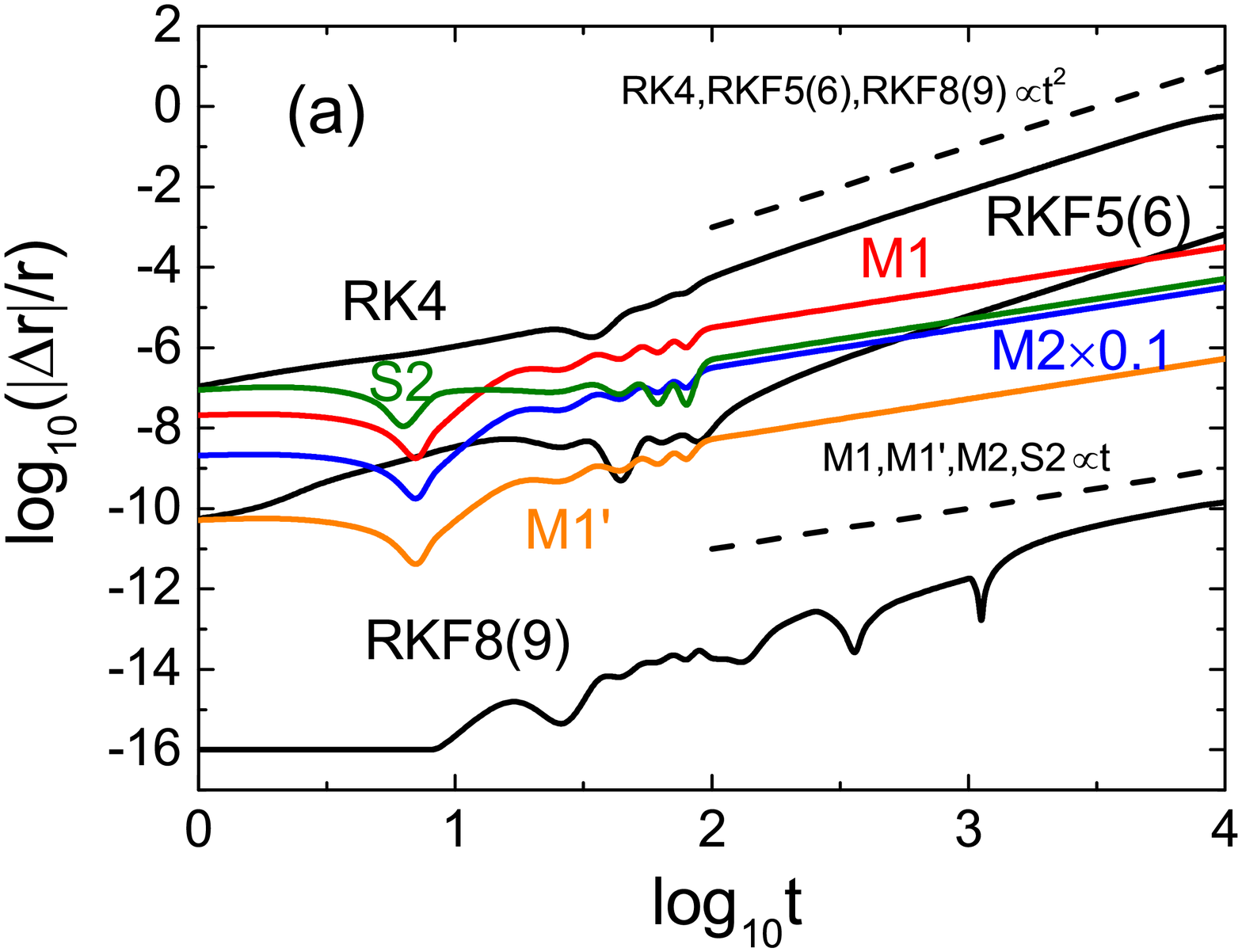}
\includegraphics[scale=0.185]{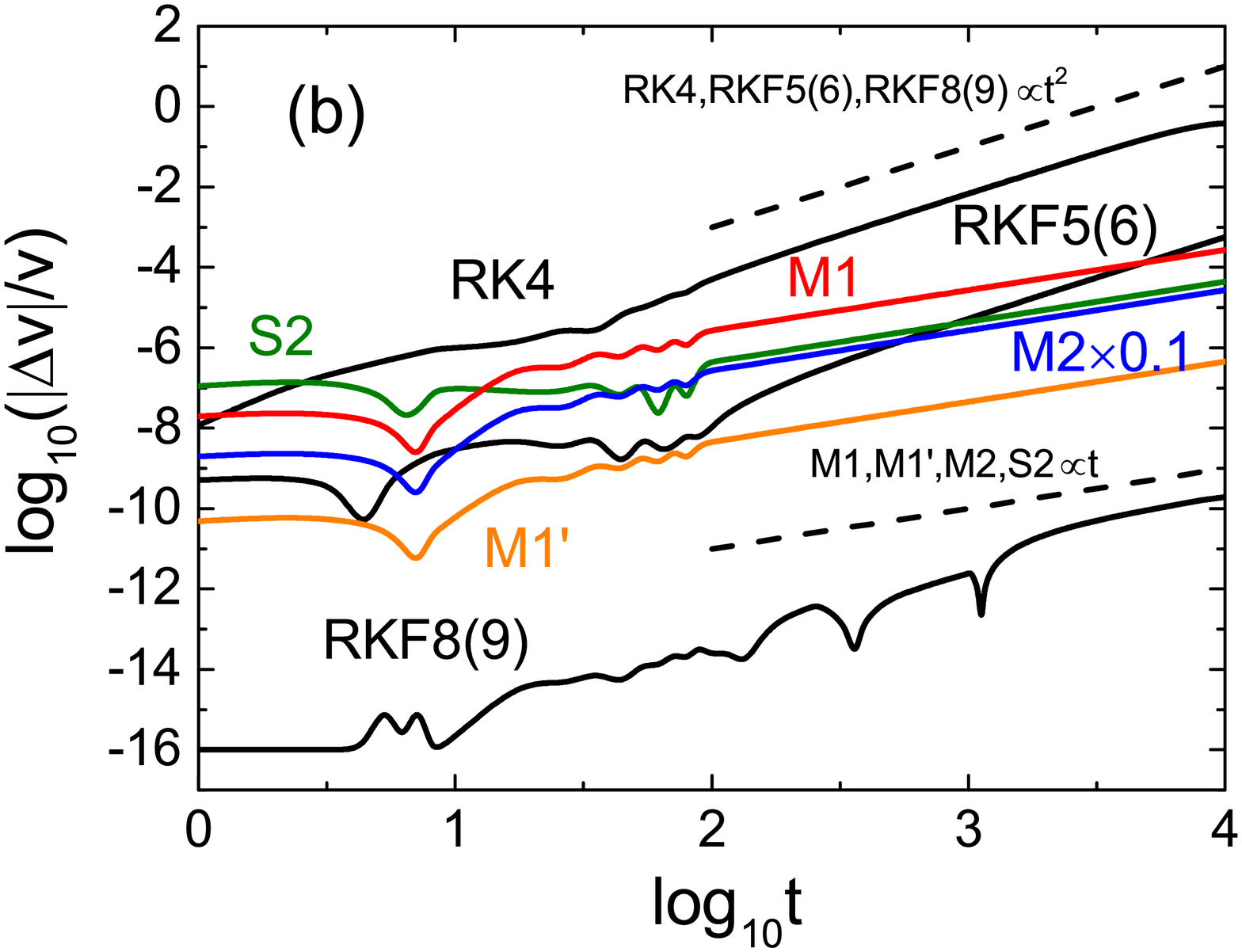}
\includegraphics[scale=0.185]{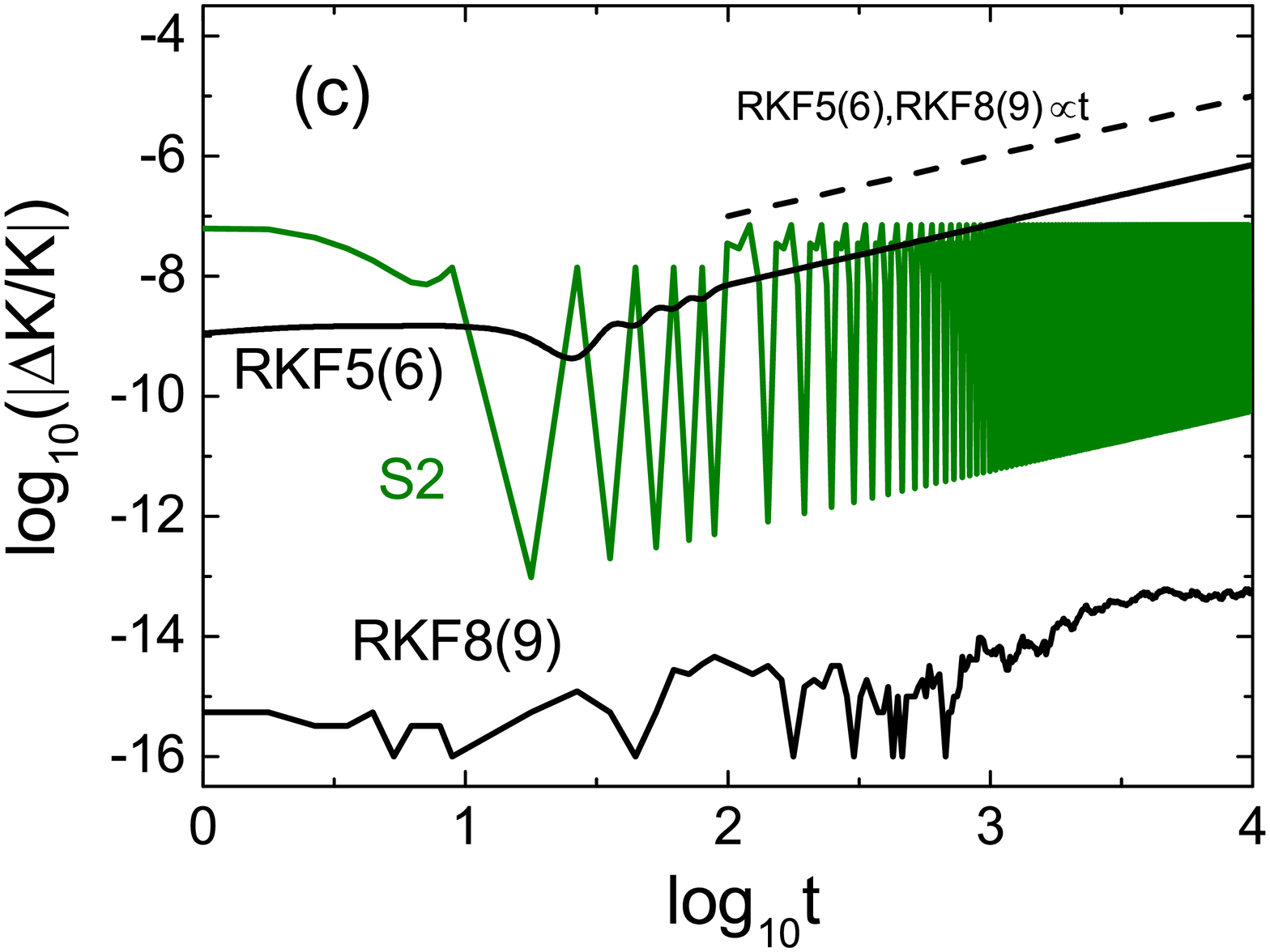}
\caption{Same as Figure 1, but relative position  errors (a) and
velocity  errors (b) are used rather than the errors of the
orbital elements. Here, M1 is the new projection method of RK4,
and M2 is  Fukushima's method of RK4. RKF5(6), its new correction
scheme M1', RKF8(9) and symplectic integrator S2 are added. The
position or velocity errors linearly grow with time for the
projection methods M1, M2 and M1' and S2, whereas they linearly
grow with a law of the square of time for RK4, RKF5(6) and
RKF8(9). The energy errors in (c) linearly grow for RKF5(6) and
RKF8(9), but remain at the machine precision level for the
projection methods (not plotted) and are bounded for S2.}}
\label{fig2}
\end{figure*}

\begin{figure*}
\center{
  \includegraphics[scale=0.26]{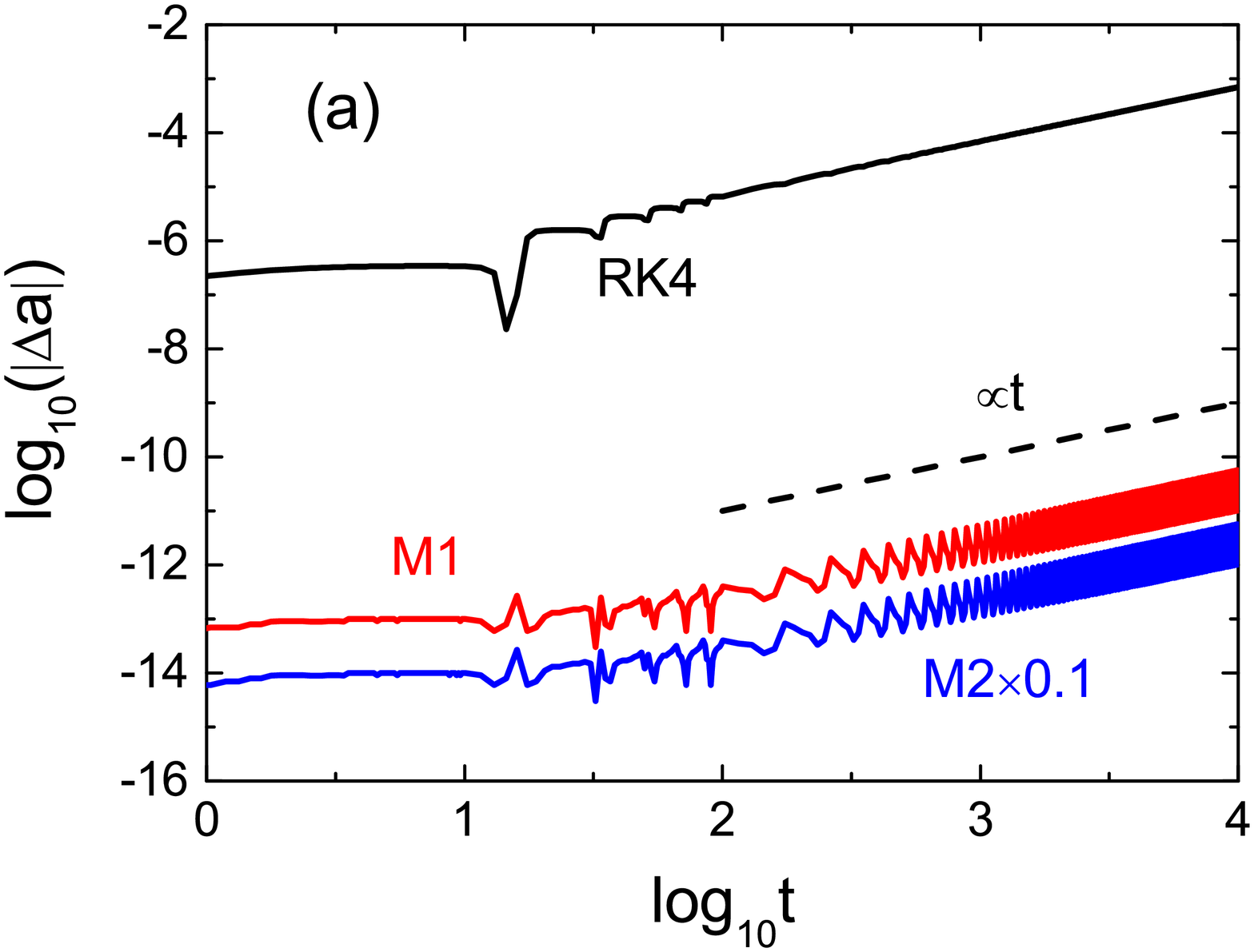}
  \includegraphics[scale=0.26]{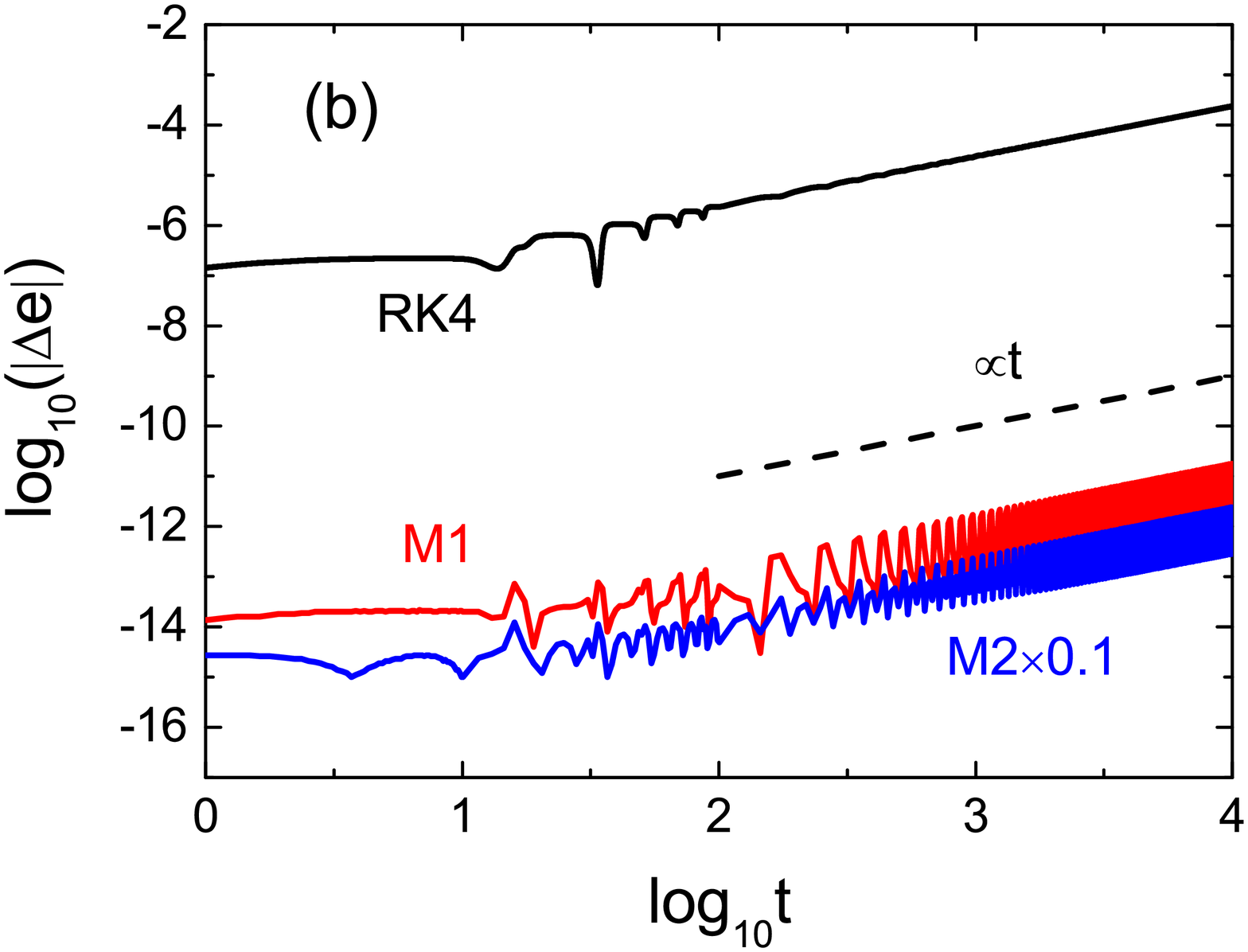}
  \includegraphics[scale=0.26]{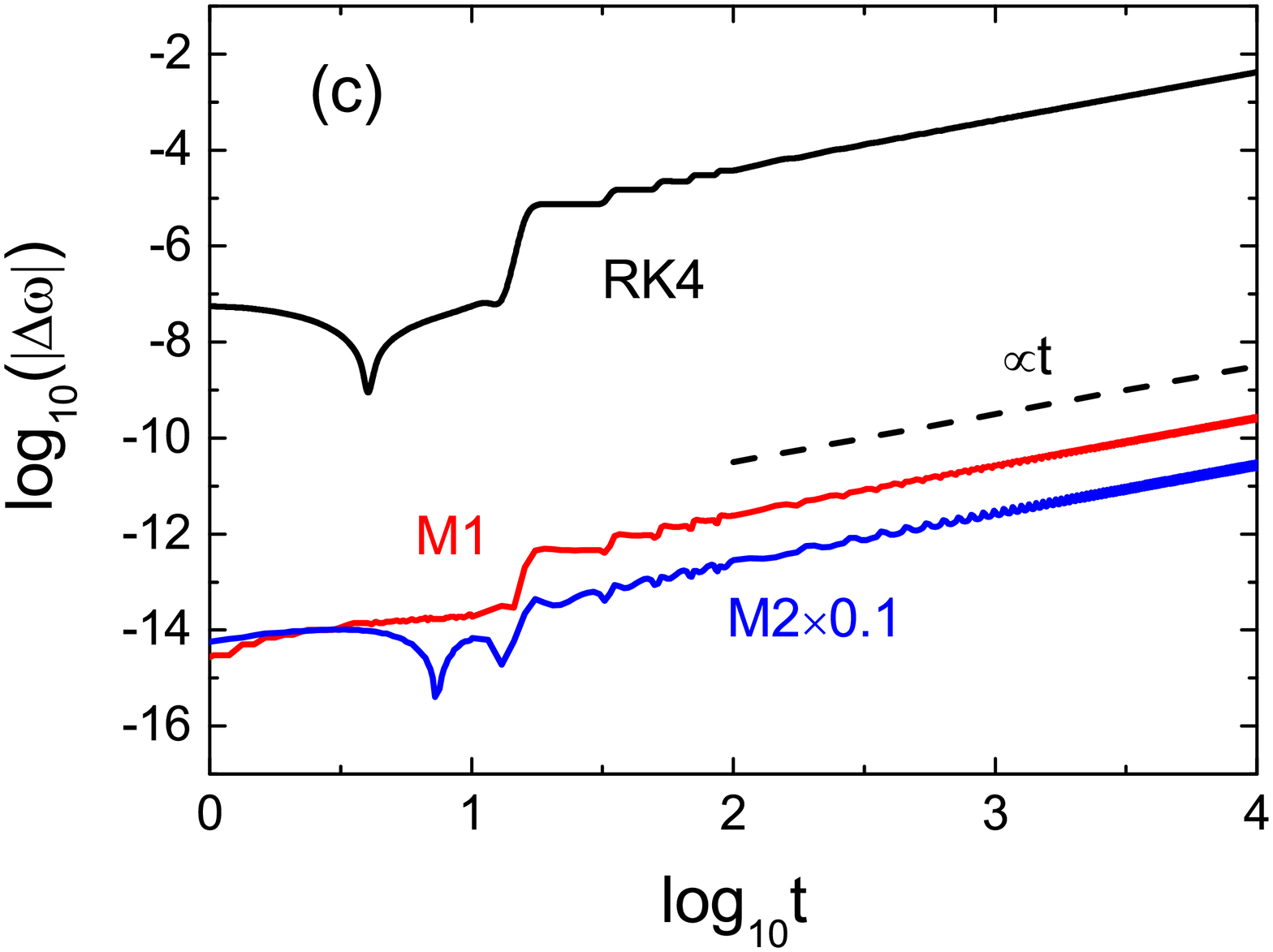}
  \includegraphics[scale=0.26]{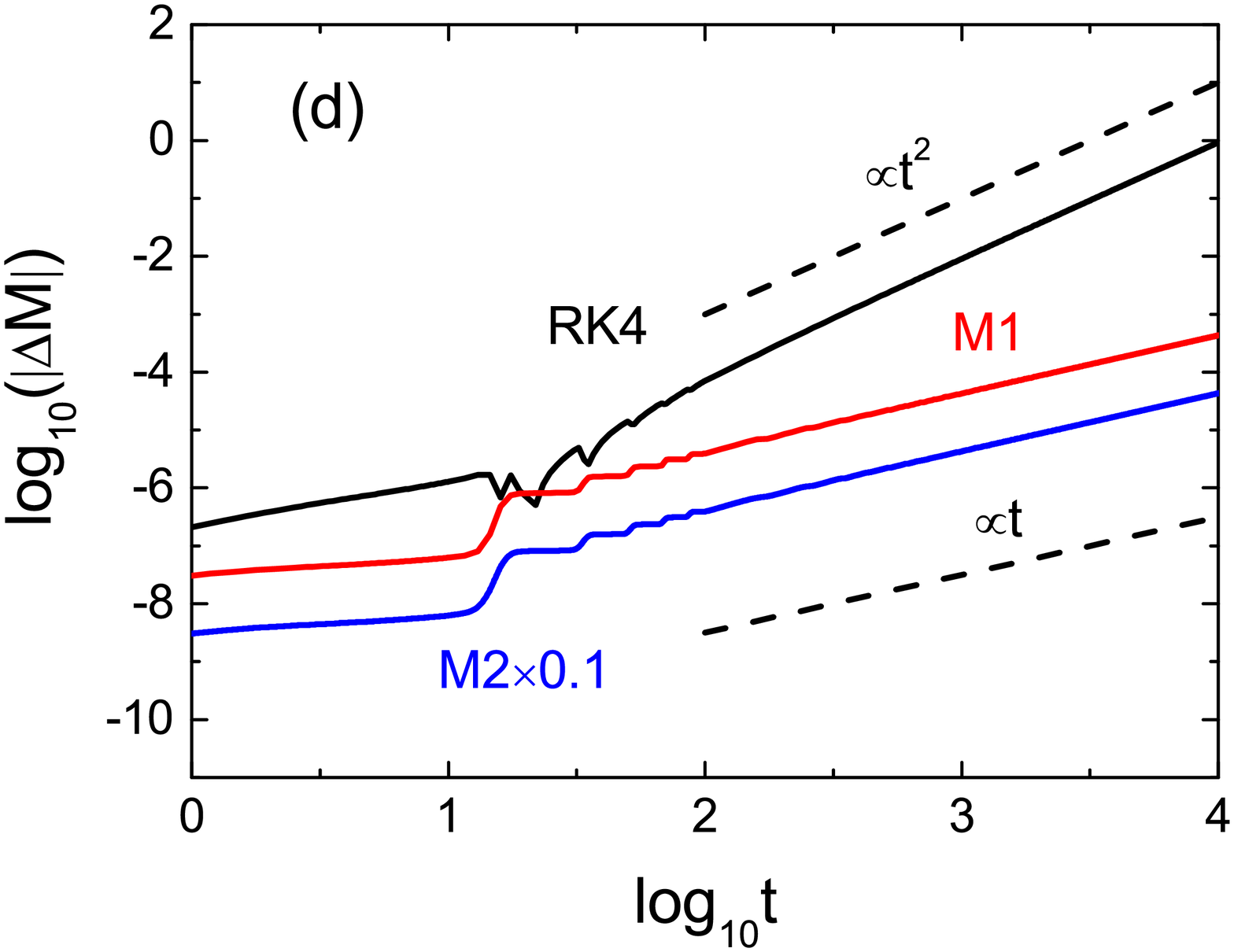}
 \caption{Errors of elements $a$, $e$, $\omega$
 and $M$ for the post-Newtonian two-body problem. The accuracies of $a$, $e$ and
 $\omega$ for the two
 correction schemes M1 and M2 are typically improved by 6 and 7 orders of
 magnitude, but cannot remain on the machine precision because the post-Newtonian effect
 makes the three elements vary with time.  The errors of $M$ are
 reduced by several orders of  magnitude.
}} \label{fig3}
\end{figure*}

\begin{figure*}
\center{
\includegraphics[scale=0.25]{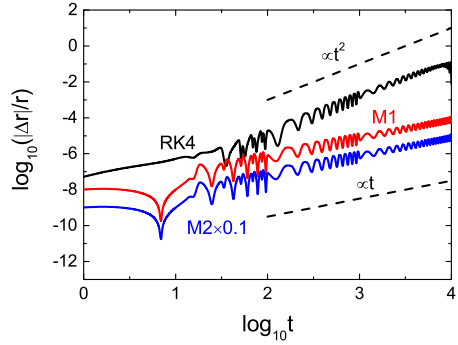}
\caption{Same as Figure 3 but relative position errors are used
rather than the errors of the elements. The error growth is linear
for M1 and M2, but is a quadratic function of time for RK4.}}
\label{fig4}
\end{figure*}

\begin{figure*}
\center{
  \includegraphics[scale=0.25]{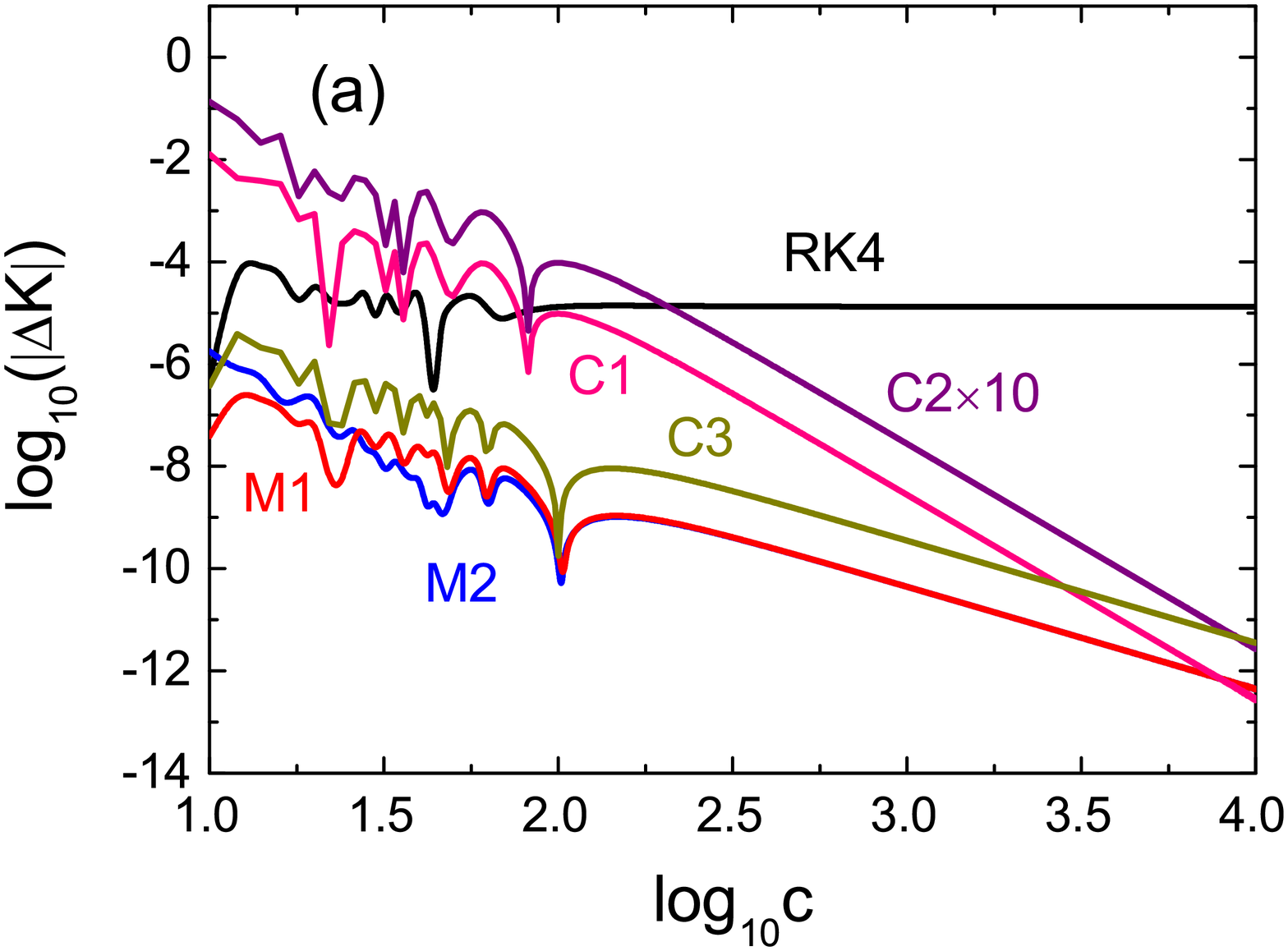}
  \includegraphics[scale=0.25]{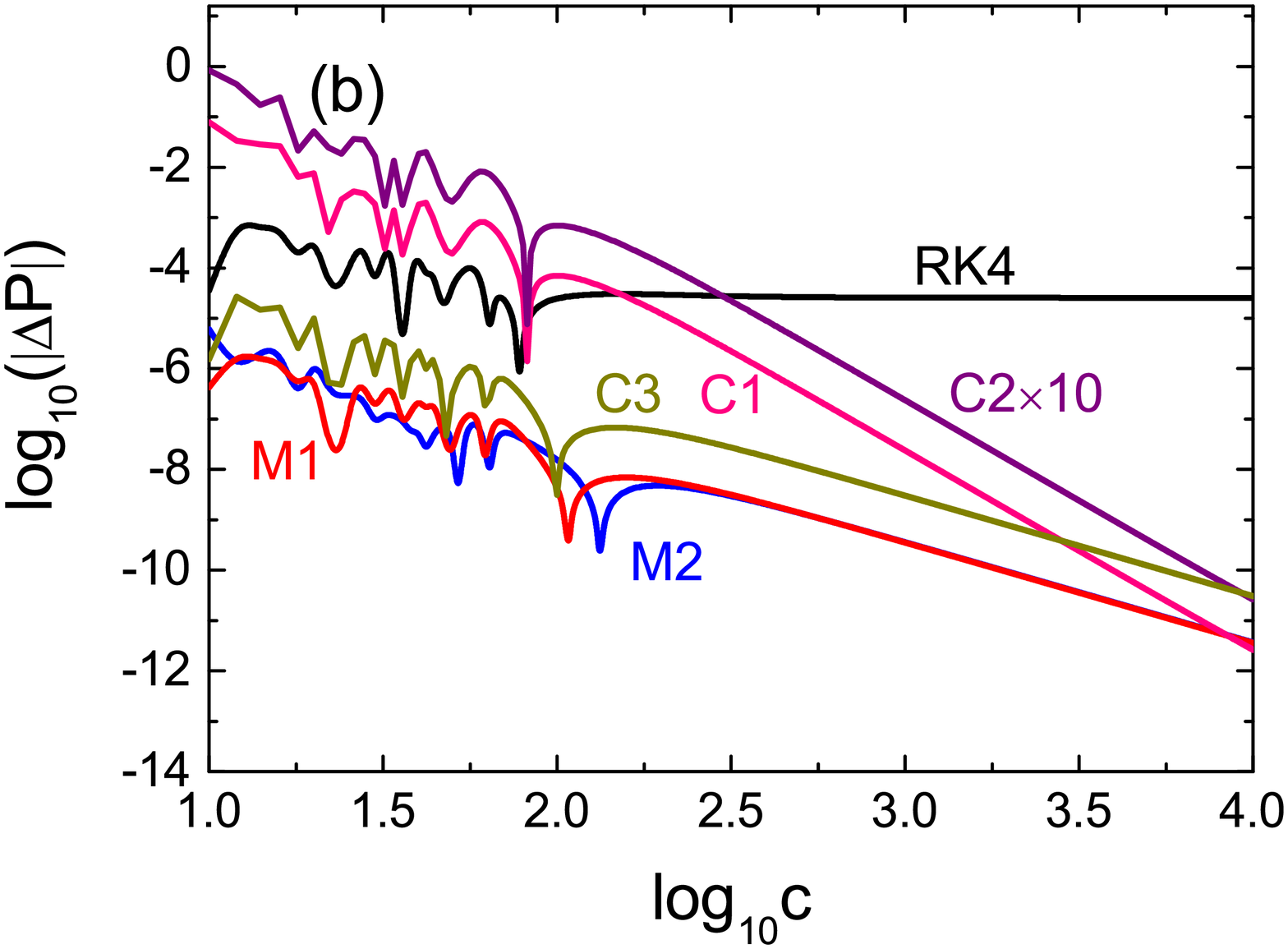}
  \includegraphics[scale=0.25]{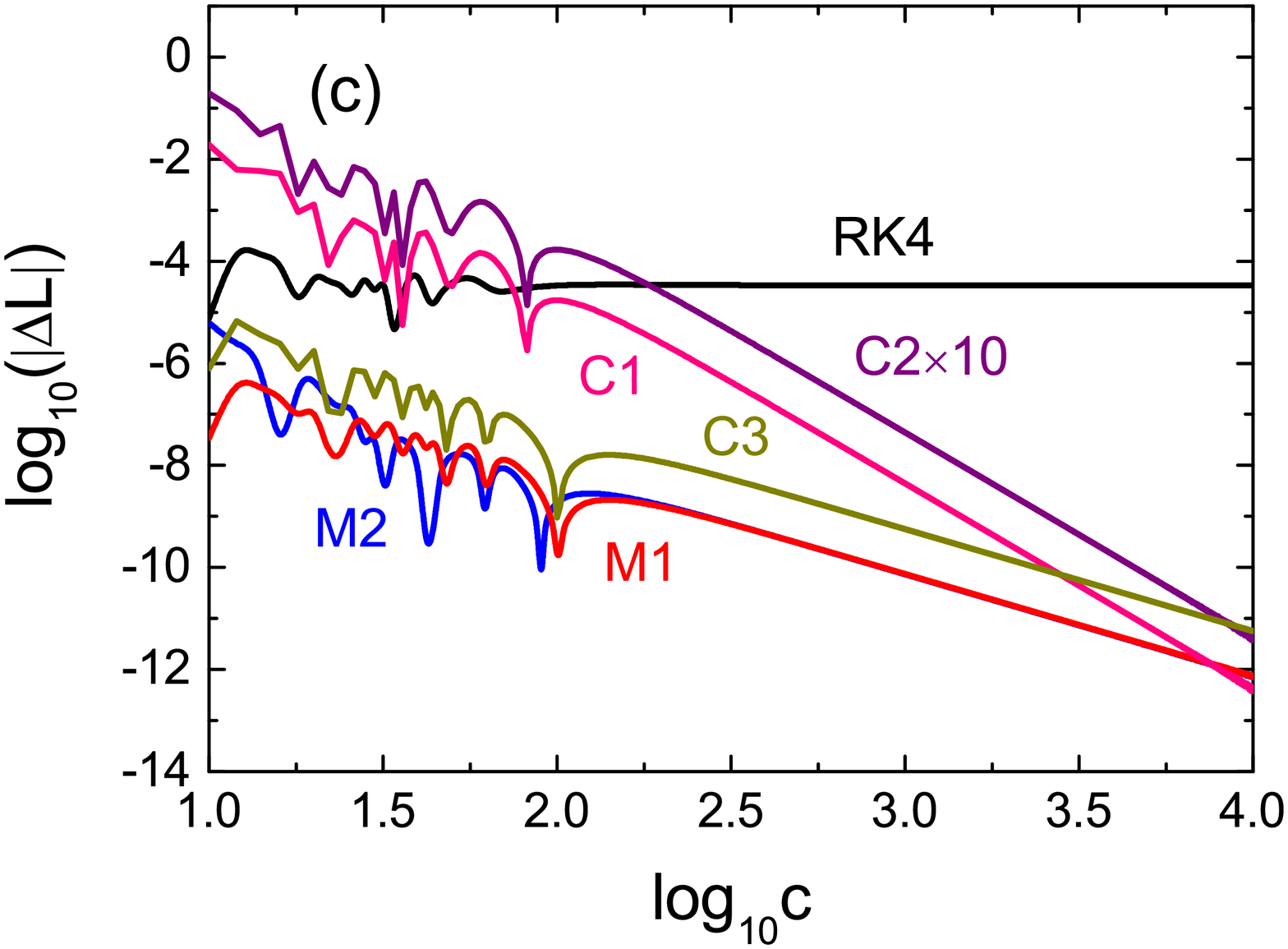}
  \includegraphics[scale=0.25]{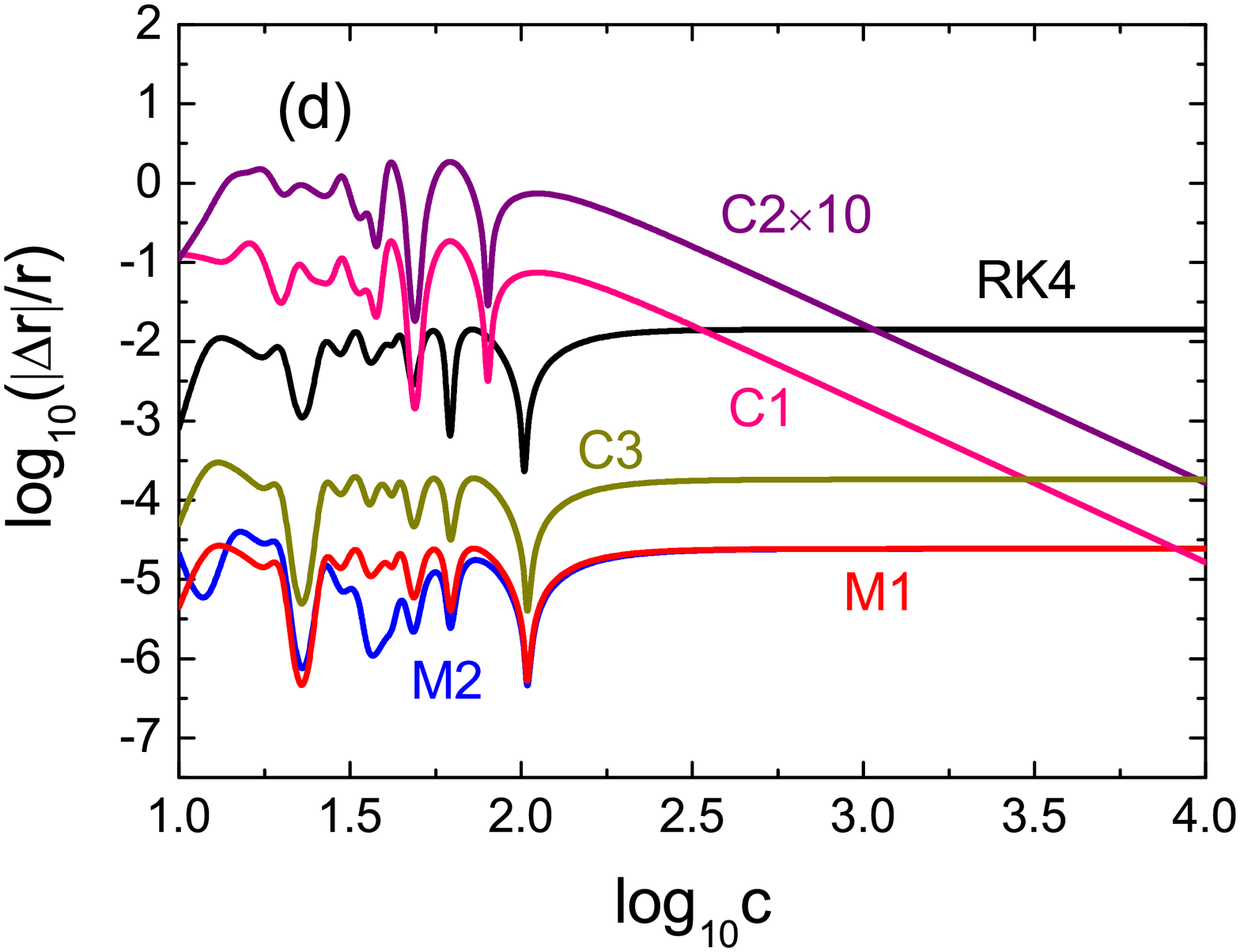}
\caption{Dependence of errors of $K$, $|\textbf{L}|$,
$|\textbf{P}|$ and $|\textbf{r}|$ on the value of $c$ in the
post-Newtonian two-body problem. The variation of $c$ indicates
that of the  perturbation. Four choices labelled as C1, C2, C3 and
M1 can be used for the calculation of the eccentric anomaly. M1 is
the best choice for various values of $c$.}} \label{fig5}
\end{figure*}

\begin{figure*}
\center{
  \includegraphics[scale=0.185]{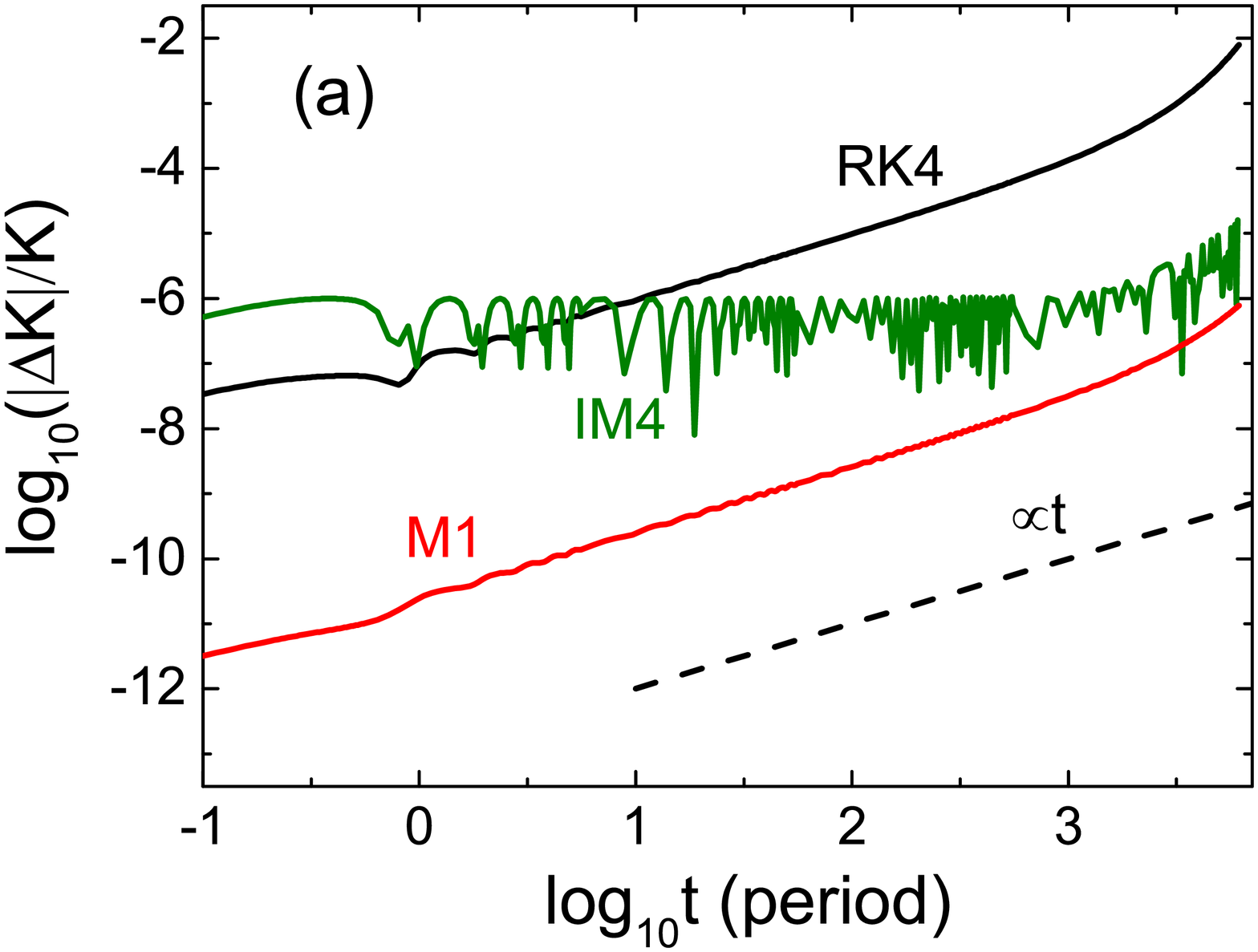}
  \includegraphics[scale=0.185]{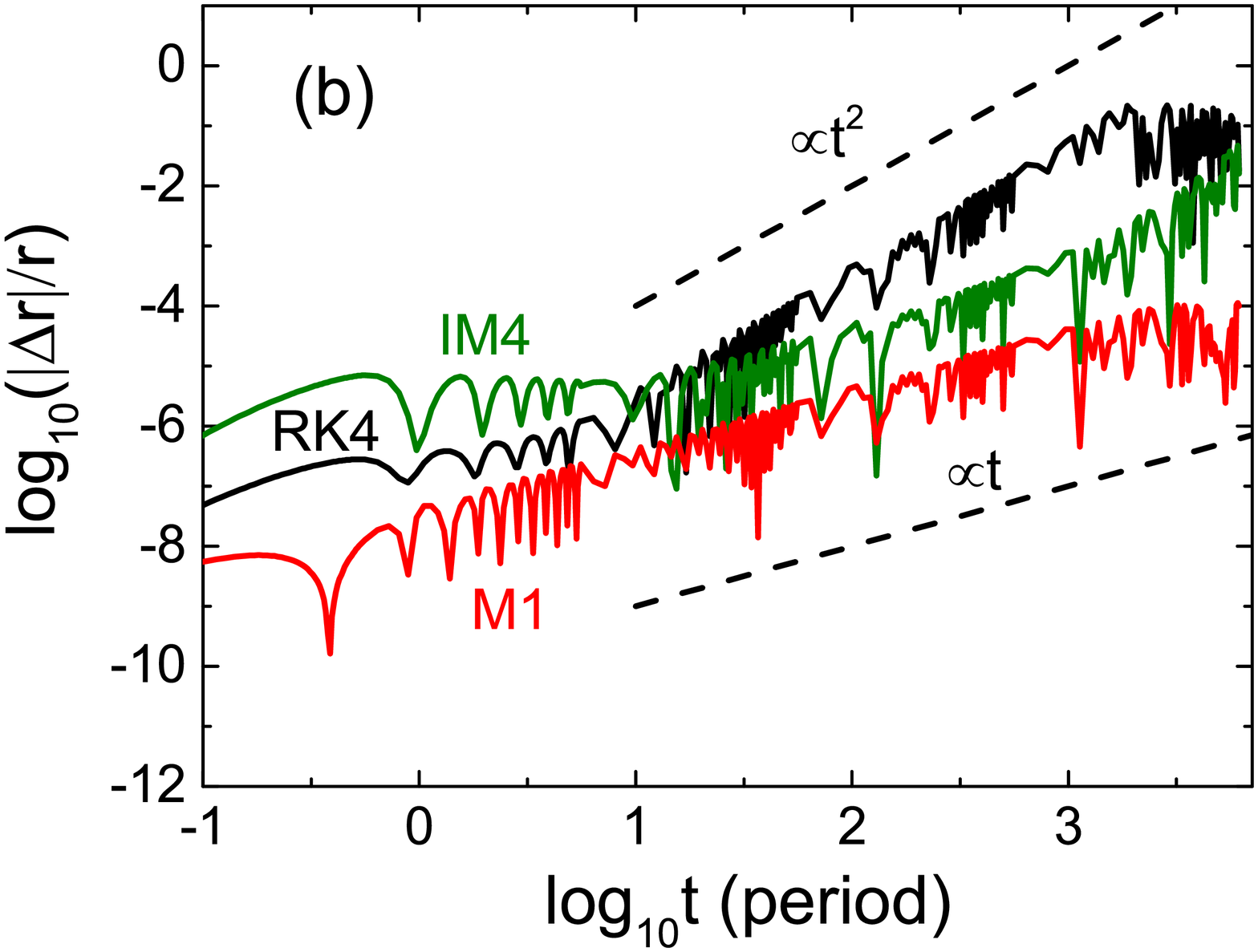}
  \includegraphics[scale=0.185]{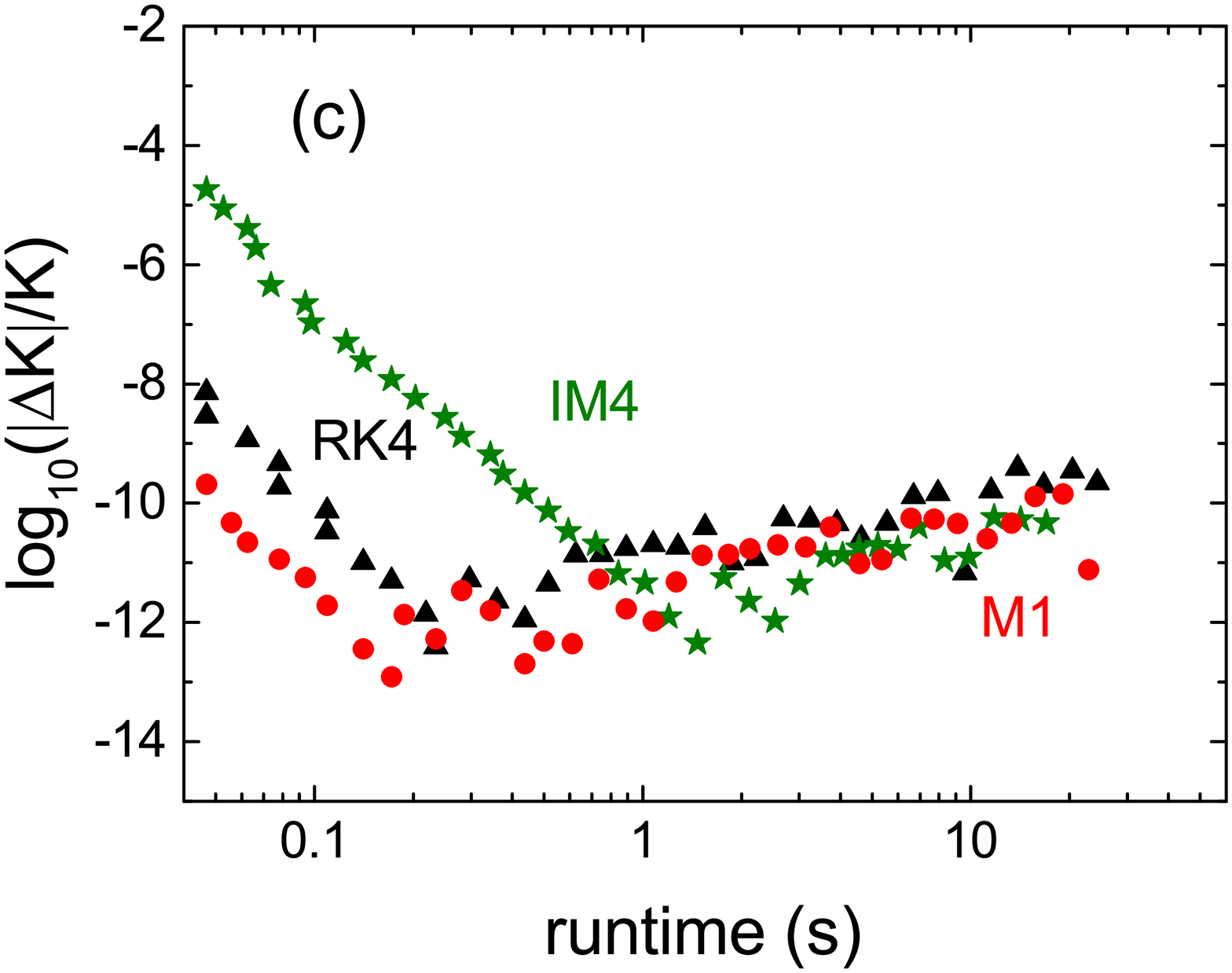}
\caption{Relative errors of the Kepler energy (a) and  position
(b) in the dissipative two-body problem. The parameters, initial
conditions and step-size are the same as those (except $e=0.3$) in
Figure 3. M2 is not considered, and the fourth-order implicit
symplectic method IM4 is used for comparison. When the time is
measured in units of unperturbed period $T$ in panels (a) and (b),
the integration time $t=10^5$ is $5630~T$. (c) Efficiencies for
the description of dependence of relative energy errors on the
computer runtime. Each point corresponds to the energy error after
integration time $t=10^4$ (i.e. $563~T$). For a given runtime, the
three algorithms use different fixed step sizes. }} \label{fig6}
\end{figure*}

\begin{figure*}
\center{
  \includegraphics[scale=0.25]{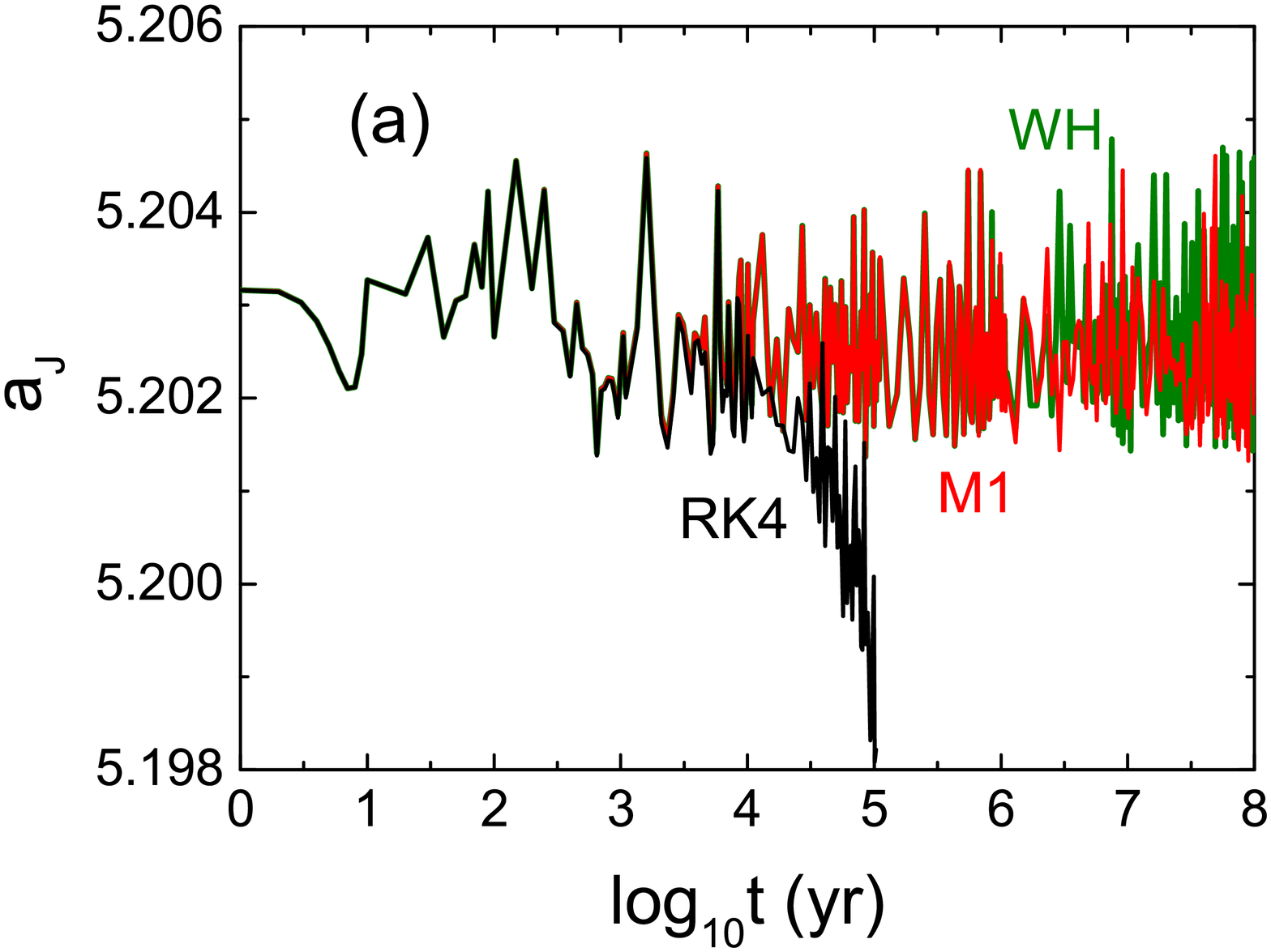}
  \includegraphics[scale=0.25]{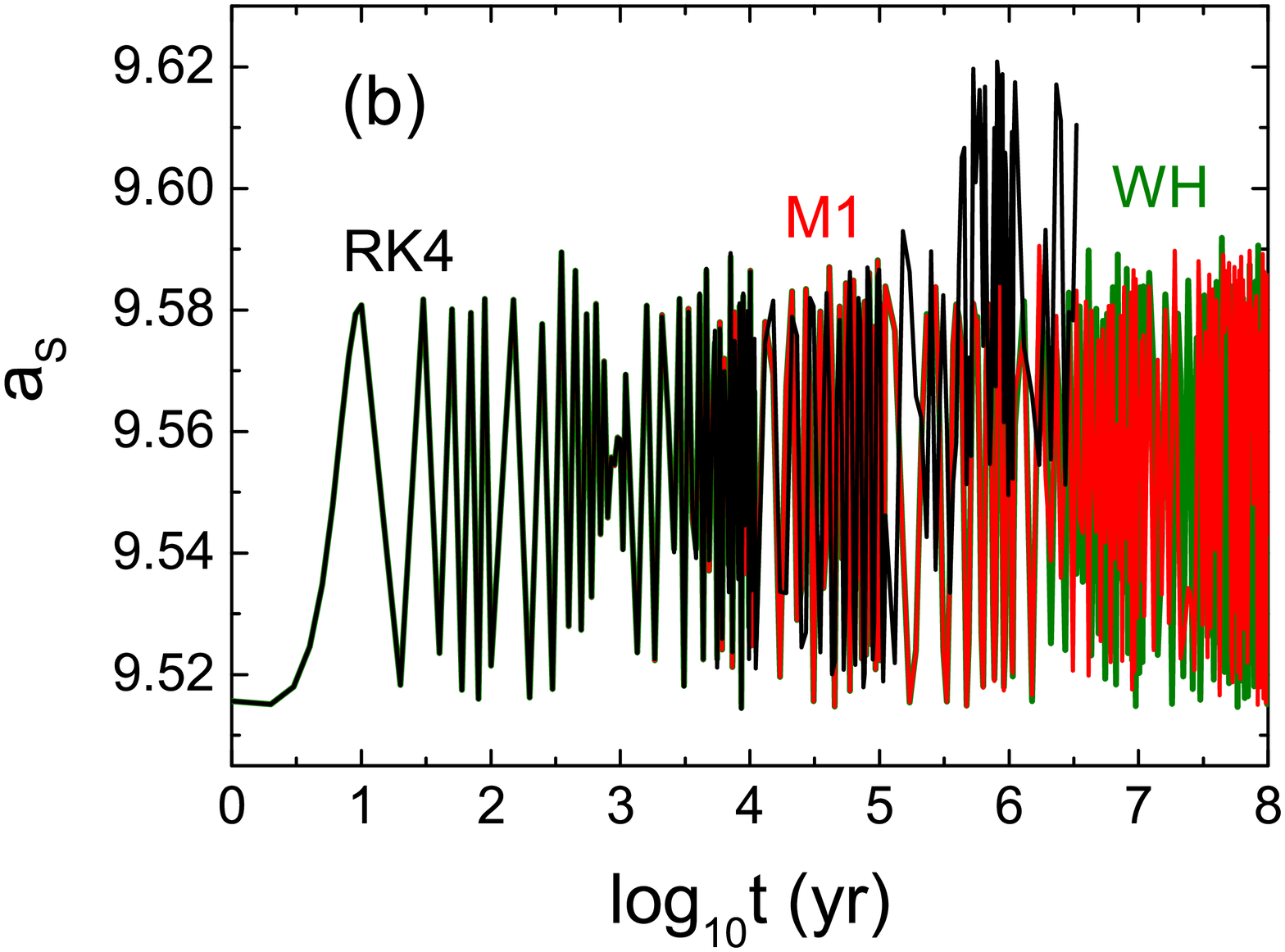}
  \includegraphics[scale=0.25]{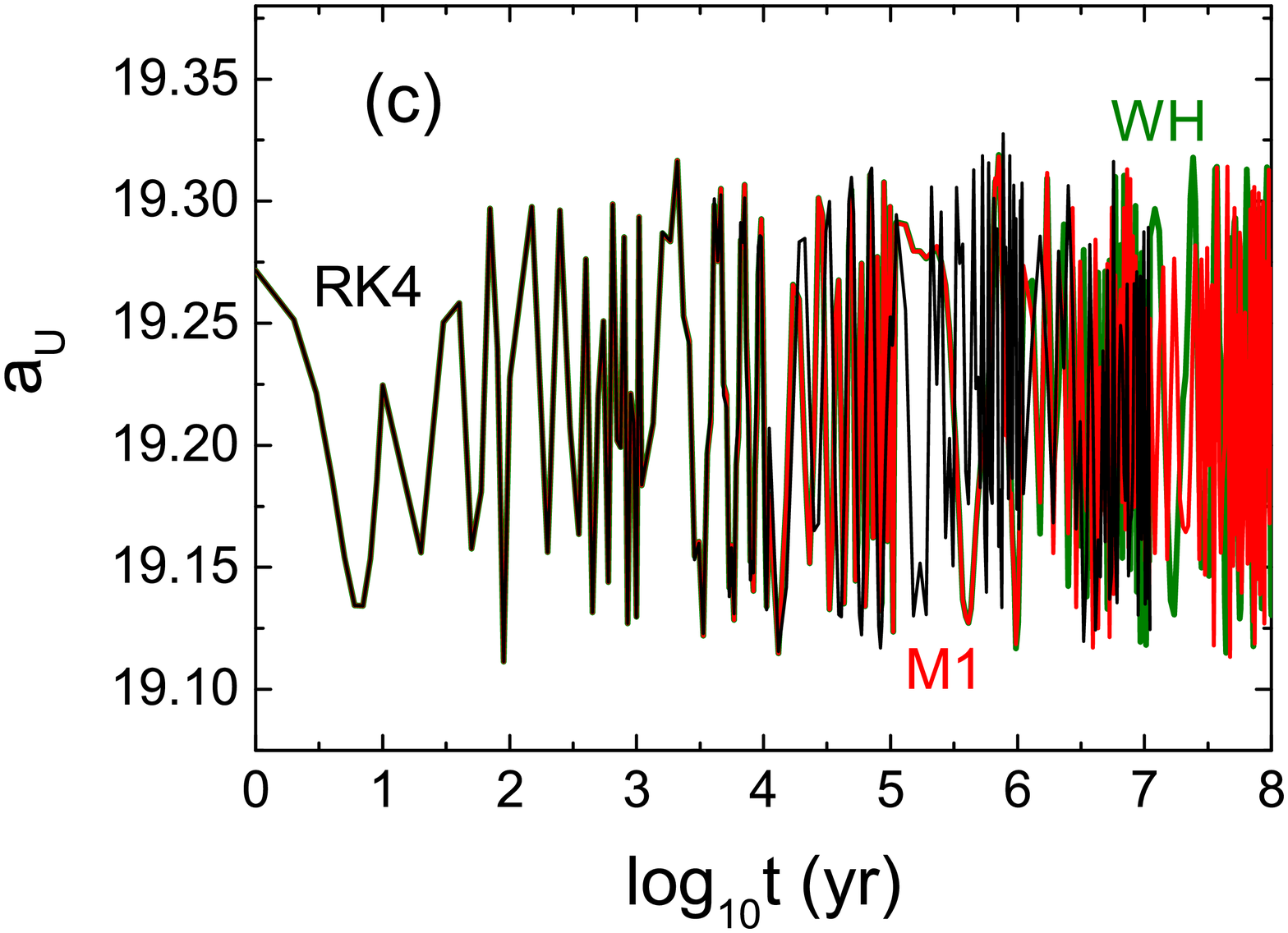}
  \includegraphics[scale=0.25]{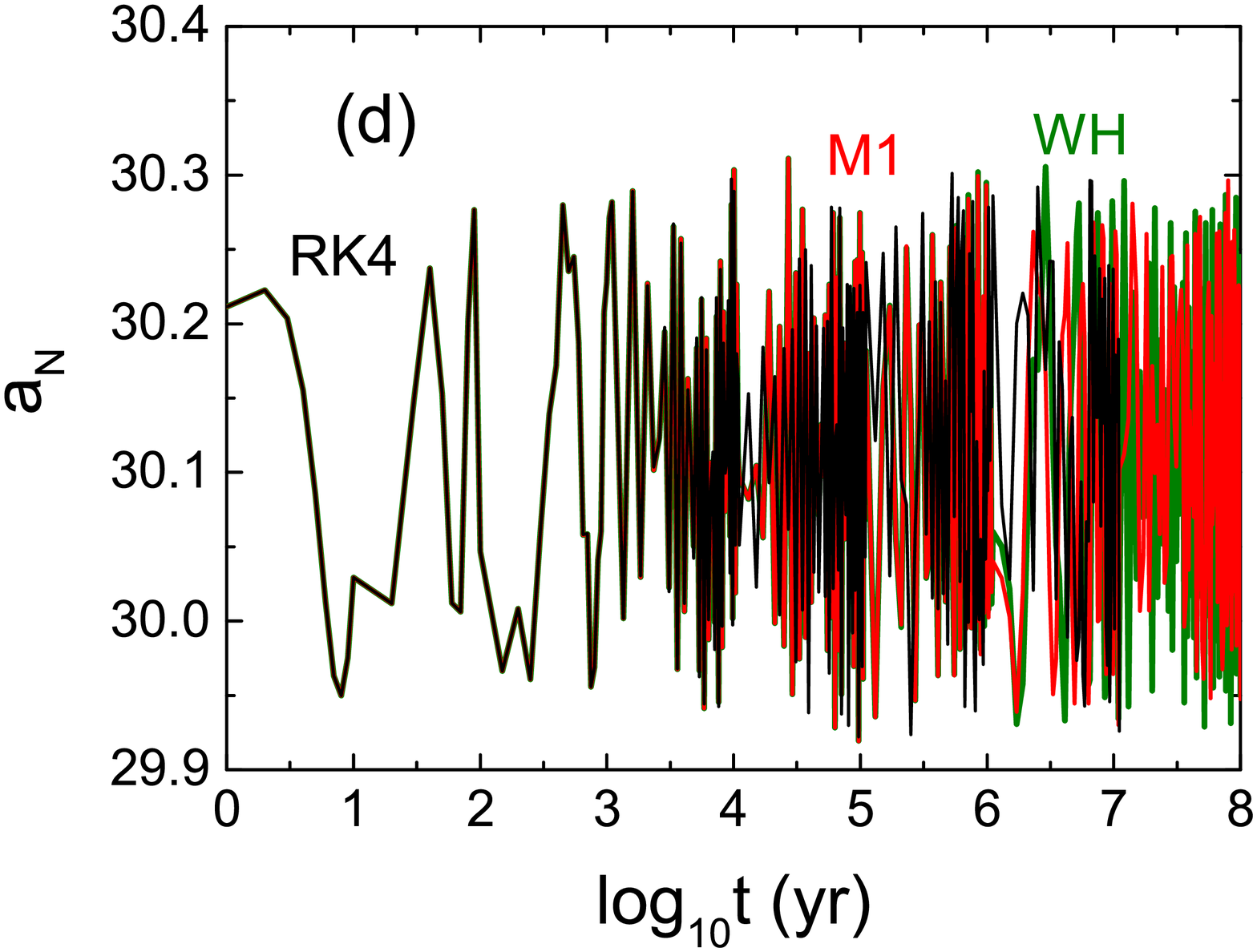}
  \includegraphics[scale=0.25]{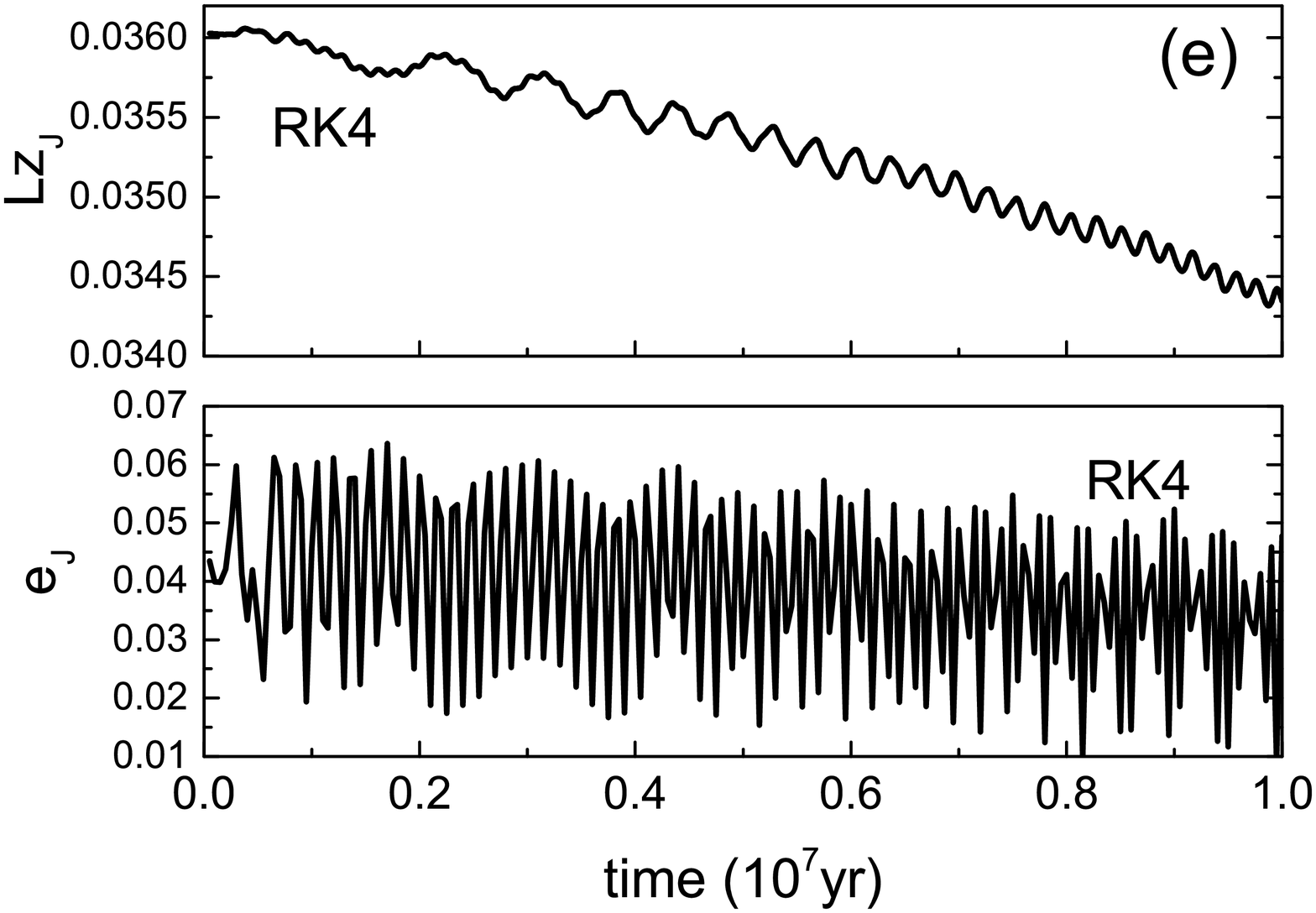}
  \includegraphics[scale=0.25]{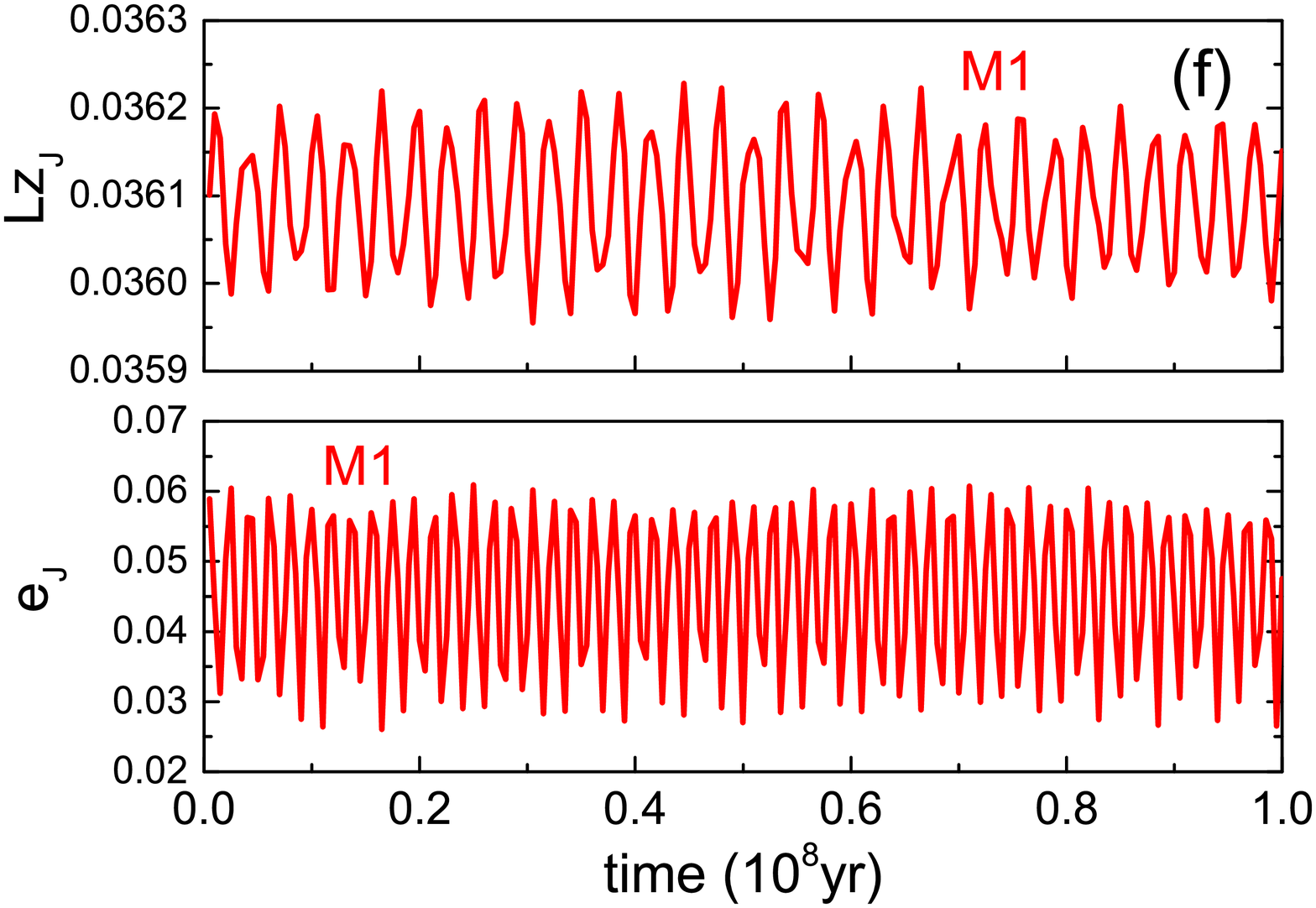}
\caption{Evolution of some orbital elements and quasi-integrals in
the five-body problem of the Sun and outer planets. The algorithms
are RK4 and its correction method M1. The WH method is  used for
comparison. Semi-major axis $a$ of Jupiter in panel (a) and
eccentricity $e$ and  $z$-direction angular momentum $L_z$ of
Jupiter in panel (e) decay with time for RK4 but remain bounded by
M1 (similar to WH) until integration time $t$ reaches $10^{8}$
years in panels (a) and (f). }} \label{fig7}
\end{figure*}

\begin{figure*}
\center{
  \includegraphics[scale=0.25]{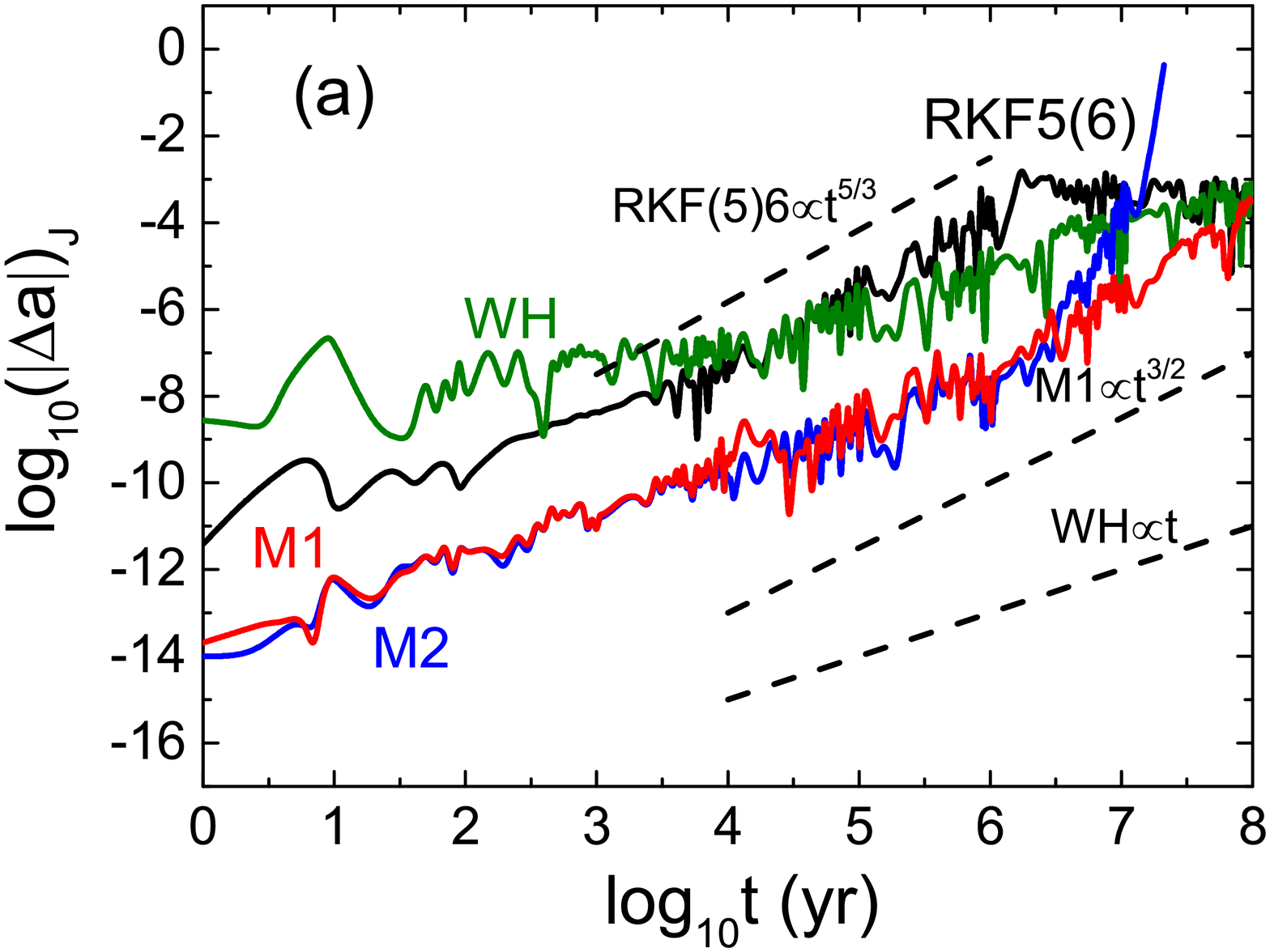}
  \includegraphics[scale=0.25]{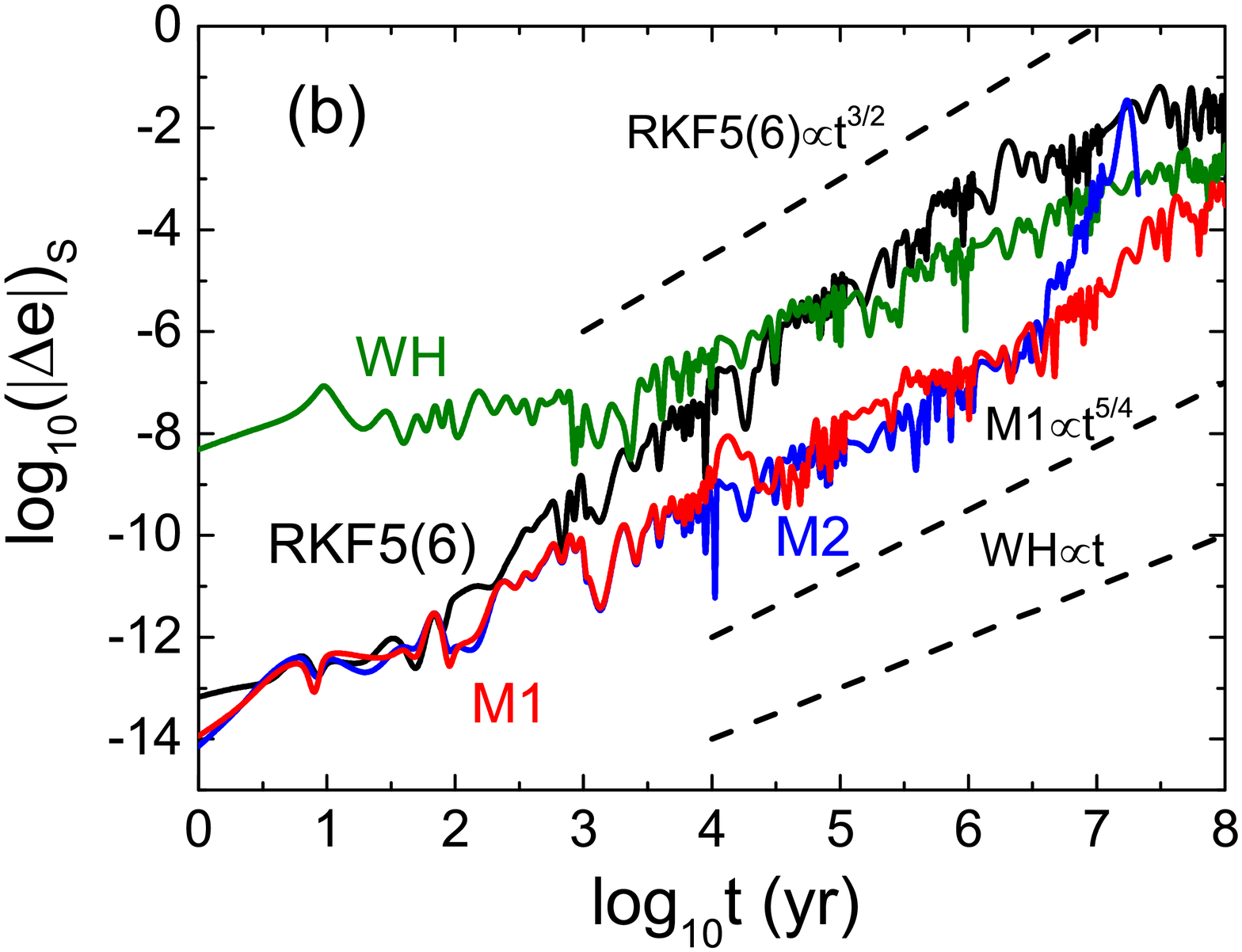}
  \includegraphics[scale=0.25]{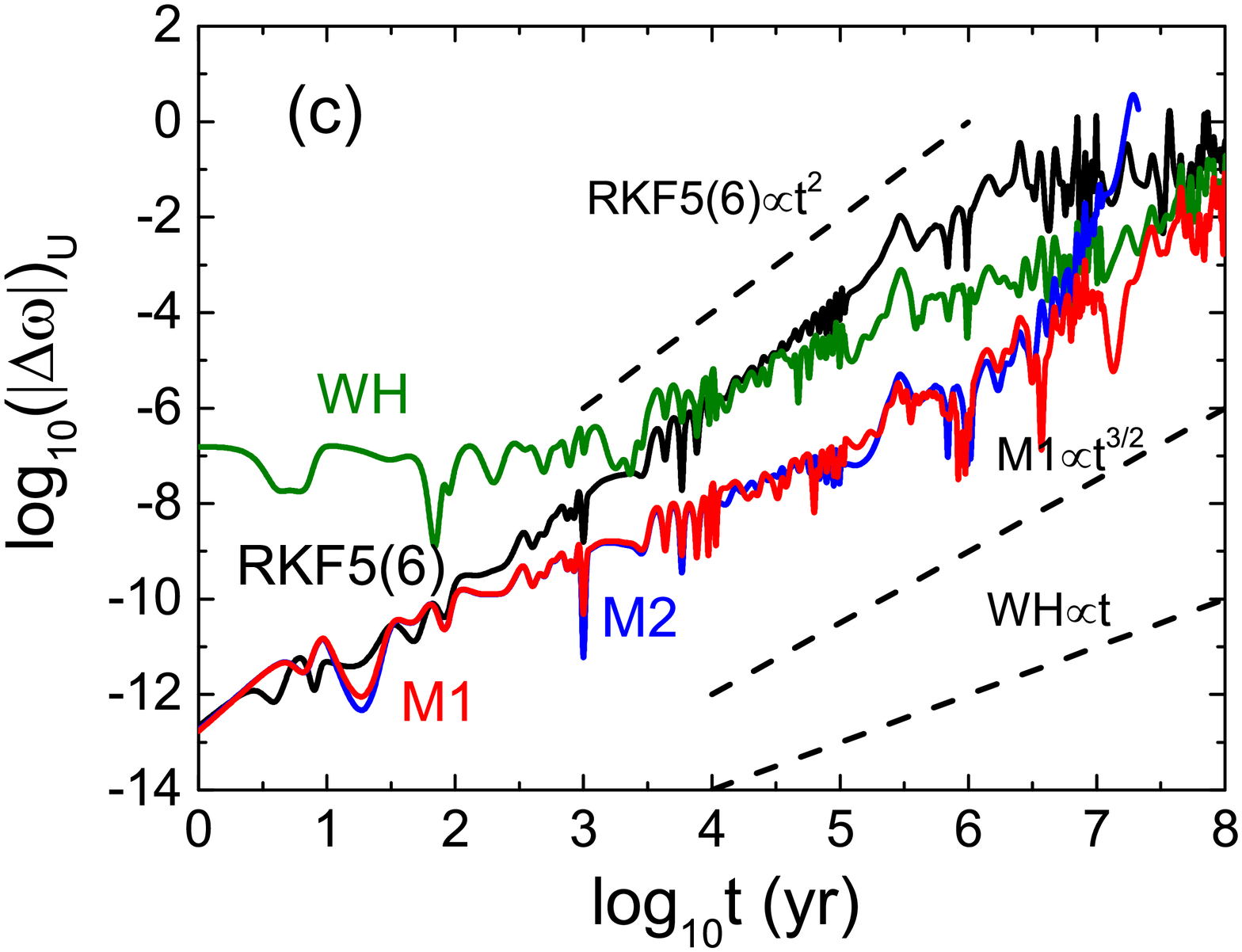}
  \includegraphics[scale=0.25]{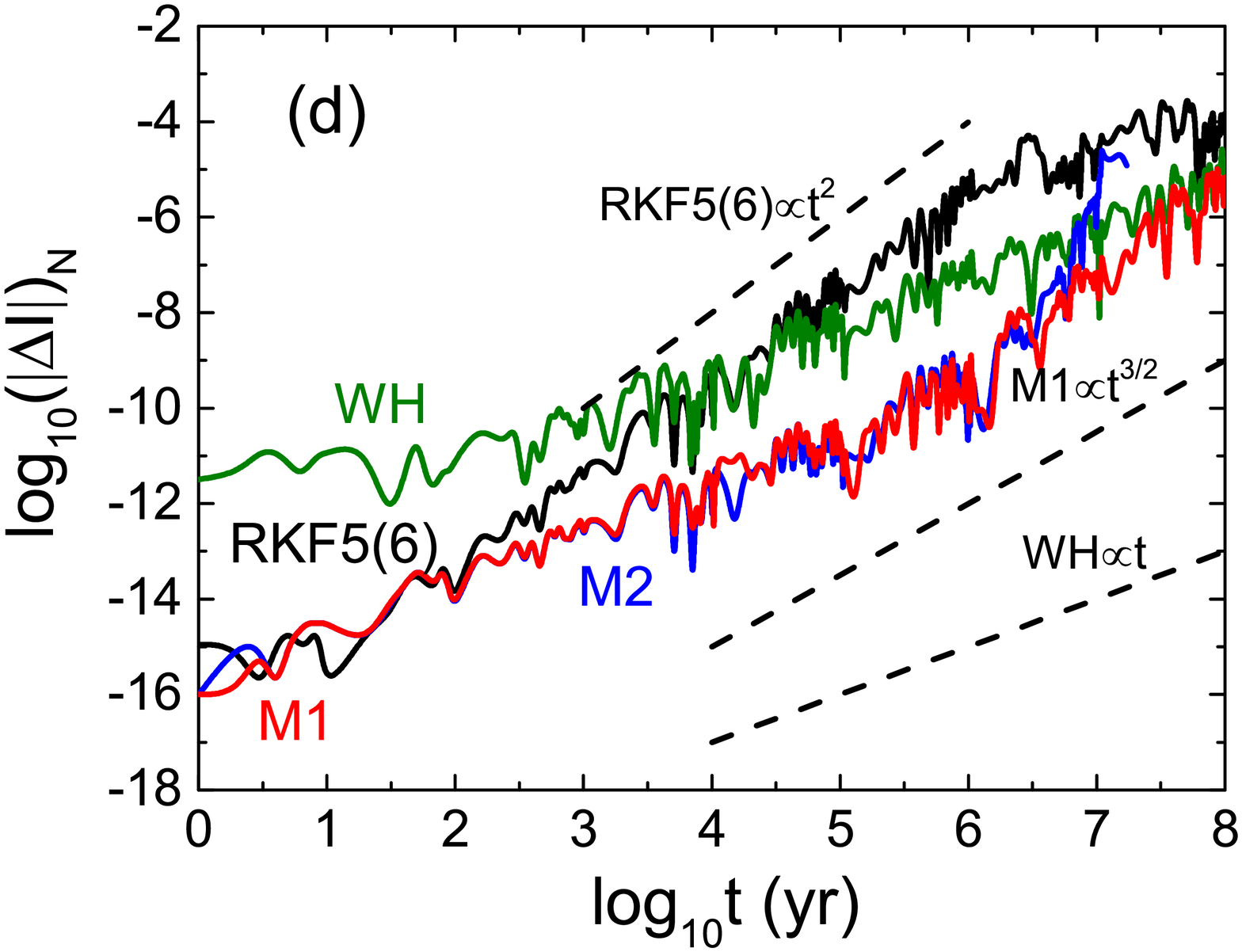}
\caption{Same as Figure 7 but errors of some orbital elements for
the other outer planets. RKF5(6) is used rather than RK4, and
Fukushima's method M2 is added. RKF8(9) is taken as a reference
integrator to provide high-precision solutions. (a) Semi-major
axis $a$ of Jupiter, (b)  eccentricity $e$ of Saturn, (c) argument
of perihelion $\omega$ of Uranus and (d)  inclination $I$ of
Neptune. When $t\approx 10^{6}$ years, M2 begins to become worse.
At this time, errors $\Delta a$ of the semi-major axis of Jupiter
remain invariant at  0.001 for RKF5(6) in panel (a), whereas they
tend to this value for M1 and WH as $t$ arrives at $10^{8}$ years.
}} \label{fig8}
\end{figure*}

\begin{figure*}
\center{
  \includegraphics[scale=0.25]{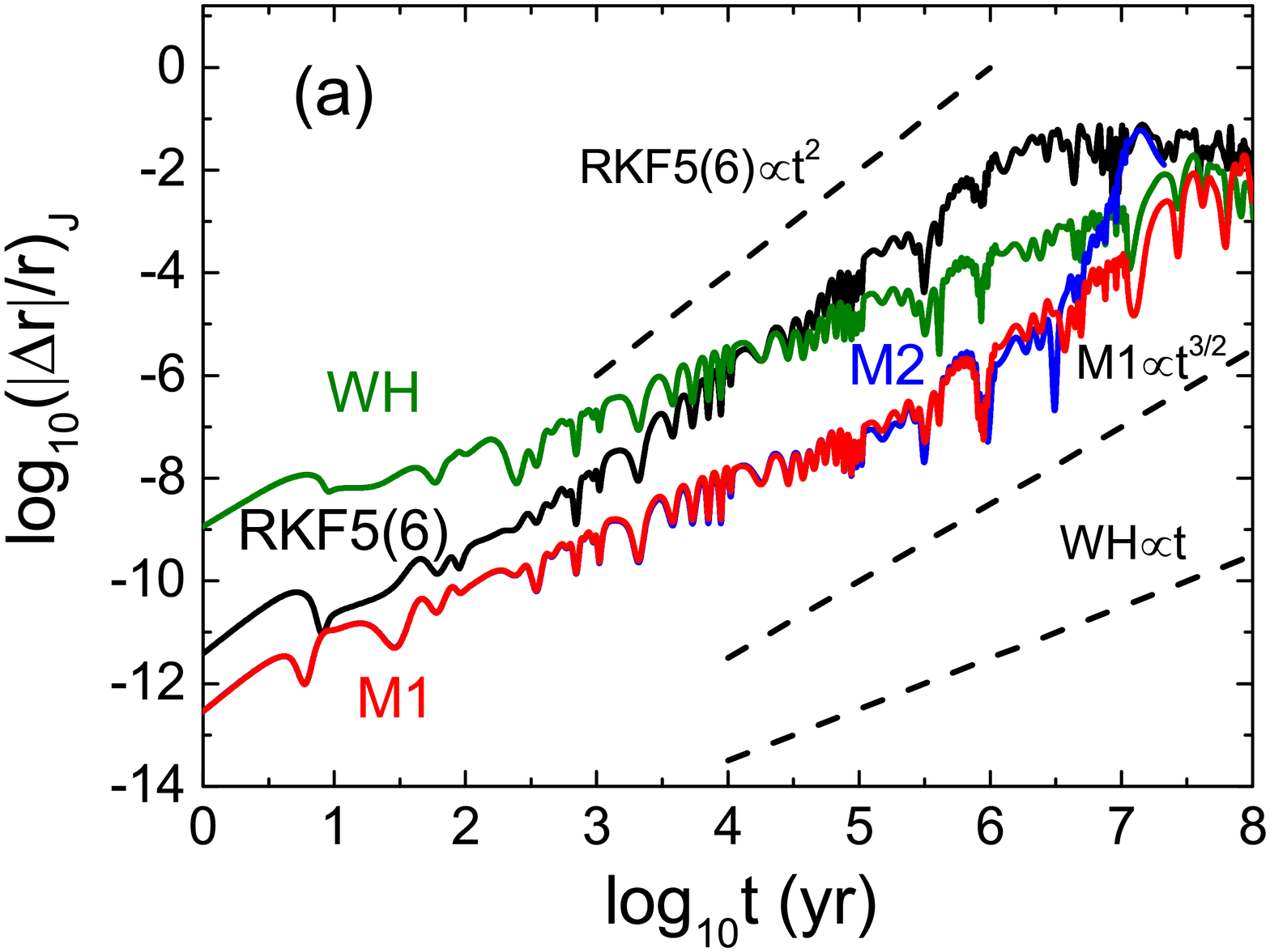}
  \includegraphics[scale=0.25]{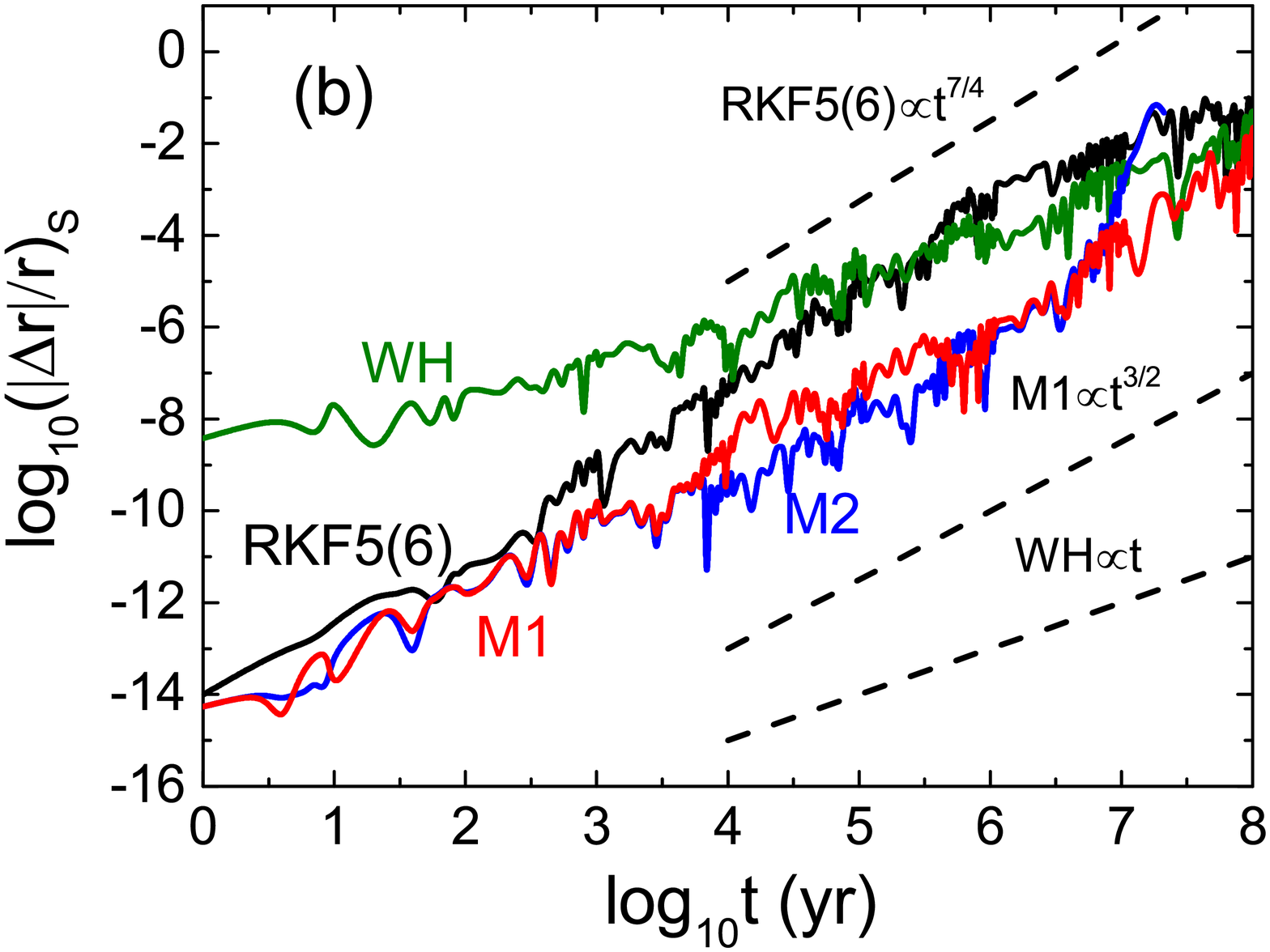}
  \includegraphics[scale=0.25]{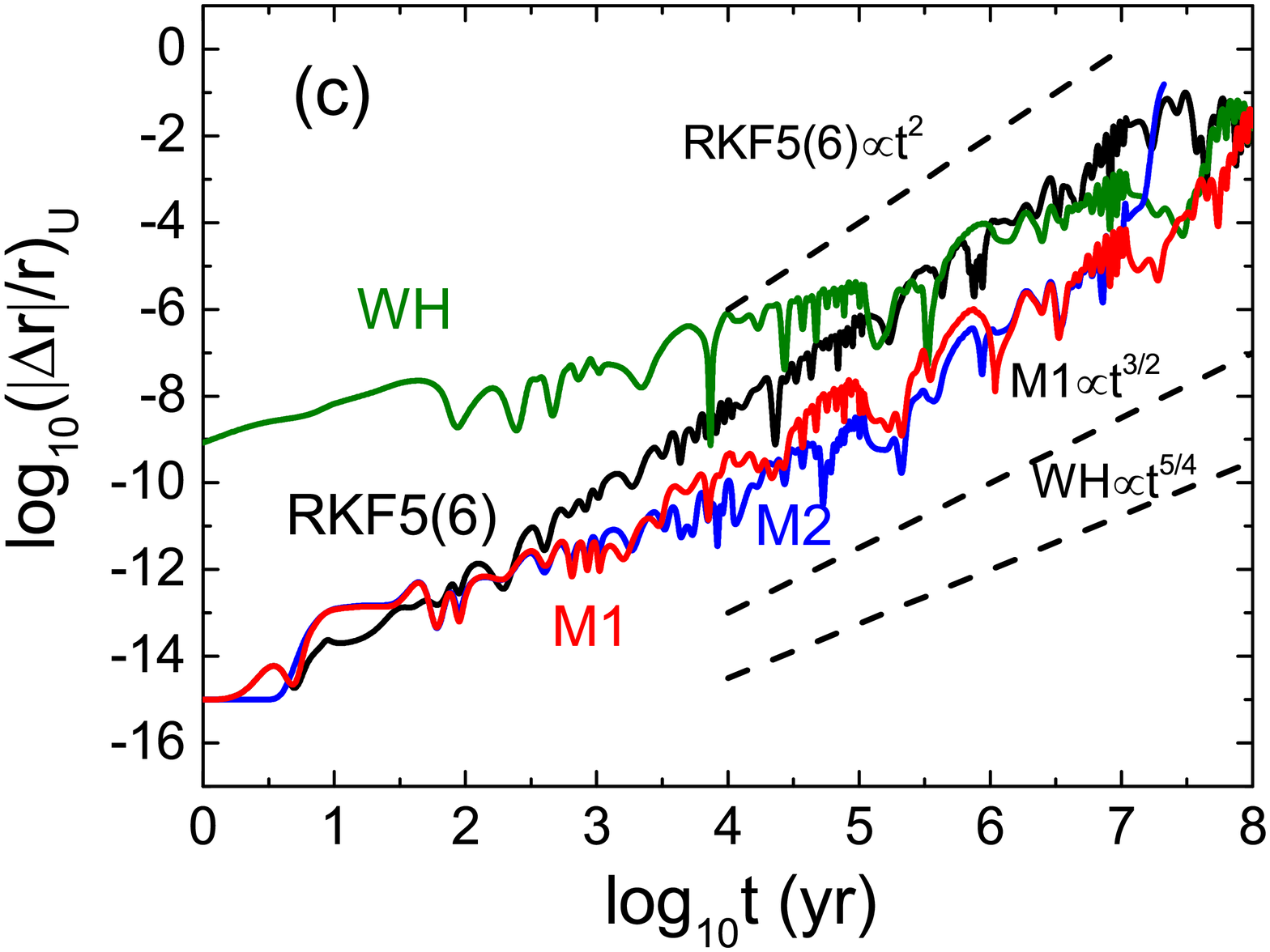}
  \includegraphics[scale=0.25]{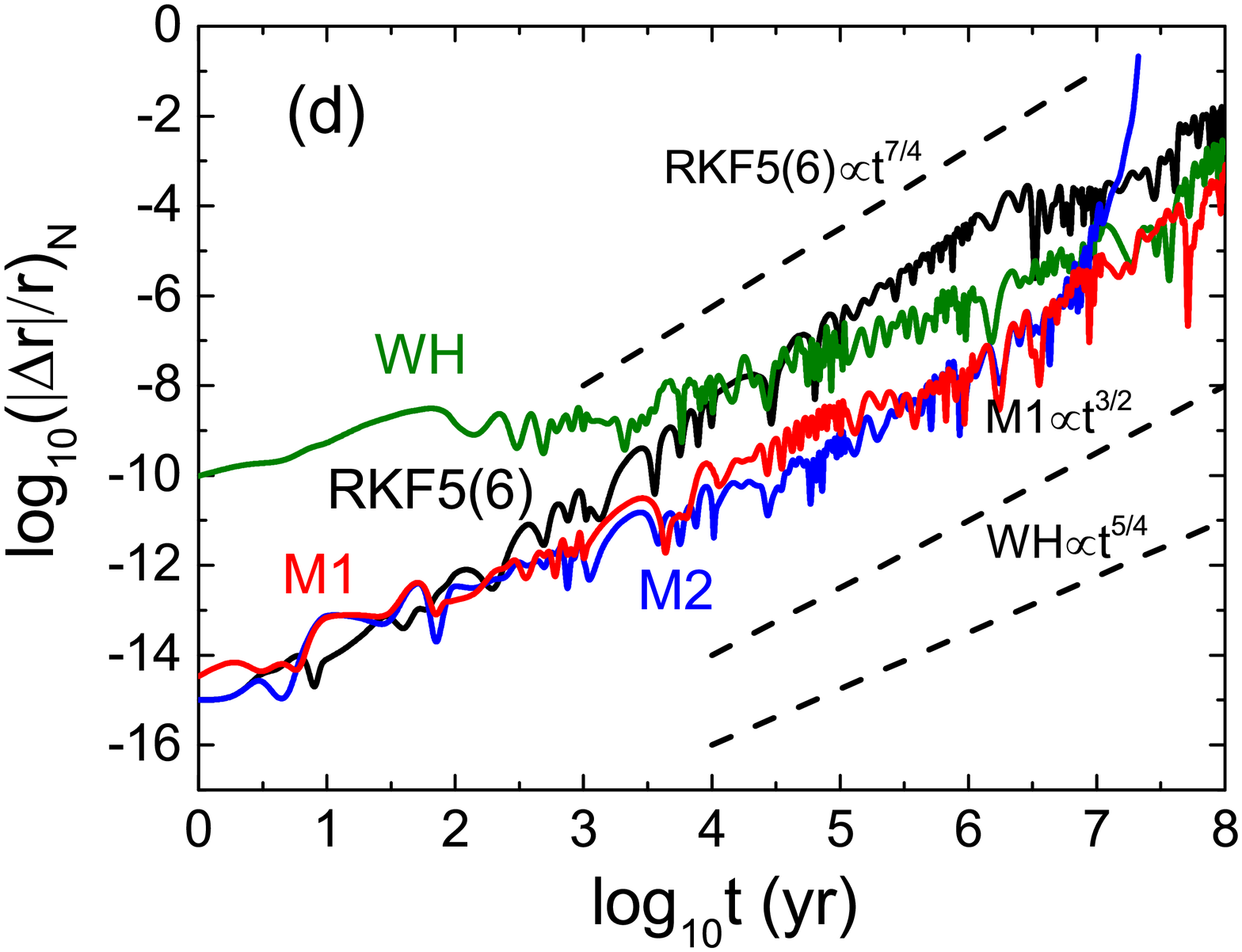}
\caption{Same as Figure 8 but relative position errors of the
outer planets. (a) Jupiter, (b) Saturn, (c) Uranus and (d)
Neptune. When $t\approx 10^{6}$ years, M2 begins to become worse.
At this time, the relative position errors in panel (a) are
stabilised at  0.01 for RKF5(6). The other errors also tend to
this value for RKF5(6), M1 and WH as $t$ reaches $10^{8}$ years.}}
\label{fig9}
\end{figure*}

\begin{figure*}
\center{
  \includegraphics[scale=0.25]{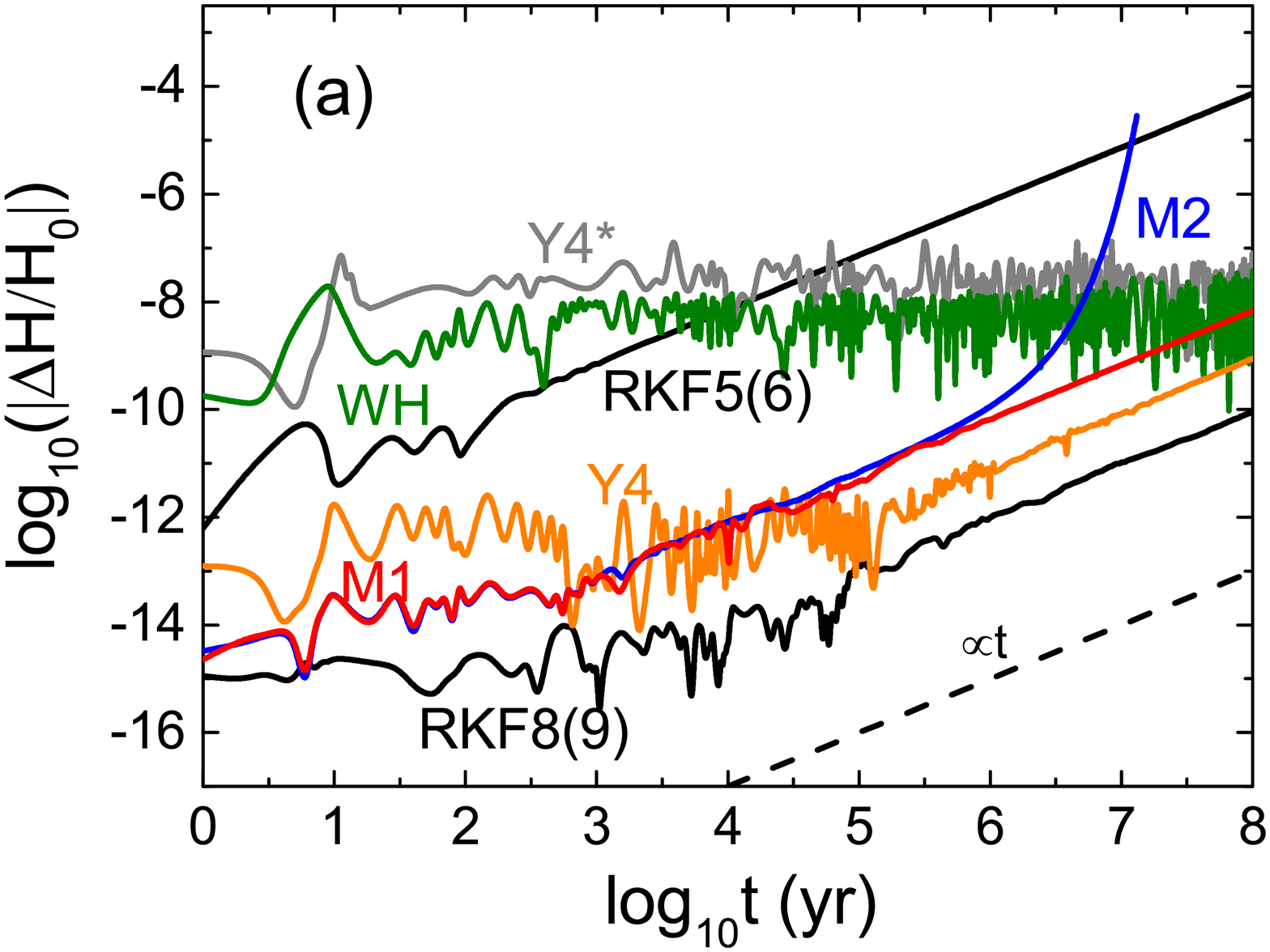}
  \includegraphics[scale=0.25]{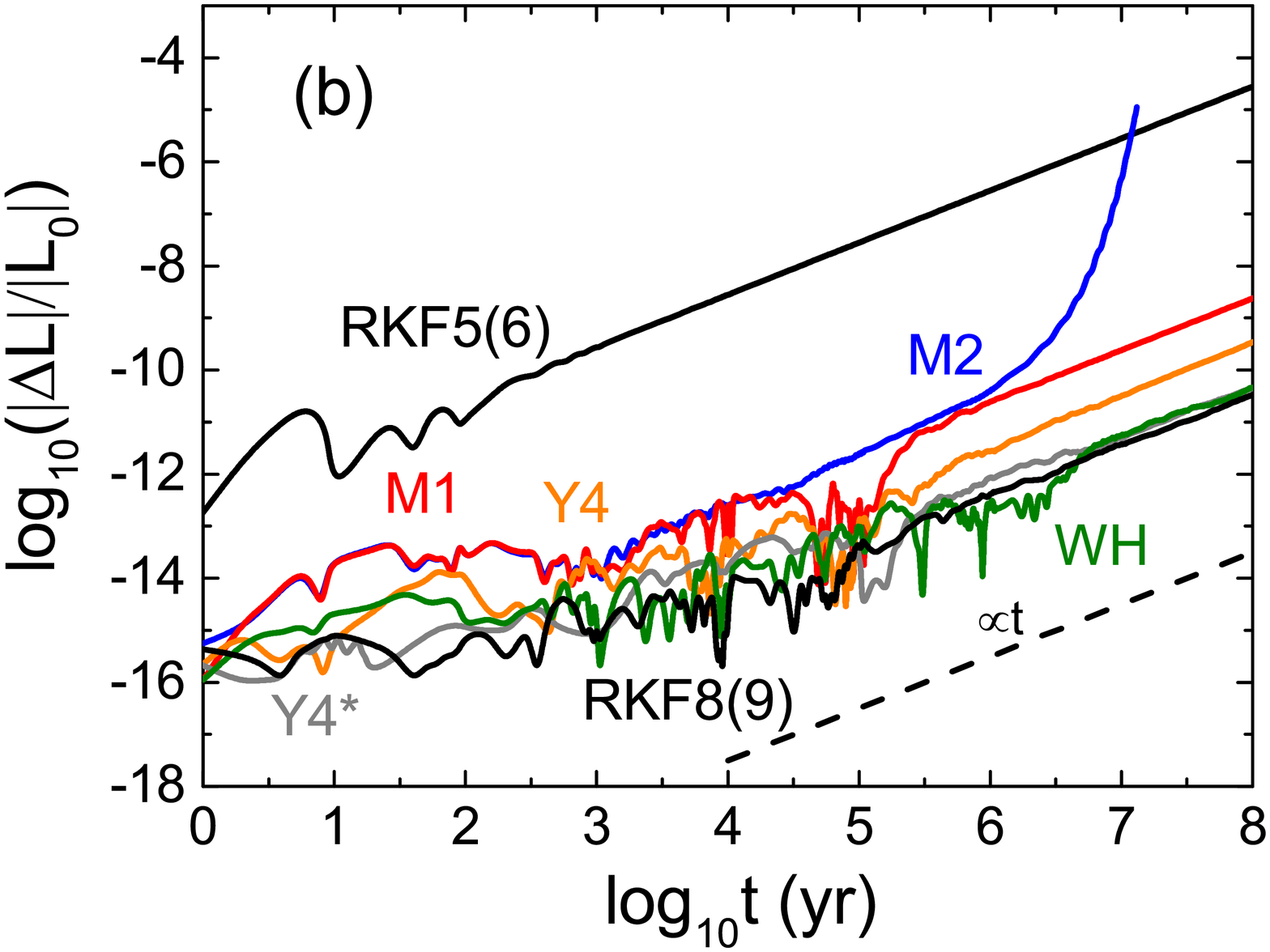}
\caption{Same as Figure 8 but relative errors of  total energy and
total angular momenta in the barycentre coordinate system. Y4 is
the fourth-order symplectic method of Yoshida, consisting of three
WH methods. WH, Y4, RKF5(6) and its projection methods M1 and M2
use a small time step $h=36.525$ days, whereas Y4* adopts a large
time step $h$*=350 days. The WH method conserves the total energy
but does not conserve the total angular momenta because of
roundoff errors. The energy errors are not bounded for Y4 adopting
the small time-step, whereas are for Y4* adopting the large
time-step. RKF8(9), M1 and M2 do not conserve the total energy. M2
begins to fail to work well when $t\approx 10^{6}$ years and is
unsuitable for such a long-term integration. M1 immensely controls
the rapid error growth compared with the uncorrected method
RKF5(6). }} \label{fig10}
\end{figure*}

\begin{figure*}
\center{
  \includegraphics[scale=0.25]{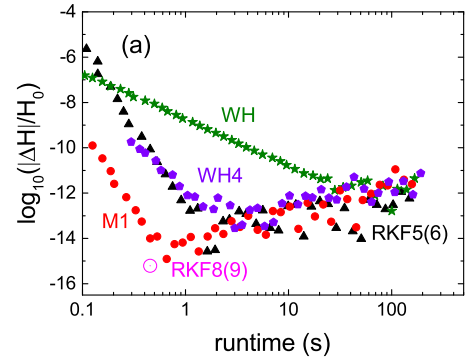}
 \caption{Same as Figure 10 but regarding efficiencies for several
 algorithms. Each point corresponds to the total energy error obtained after the integration time reaches $10^4$
years. The same runtime indicates that the algorithms (except
RKF8(9)) use different fixed step-sizes. RKF8(9) with varying
time-steps has only one point, corresponding to 0.45 s CPU times.
}} \label{fig11}
\end{figure*}


\label{lastpage}

\end{document}